\documentclass[12pt]{article}

\usepackage{amssymb}

\newcommand{\acl}{\ensuremath{\mathcal{A}_{cl}}}

\newcommand{\sca}{\ensuremath{\mathcal{A}}}
\newcommand{\scb}{\ensuremath{\mathcal{B}}}
\newcommand{\scd}{\ensuremath{\mathcal{D}}}
\newcommand{\scg}{\ensuremath{\mathcal{G}}}
\newcommand{\sch}{\ensuremath{\mathcal{H}}}
\newcommand{\scs}{\ensuremath{\mathcal{S}}}
\newcommand{\scx}{\ensuremath{\mathcal{X}}}
\newcommand{\scy}{\ensuremath{\mathcal{Y}}}

\newcommand{\sone}{\ensuremath{\mathcal{S}_{1}}}

\newcommand{\cinfm}{\ensuremath{C^{\infty}(M)}}
\newcommand{\xm}{\ensuremath{\mathcal{X}(M)}}

\newcommand{\sdera}{\ensuremath{SDer(\mathcal{A})}}
\newcommand{\za}{\ensuremath{Z(\mathcal{A})}}

\newcommand{\phst}{\ensuremath{\Phi_*}}
\newcommand{\phstup}{\ensuremath{\Phi^*}}

\newcommand{\oa}{\ensuremath{\mathcal{O}(\mathcal{A})}}

\newcommand{\omcl}{\ensuremath{\omega_{cl}}}

\newcommand{\gstar}{\ensuremath{\mathcal{G}^*}}

\begin{document}
\noindent \textbf{\large Supmech: the Geometro-statistical Formalism
Underlying Quantum Mechanics}

\vspace{.15in}\noindent \textbf{Tulsi Dass}

\noindent Indian Statistical Institute, Delhi Centre, 7, SJS
Sansanwal Marg, New Delhi, 110016, India. \\ Centre for Theoretical
Physics, Jamia Millia Islamia, Jamia Nagar, New Delhi-110025, India.

\noindent  E-mail: tulsi@isid.ac.in; tulsi@iitk.ac.in

\vspace{.18in}  \noindent \textbf{Abstract.} As the first step in an
approach to the solution of Hilbert's sixth problem, a general
scheme of mechanics, called `supmech', is developed integrating
noncommutative symplectic geometry and noncommutative probability
theory in an algebraic framework; it has quantum mechanics (QM) and
classical mechanics as special subdisciplines and facilitates an
autonomous development of QM and  satisfactory treatments of
quantum-classical correspondence and quantum measurements (including
a straightforward \emph{derivation} of the von Neumann reduction
rule). The scheme associates, with every `experimentally accessible'
system, a symplectic superalgebra and operates essentially as
noncommutative Hamiltonian mechanics incorporating the extra
condition that the sets of observables and pure states be mutually
separating. The latter condition serves to smoothly connect the
algebraically defined quantum systems to Hilbert space-based ones;
the rigged Hilbert space - based Dirac bra-ket formalism naturally
appears. The formalism has a natural place for commutative
superselection rules. Noncommutative analogues of objects like the
momentum map and the Poincar$\acute{e}$-Cartan form are introduced
and some related symplectic geometry developed.

\vspace{.18in} \noindent \textbf{Key Words:} supmech; Hilbert's
sixth problem; noncommutative symplectic geometry; measurement
problem; autonomous quantum mechanics; noncommutative probability.

\newpage
 \noindent
\emph{Underlying everything you see \\
There is motion \\
Governed by \\
Noncommutative symplectics.}

\vspace{.3in} \noindent \textbf{1. INTRODUCTION}

\vspace{.12in}
  The statement of the famous sixth problem of Hilbert (Hilbert
1902; Wightman 1976), henceforth referred to as H6, reads :

\vspace{.1in} \noindent  ``To treat in the same manner, by means of
axioms, the physical sciences in which mathematics plays an
important part; in the first  rank are the theory of probabilities
and mechanics.''

\vspace{.1in} It appears reasonable to have a somewhat augmented
version of H6; the following formal statement is being hereby
proposed for this :

\vspace{.1in} ``To evolve an axiomatic scheme covering all physics
including the probabilistic framework employed for the treatment of
statistical aspects of physical phenomena.''

\vspace{.1in}A solution of this problem must include a satisfactory
treatment of the dynamics of the universe and its subsystems. Since
all physics is essentially mechanics, the formalism underlying such
a solution must be an elaborate scheme of mechanics (with elements
of probability incorporated). Keeping in view the presently
understood place of quantum mechanics (QM)  in the description of
nature, such a scheme of mechanics must incorporate, at least as a
subdiscipline or in some approximation, an ambiguity-free
development of QM without resorting to the prevalent practice of
\emph{quantization} of classical systems. One expects that, such a
development will, starting with some appealing basics, connect
smoothly to the traditional Hilbert space QM and facilitate a
satisfactory treatment of measurements.

Since a large class of systems in nature admit a classical
description to a very high degree of accuracy, the envisaged
mechanics must also facilitate a transparent treatment of
quantum-classical correspondence. For this to be feasible, the
underlying framework must be such that both quantum and classical
mechanics can be described in it (Dass 2002).

  The literature on the foundations of QM (Dirac 1958; von Neumann
1955; Jordan, von Neumann and Wigner 1934; Segal 1947; Mackey 1963;
Jauch 1968; Jammer 1974; Holevo 1982; Ludwig 1985; Bell 1988; Bohm
and Hiley 1993; Peres 1993; Busch, Grabouski and Lahti 1995) is
quite rich and full of valuable insights; it appears, however, fair
to say that, at present, there does not exist a formalism satisfying
the conditions stated above.  In this paper, we shall evolve a
formalism of the desired sort which does the needful relating to
(non-relativistic) QM and quantum-classical correspondence mentioned
above and holds promise to provide a base for a solution of (the
augmented) H6.

The desired formalism must have an all-embracing underlying
geometry. An appealing choice for the same is noncommutative
geometry (NCG) (Connes 1994; Dubois-Violette 1991; Madore 1995;
Landi 1997; Gracia Bondia, Varilly and Figuerra 2001).
Noncommutativity is the hallmark of QM. Indeed, the central point
made in Heisenberg's (1925) paper that marked the birth of QM was
that the kinematics underlying QM must be based on a non-commutative
algebra of observables. This idea was developed into a scheme of
mechanics --- called matrix mechanics --- by Born, Jordan, Dirac and
Heisenberg (Born and Jordan 1925; Dirac 1926; Born, Heisenberg and
Jordan 1926). The proper geometrical framework for the construction
of the quantum Poisson brackets of matrix mechanics is provided by
non-commutative symplectic structures (Dubois-Violette
1991,1995,1999; Dubois-Violette, Kerner and Madore 1994; Djemai
1995). The NCG scheme employed in these works is a straightforward
generalization of the scheme of commutative differential geometry in
which the algebra $C^{\infty}(M)$ of smooth functions on a manifold
M is replaced by a general (not necessarily commutative) complex
associative *-algebra \sca \ and the Lie algebra of smooth vector
fields on M by that of derivations on \sca. \

\begin{sloppypar}
  While Heisenberg presented the quantum view of  observables,
Schr$\ddot{o}$dinger (1926) dealt with wave functions which, through
the Born (1926) interpretation, brought out the important aspect of
QM as an intrinsically probabilistic theory. Noncommutativity of the
algebra of observables has important implications relating to the
basic character of the operative probability theory -- the so-called
`quantum probability' of which a variety of versions/formulations
have appeared in literature (von Neumann 1955, Jordan, von neumann
and Wigner 1934; Segal 1947; Mackey 1963; Jauch 1968; Varadarajan
1985; Accardi 1981; Parthasarathy 1992; Meyer 1995). Among these
formulations, the one that suits our needs best is the one provided
by the observable-state framework based on complex associative,
unital (topological) *-algebras (Meyer 1995; Dubin and Hennings
1990; Lassner 1984; Inoue 1998). This choice serves the important
purpose of allowing us to adopt the strategy of combining elements
of noncommutative symplectic geometry and noncommutative probability
in an algebraic framework.
\end{sloppypar}

The scheme based on normed algebras (Jordan, von Neumann and Wigner
1934; Segal 1947, 1963; Haag and Kastler 1964; Emch 1972, 1984;
Bratteli and Robinson 1979, 1981; Haag 1992; Araki 1999; Bogolubov
et al. 1990), although it makes use of observables and states, does
not serve our needs because it is not suitable for the treatment of
noncommutative symplectic geometry. Iguri and Castagnino (1999) have
analyzed the prospects of a more general class of algebras (nuclear,
barreled b*-algebras) as a mathematical framework for the
formulation of quantum principles prospectively better than that of
the normed algebras. These algebras accommodate unbounded
observables at the abstract level. Following essentially the
`footsteps' of Segal (1947), they obtain results parallel to those
in the C*-algebra theory --- an extremal decomposition theorem for
states, a functional representation theorem for commutative
subalgebras of observables and an extension of the classical GNS
theorem. In a sense, this work is complementary to the present one
where the emphasis is on the development of noncommutative
Hamiltonian mechanics. We have employed locally convex
(super-)algebras restricted by a condition of `compatible
completeness' (referred to as the `CC condition') on the collections
of observables and pure states (it is satisfied in classical
Hamiltonian mechanics and the traditional Hilbert space QM) which
plays a crucial role in connecting the basic algebraic scheme of
mechanics  to the traditional Hilbert space QM.

  In section 4 of (Dass 2002), a scheme of mechanics based on
noncommutative symplectic geometry was introduced; it was designed
to provide a proper geometrical setting for the matrix mechanics
mentioned above. States were, however, not introduced in the
algebraic setting. This work, therefore, falls short of a proper
realization of the strategy mentioned above. In the present work,
this deficiency has been removed and a proper integration of
noncommutative symplectic geometry and noncommutative probability
has been achieved.  The improvement in the definition of
noncommutative differential forms introduced in (Dubois-Violette
1995,1999) [i.e. demanding $\omega(...,KX,...) = K
\omega(...,X,...)$ where K is in the center of the algebra; for
notation, see section 3] is also incorporated. Moreover, to
accommodate fermionic objects on an equal footing with the bosonic
ones, the scheme developed here is based on superalgebras. The
scheme developed along the above lines is given, for easy reference,
the name `Supmech'.

  Supmech has quantum and classical mechanics as special
subdisciplines. This fact appears to open the prospects of a
consistent treatment of the interaction of a quantum and a classical
system. In (Dass 2006), the author applied such a formalism to the
treatment of measurements in QM providing what appeared to be the
most natural solution to the measurement problem in QM. An important
ingredient in this work was the Poisson bracket on the tensor
product of two algebras [the non-super version of the formula (154)
below]. Shortly after that paper  appeared in the arXiv, M.J.W. Hall
pointed out, in a private communication to the author, that the
`Poisson bracket' mentioned above does not satisfy the Jacobi
identity in some cases as shown, for example, in (Caro and Salcedo
1999). A revised calculation by the author [correcting the mistake
resulting from not realizing that an equation of the form (136)
below need not always represent a derivation of the tensor product
algebra] produced results which were partly discouraging [in that a
`natural'/`canonical' symplectic structure on the tensor product of
a (super-)commutative and a non- (super-)commutative (super-)algebra
(both the (super-)algebras being of the above mentioned type) does
not exist] and partly very very interesting : a symplectic structure
of the above sort on the tensor product of two
non-(super)commutative (super-)algebras exists if and only if each
of the (super-)algebras has a `quantum symplectic structure' [i.e.
one which gives a Poisson bracket which is a (super-)commutator up
to multiplication by a constant $(i \lambda^{-1})$ where $\lambda$
is a real-valued constant of the dimension of action] characterized
by a \emph{universal} parameter $\lambda$. The formalism, therefore,
has a natural place for the Planck constant as a universal constant
--- just as special relativity has a natural place for a universal
speed. In fact, the situation in supmech is somewhat better because,
whereas, in special relativity, the existence of a universal speed
is \emph{postulated}, in supmech, the existence of a universal
Planck-like constant is \emph{dictated/predicted} by the formalism.

The negative result about the possibility of a consistent
quantum-classical interaction in the supmech framework is by no
means `fatal' for the treatment of measurement problem in supmech.
It turns out that it is adequate to treat the apparatus as a
\emph{genuine} quantum system approximated well by a classical
system (in the setting of, for example, phase space descriptions of
quantum and classical dynamics); the fact that supmech has both
quantum and classical mechanics as special subdisciplines
facilitates such a treatment.

The detailed plan of the rest of the paper is as follows. In section
2, we present arguments, based on some fundamentals relating to
theory construction, for adopting the kind of formalism that we do.
These arguments make it quite plausible that the formalism being
evolved is  appropriate  for doing physics at the deepest level. In
section 3, essential developments in the (super-) derivation -based
noncommutative differential calculus and symplectic structures are
presented. The induced mappings on (super-)derivations and
differential forms (\phst \ and \phstup --- analogues of the
push-forward and pull-back mappings induced by diffeomorphisms in
the traditional differential geometry) are described; they play
important roles in supmech. Section 4 is devoted to the development
of the formalism of supmech; it includes, besides the basic
formalism as noncommutative Hamiltonian mechanics (incorporating the
CC condition), a treatment of symplectic actions of Lie groups  and
 noncommutative avatars of the momentum map,
 Poincar$\acute{e}$ - Cartan form and
the symplectic version of Noether's theorem. A general treatment of
localizable systems (more general and simpler than that in the
traditional approaches) is also given. In section 5, elementary
systems are defined in supmech and the special cases of
nonrelativistic and relativistic elementary systems are treated. The
role of relativity groups in the identification of fundamental
observables of elementary systems is emphasized. Particles are
treated as localizable elementary systems. In section 6, coupled
systems are treated and the results about the symplectic structure
on the tensor products of the superalgebras mentioned above are
obtained. [A reasonably self-contained presentation of the non-super
version of these results was given in (Dass 2007).]

Section 7 is devoted to the treatment of quantum and classical
systems as special categories of systems in supmech. Quantum systems
are taken up before classical systems to emphasize the autonomous
nature of the treatment of QM. `Standard quantum systems' are
defined in the algebraic setting; the CC condition ensures the
existence of their Hilbert space - based faithful realizations.
[\emph{An advantage of adopting the plan outlined above is that
when, treating material particles as localizable elementary quantum
systems, the Schr$\ddot{o}$dinger wave functions are introduced at
an appropriate place, their traditional Born interpretation is
obvious and the  Schr$\ddot{o}$dinger equation appears as a matter
of course --- without ever using the classical Hamiltonian or
Lagrangian in the process of obtaining it.}] The formalism is shown
to have a natural place for commutative superselection rules. A
transparent treatment of quantum - classical correspondence is given
emphasizing some formal aspects. The superclassical extension of
classical mechanics (incorporating fermionic objects in the setting
of a supercommutative superalgebra) is treated and is shown (for the
case of a finite number of fermion generators) to generally violate
the CC condition which disqualifies it from being a bonafide
subdiscipline of supmech. In section 8, measurements in quantum
systems are treated and a straightforward solution of the
measurement problem is given along the lines mentioned above; the
von Neumann reduction rule, traditionally \emph{postulated}, is
\emph{derived}. The treatment of the apparatus (properly as a
system) automatically incorporates the desirable decoherence effects
to suppress the unwanted macroscopic quantum interference terms. In
section 9, a set of axioms underlying the work presented in sections
4-8 is given. The last section contains some concluding remarks.

As the subject matter treated here is of considerably wide interest
and some mathematical techniques employed are not sufficiently well
known, the author has chosen to keep the presentation below somewhat
easy paced.

This paper is a refined version of the one which appeared, with the
same title, in arXiv : 0807.3604 v1 (quant-ph).

\vspace{.15in} \noindent \textbf{2. FROM BASICS TO ALGEBRAS}

\vspace{.12in}
In this section, we shall present arguments based on fundamentals relating to
physical theories, for making the choice of the type of formalism for supmech.

We look for the ingredients that should go into the formalism that
is intended to cover \emph{all physics}. To this end, we start by
considering the primitive elements which every physical theory
--- classical, quantum, or more general --- is expected to possess
(explicitly or implicitly). The author came across the term
`primitive elements of physical theory' (PEPT) in a not so well
known but an instructive and insightful paper by Houtappel, Van Dam
and Wigner (HVW)(1965) which aimed at a treatment of symmetries
[especially the `geometric invariance principles' (space-time
symmetries)] in a very general setting involving the PEPT which,
according to HVW, were `measurements and their results'. We,
however, would like to have a reasonably `complete'(in the intuitive
sense of `good enough for doing some concrete physics') minimal set
of such primitive elements. To this end, it serves well to have a
close look at the ingredients going into the construction of the
mathematical objects employed by HVW (the $\Pi$-functions --- the
`forefathers' of the objects presently known as histories (Griffiths
1984, 2002; Omnes 1992; Dass 2005). Doing this, one finds that such
a minimal set may be taken as (Dass and Joglekar 2001, last section)

\vspace{.1in} \noindent (i) observations/measurements;

\noindent (ii) description of evolution of systems (typically in
terms of a discrete or continuous parameter called `time');

\noindent (iii) conditional predictions about systems : given some
information about a system (typically in terms of values of
appropriate measurable quantities at one or more instants of time),
to make predictions/retrodictions about its behavior.

\vspace{.1in}  Item (iii) above generally involves elements of
probability. It is not difficult to see why Hilbert, while
formulating his VIth problem, chose to put probability along with
physics. Any formal axiomatization of probability must include
hypotheses about the way uncontrollable random influences affect
outcomes of experiments. Since these influences have physical
causes, any comprehensive (theory construction)/axiomatization of
physics must appropriately treat these causes. In any physical
theory, the theoretical apparatus employed to cover item (iii) above
is very much a part of the theory and, in case a (partial or total)
failure of the theory occurs, may well have to be subjected to
scrutiny along with other ingredients of the theory.

The desired formalism must incorporate the above three ingredients
in as general a setting as possible. Introduction of appropriate
mathematical objects to represent \emph{observables} (measurable
quantities) appears to be a must. When that has been done, a useful
concept that serves to introduce elements of probability in a
sufficiently general way and integrate items (i)-(iii) above is that
of \emph{state}. A state of a system encodes available information
about the system in terms of values of appropriate observables and
concretely describes preparation of a system before a measurement.
Evolution of systems can be described in terms of change of state
with time. Problems of conditional predictions can be formulated in
terms of probabilities of transitions of systems prepared in given
states to various possible states.

This reasoning leads to the prospects of a reasonably economical and
general description of systems in terms of observables and states.
In contrast to, for example, Araki's (1999) book, where these
objects were introduced through analysis of experiments, we have
introduced them by considering some basics of theory construction.
This is obviously more in tune with the general theme of the present
work.

The traditional algebraic schemes (generally based on C*-algebras)
have employed these objects as basic structures and have achieved
some good results. They, however, do not realize the true potential
of such an approach. In these works, one generally puts the Weyl
form of commutation relations (for finite as well infinite number of
degrees of freedom) `by hand' without connecting them to some
underlying geometry. This deficiency can be overcome by dropping the
restriction to normed algebras. As pointed out in the previous
section, the underlying geometry of QM is noncommutative symplectic
geometry whose vehicles are complex associative algebras. A program
based on such algebras, integrating noncommutative symplectic
geometry with noncommutative probability (arrived at in the previous
section as an intuitively appealing step taken after noting the two
roles of these algebras), therefore, appears to be the most
appropriate for the desired formalism also from considerations based
on  physics fundamentals.

Taking an appropriate class of (super-)algebras as the basic
objects, we shall define noncommutative symplectic structure on
them. The promised  mechanics (Supmech)  will be developed in the
form of Hamiltonian mechanics in the setting of this structure. We
shall try to exploit the underlying noncommutative symplectic
geometry as much as possible . For systems admitting a space-time
description, for example, we shall insist on the action of the
relativity group on the system algebra to be a hamiltonian action
extendable, in favourable situations, to a Poisson action (for the
definitions of these actions, see section 4.4) so that their
infinitesimal actions are generated, through Poisson brackets, by
some observables. We shall use this feature to identify the
fundamental observables for appropriately defined elementary systems
(material particles will be in the localizable subclass of these
systems); their system algebras may then be taken as those generated
by the fundamental observables. Employing appropriate tensor
products, one then has a \emph{canonical} procedure for setting up
the system algebras for systems of a finite number of particles.

A point worth noting is the generality of the reasoning employed
above. We did not restrict ourselves to any distinguished class of
physical systems (particles, fields,\ldots) nor did we talk about
space as the arena for all dynamics. In fact, the formalism evolved
will be general enough to permit, in principle, construction of
theories in which one starts (for the treatment of the dynamics of
the universe as a whole) with matter and its dynamics and space as
an arena appears in the description of a later stage in its
evolution.

\vspace{.15in} \noindent \textbf{3. SUPERDERIVATION-BASED
DIFFERENTIAL CALCULUS; SYMPLECTIC STRUCTURES}

\vspace{.12in} \begin{sloppypar} In this section the geometrical
setting underlying the main work is presented. It is a scheme of
non-commutative differential calculus which is a superalgebraic
version of  Dubois-Violette's scheme of noncommutative geometry
(henceforth referred to as DVNCG) along with some supplementary
developments by the author (Dass 1993a,2002). The induced mappings
on (super-)derivations and differential forms (\phst \ and \phstup),
which play an important role in the scheme of mechanics to be
developed,  are treated in some detail. A generalization of DVNCG,
which replaces a superalgebra \sca \ by a pair (\sca, \scx) (where
\scx \ is a Lie sub-superalgebra of the Lie superalgebra SDer(\sca)
of superderivations of \sca) as the basic entity, is also described.
This generalization will be employed in the treatment of general
quantum systems  admitting superselection rules.\end{sloppypar}

\vspace{.1in}  \noindent \emph{Note}. In most applications of
supmech, the non-super version of the formalism developed below is
adequate; this can be obtained by simply putting,in the formulas
below, all the epsilons representing parities equal to zero. The
reference (Dass 2007) contains a brief account of the non-super
version.

\vspace{.12in} \noindent \textbf{3.1. Superalgebras and
superderivations}

In superalgebra (Manin 1988; Berezin 1987; Leites 1980;Scheunert
1979), all mathematical structures for which addition is defined
(vector spaces, algebras, derivations, differential forms etc), have
a $Z_2$-grading where $Z_2 = Z/2Z = \{0,1\}$. This means that each
element of  such a structure can be uniquely written as a sum of two
parts each of which is assigned a definite parity  (0 or 1;
correspondingly it is called even or odd). Elements with definite
parity are called homogeneous. When a multiplicative operation is
defined between homogeneous elements of the same or different
mathematical types, the product is a homogeneous element  (of
appropriate mathematical type) and its parity is the sum (mod 2) of
the parities of the multiplicands. We shall denote the parity of a
homogeneous object $w$ by $\epsilon(w)$ or $\epsilon_w$ according to
convenience.

A \emph{supervector space} is a (complex) vector space V admitting a
direct sum decomposition  $V = V^{(o)} \oplus V^{(1)} $ into spaces
of even and odd vectors; a vector $v \in V$ can be uniquely
expressed as a sum $v = v_0 + v_1$ of even and odd vectors. A
\emph{superalgebra} \sca \ is a supervector space which is an
associative algebra with identity; it becomes a
\emph{*-superalgebra} if an antilinear
*-operation or involution $*:\sca \rightarrow \sca$ \ is defined
which satisfies the relations

\begin{center}
$ (AB)^* = \eta_{AB}B^*A^*, \hspace{.12in} (A^*)^* = A , \hspace{.12in}
I^* =I $
\end{center}
where I is the identity/unit element and $\eta_{AB} =
(-1)^{\epsilon_A \epsilon_B}$. An element $A \in \sca$ \  will be
called \emph{hermitian} if $ A^* = A.$

The \emph{supercommutator} of two elements A,B of a superalgebra is
defined as $[A,B] = AB -\eta_{AB}BA.$ For ordinary (anti-)commutators, we
shall employ the notations $[A,B]_{\mp} = AB \mp BA.$ A superalgebra \sca \
is said to be
\emph{supercommutative} if the supercommutator of every pair of its elements
vanishes.

The \emph{graded center} of \sca \ , denoted as $Z(\sca)$, consists of those
elements of \sca \ which have vanishing supercommutators with all elements
of \sca; it is clearly a supercommutative superalgebra. Writing
$Z(\sca) = Z_0(\sca) \oplus Z_1(\sca)$, the object $Z_0(\sca)$
is the traditional center of \sca.

A \emph{(*-)homomorphism} of a superalgebra \sca \ into \scb \ is a linear
mapping $ \Phi : \sca \rightarrow \scb $ which preserves products,
identity elements, parities (and involutions) :
\begin{eqnarray*}
\Phi (AB) = \Phi (A) \Phi (B), \hspace{.12in} \Phi (I_{\sca}) = I_{\scb},
\hspace{.12in} \epsilon (\Phi (A)) = \epsilon (A), \hspace{.12in}
\Phi (A^*) = (\Phi (A))^*;
\end{eqnarray*}
if it is, moreover, bijective, it is called a \emph{(*-)isomorphism}.

A \emph{Lie superalgebra} is a supervector  space $\mathcal{L}$ with a
\emph{superbracket} operation [ \ , \ ] : $\mathcal{L} \times \mathcal{L}
\rightarrow \mathcal{L}$ which is (i) bilinear, (ii) graded skew-symmetric
which means that, for any two homogeneous elements $a,b \in \mathcal{L},
[a,b] = -\eta_{ab}[b,a]$
and (iii) satisfies the \emph{super
Jacobi identity}
\begin{eqnarray*}
[a,[b,c]] = [[a,b],c] + \eta_{ab}[b,[a,c]].
\end{eqnarray*}

A (homogeneous) \emph{superderivation} of a superalgebra \sca \ is a
linear map $X: \sca \rightarrow \sca$ such that $X(AB) = X(A)B +
\eta_{XA} AX(B)$; this is the superalgebraic generalization of the
concept of derivation of an algebra. Introducing the multiplication
operator $\mu$ on \sca \ defined as $\mu (A)B = AB$, a necessary and
sufficient condition that a linear map $X: \sca \rightarrow \sca$ is
a superderivation is
\begin{eqnarray}
X \circ \mu(A) - \eta_{XA} \mu(A) \circ X = \mu(X(A)).
\end{eqnarray}
\emph{Proof}. In the equation defining the superderivation X above,
express every term as a sequence of mappings acting on the element
B; the resulting equation is precisely the equation obtained by
operating each side of Eq.(1) on B. \  $ \Box $

\noindent The set of all superderivations of \sca \ constitutes a
Lie superalgebra SDer(\sca) [$ = \sdera^{(0)} \oplus \sdera^{(1)}$];
this is the superalgebraic generalization of the Lie algebra
Der(\sca) of all derivations of the algebra \sca. \ The \emph{inner
superderivations} $D_A$ defined by $D_AB = [A,B]$ are easily seen to
 satisfy the relation
\begin{eqnarray*}
[D_A,D_B] = D_{[A,B]}
\end{eqnarray*}
and constitute a Lie sub-superalgebra ISDer(\sca) of SDer(\sca).

As in DVNCG, it will be implicitly assumed that the superalgebras
being employed have a reasonably rich supply of superderivations so
that various constructions involving them have a nontrivial content.
Some discussion  and useful results relating to the precise
characterization  of the relevant class of algebras may be found in
(Dubois-Violette et al. 2001).

The following two facts involving the graded center and the
superderivations will be useful  [in proving the subcomplex property
of $\Omega(\sca)$ below]:

\vspace{.1in} \noindent (i) If $ K\in Z(\sca) $, then $ X(K) \in
Z(\sca) $ for all $ X \in \sdera. $

\vspace{.1in} \noindent (ii) For any $ K \in \za $ and $ X,Y \in
\sdera $, we have
\begin{eqnarray}
[X,KY] = X(K)Y + \eta_{XK}K[X,Y].
\end{eqnarray}
\emph{Proof}. (i) Expand the two sides of the  the relation $ X(AK)
= \eta_{AK}X(KA)$ (for any $A \in \sca$) and cancel the terms
containing  K on the two sides.

\noindent
(ii) Expand $[X, KY](A)$ (for any $A \in \sca$).  $\Box $

\vspace{.1in} \noindent \emph{Corollary}. \sdera \ is a \za-module.

\vspace{.1in}  An involution * on \sdera \ is defined by the
relation $ X^*(A) = [X(A^*)]^*.$  We have

\vspace{.12in} \noindent (i) $[X,Y]^* = [X^*,Y^*]; \ \ (ii) (D_A)^*
= -D_{A^*}.$

\vspace{.12in} \noindent \emph{Proof}. In each case, apply the left
hand side to a general element $B\in \sca$ and follow the
definitions. [For an illustration of the kind of steps involved, see
the proof of the equations (4) below.] $ \Box $

\vspace{.1in}  A superalgebra-isomorphism  $\Phi : \sca \rightarrow
\scb$ induces a mapping \begin{eqnarray} \phst : \sdera \rightarrow
SDer(\scb)
 \ \ \textnormal{given by} \ \  (\phst X)(B) = \Phi(X[\Phi^{-1}(B)])
\end{eqnarray}
for all $X\in \sdera $ and $B \in \scb.$ It is the analogue (and a
generalization) of the push-forward mapping induced by a
diffeomorphism between two manifolds on the vector fields and
satisfies the expected relations (with $\Psi : \scb \rightarrow
\mathcal{C}$)
\begin{eqnarray}
(\Psi \circ \Phi)_* = \Psi_* \circ \Phi_*; \hspace{.12in}
\Phi_* [X,Y] = [\phst X, \phst Y].
\end{eqnarray}
\emph{Proof} : (i) For any $X \in Sder(\sca)$ and  $C \in
\mathcal{C}$,
\begin{eqnarray*}
[(\Psi \circ \Phi)_*X](C) & = & (\Psi \circ \Phi)(X[(\Psi \circ \Phi)^{-1}
                                 (C)]) \\
              & = & \Psi [\Phi (X [\Phi^{-1}(\Psi^{-1}(C))])] \\
                          & = & \Psi [ (\Phi_*X)(\Psi^{-1}(C)] \\
                          & = & [\Psi_*(\Phi_*X)](C).
\end{eqnarray*}
(ii) For any $B \in \mathcal{B}$
\begin{eqnarray*}
( \phst[X,Y])(B) & = & \Phi ([X,Y](\Phi^{-1}(B))) \\
                 & = & \Phi[X (Y (\Phi^{-1}(B))) - \eta_{XY} Y(X(\Phi^{-1}(B))].
\end{eqnarray*}
Now insert $\Phi^{-1} \circ \Phi$ between X and Y in each of the two terms on
the right and follow the obvious steps. $ \Box $.

\vspace{.1in}  Note that  \phst \ is a Lie superalgebra isomorphism
(i.e. it is bijective and linear and preserves superbrackets).

\vspace{.12in} \noindent \textbf{3.2. The cochain complex C(\sdera,
\sca)}

In DVNCG, one starts with  a complex associative algebra \sca \ and
constructs a \emph{differential calculus} on it which means a
formalism involving differential form like objects on \sca \ with
analogues of exterior product, exterior derivative and involution
defined on them. For noncommutative \sca, \ the choice of
differential calculus is not unique; a systematic discussion of the
variety of choices may be found in (Dubois-Violette 1995). In
applications of NCG, one makes a choice according to convenience.
Our needs are best served by a DVNCG type formalism.

\begin{sloppypar}
For the constructions involving the superalgebraic generalization of
DVNCG given in this subsection, some relevant background is provided
in (Dubois-Violette 1999; Grosse and Reiter 1999; Scheunert
1979a,1979b,1983,1998). Grosse and Reiter (1999) have given a
detailed treatment of the differential geometry of graded matrix
algebras [generalizing the treatment of differential geometry of
matrix algebras in (Dubois-Violette, Kerner and Madore 1994)]. Some
related work on supermatrix geometry has also appeared in
(Dubois-Violette, Kerner and Madore 1991; Kerner 1993); however, the
approach adopted below is closer to (Grosse and Reiter 1999).
\end{sloppypar}

The central object in the developments presented below is a
superalgebra \sca \ [complex, associative, unital (i.e. possessing a
unit element), not necessarily supercommutative]; it is the
counterpart of \cinfm, the commutative algebra of complex smooth
functions on the  manifold M, in commutative geometry. The Lie
superalgebra \sdera \  is the analogue of the Lie algebra \xm \ of
smooth vector fields on M.

Recalling that, in the commutative differential geometry, the
differential p-forms are defined as skew-symmetric multilinear maps
of $\mathcal{X}(M)^p $ into $C^{\infty}(M)$, the natural first
choice for the space of (noncommutative) differential p-forms is the
space
\begin{eqnarray*}  C^p(SDer\sca,\sca) [= C^{p,0}(\sdera,\sca) \oplus
C^{p,1}(\sdera,\sca)] \end{eqnarray*} of graded skew-symmetric
multilinear maps (for $p \geq 1$) of $[\sdera]^p$ into \sca \ (the
space of \sca-valued p-cochains of \sdera; it is the super-analogue
of the Chevalley-Eilenberg p-cochain space (Cartan and Eilenberg
1956; Weibel 1994). We have $C^0 (\sdera, \sca) = \sca$. For $\omega
\in C^{p,s}(\sdera, \sca),$ we have
\begin{eqnarray}
\omega(..,X,Y,..) = - \eta_{XY} \omega(..,Y,X,..).
\end{eqnarray}
For a general permutation $\sigma$ of the arguments of $\omega$, we
have
\begin{eqnarray}
\omega(X_{\sigma(1)},..,X_{\sigma(p)}) = \kappa_{\sigma}
\gamma_p(\sigma; \epsilon_{X_1},..,\epsilon_{X_p})
\omega(X_1,..,X_p)
\end{eqnarray}
where $\kappa_{\sigma}$ is the parity of the permutation $\sigma$ and
\begin{eqnarray}
\gamma_p(\sigma; s_1,..,s_p) = \prod_{\begin{array}{c}
j,k= 1,..,p; \\
j<k,\sigma^{-1}(j) > \sigma^{-1}(k) \end{array}} (-1)^{s_j s_k}.
\end{eqnarray}

An involution * on the cochains is defined by the relation
$\omega^*(X_1,..,X_p) = [\omega(X_1^*,..,X_p^*)]^*;$
 $\omega$ is said to be real (imaginary) if $ \omega^* = \omega (- \omega)$.

The \emph{exterior product}
\begin{eqnarray*} \wedge : C^{p,r}(\sdera,\sca) \times C^{q,s}(\sdera,\sca)
\rightarrow C^{p+q,r+s}(\sdera,\sca) \end{eqnarray*} is defined as
\begin{eqnarray}
(\alpha \wedge \beta)(X_1,..,X_{p+q}) =
\frac{1}{p!q!}\sum_{\sigma \in \mathcal{S}_{p+q}} \kappa_{\sigma}
\gamma_{p+q}(\sigma; \epsilon_{X_1},..,\epsilon_{X_{p+q}})
(-1)^{s \sum_{j=1}^p \epsilon_{X_{\sigma(j)}}}  \nonumber \\
\alpha(X_{\sigma(1)},..,X_{\sigma(p)}) \beta(X_{\sigma(p+1)},..,
X_{\sigma(p+q)}).
\end{eqnarray}
With this product, the graded
vector space
\begin{eqnarray*}
C(\sdera,\sca) = \bigoplus_{p\geq0}C^p(\sdera,\sca)
\end{eqnarray*}
 becomes an $ N_0 \times Z_2$-bigraded complex algebra. (Here
$ N_0 $ is the set of non-negative integers.)

The Lie superalgebra \sdera \  acts on itself and on C(\sdera,\sca) through
\emph{Lie derivatives}. For each $Y \in \sdera^{(r)}$, one defines linear
mappings
$L_Y :\sdera^{(s)} \rightarrow \sdera^{(r+s)}$ and
$L_Y : C^{p,s}(\sdera, \sca)\rightarrow
C^{p,r+s}(\sdera,\sca) $ \ such that the following three conditions hold :
\begin{eqnarray}
 L_Y(A) = Y(A) \hspace{.12in} \mbox{for all}\  A \in \sca
\end{eqnarray}
\begin{eqnarray}
 L_Y[X(A)] = (L_YX)(A) + \eta_{XY}X[L_Y(A)]
\end{eqnarray}
\begin{eqnarray}
 L_Y [\omega(X_1,..,X_p)]  =  (L_Y\omega)(X_1,..,X_p)
                           +  \sum_{i=1}^p (-1)^{\epsilon_Y(\epsilon_{\omega} + \epsilon_{X_1} + ..
+\epsilon_{X_{i-1}})}. \nonumber \\
.\omega(X_1,..,X_{i-1},L_YX_i,X_{i+1},..,X_p).
\end{eqnarray}
The first two conditions give
\begin{eqnarray}
L_Y X = [Y,X]
\end{eqnarray}
which, along with the third, gives
\begin{eqnarray}
(L_Y \omega)(X_1,..,X_p)  =  Y[\omega(X_1,..,X_p)]
                          - \sum_{i=1}^p (-1)^{\epsilon_Y(\epsilon_{\omega} + \epsilon_{X_1} + .. +
\epsilon_{X_{i-1}})}. \nonumber \\
.\omega (X_1,..,X_{i-1}, [Y,X_i],X_{i+1},..,X_p).
\end{eqnarray}
Two important properties of the Lie derivative are, in obvious
notation,
\begin{eqnarray}
[L_X,L_Y] = L_{[X,Y]}
\end{eqnarray}
\begin{eqnarray}
L_Y(\alpha \wedge \beta) = (L_Y \alpha) \wedge \beta + \eta_{Y
\alpha} \ \alpha \wedge (L_Y \beta).
\end{eqnarray}

For each $X \in \sdera^{(r)} $, we define the \emph{interior
product} \\
$ i_X : C^{p,s}(\sdera,\sca) \rightarrow
C^{p-1,r+s}(\sdera,\sca) $
 ( for $p \geq 1$) by
\begin{eqnarray}
(i_X \omega)(X_1,..,X_{p-1}) = \omega (X,X_1,..,X_{p-1}).
\end{eqnarray}
One defines $i_X(A) = 0 $ for all $A \in \sca$. Note that there is
no $\eta_{X \omega}$ factor on the right in Eq.(16). A more
appropriate notation (from the point of view of proper/unambiguous
placing of symbols) for $i_X \omega$ is $ \omega_X$. [See (Hochscild
and Serre 1953).] Some important properties of the interior product
are :
\begin{eqnarray}
i_X \circ i_Y + \eta_{XY} i_Y \circ i_X = 0
\end{eqnarray}
\begin{eqnarray}
i_X(\alpha \wedge \beta) = \eta_{X \beta} (i_X \alpha) \wedge \beta +
(-1)^p \alpha \wedge (i_X \beta)
\end{eqnarray}
\begin{eqnarray}
(L_Y \circ i_X - i_X \circ L_Y) = \eta_{X \omega} i_{[X,Y]} \omega.
\end{eqnarray}

The \emph{exterior derivative} $ d: C^{p,r}(\sdera,\sca) \rightarrow
C^{p+1,r}(\sdera,\sca) $ is
defined through the relation
\begin{eqnarray}
(i_X \circ d + d \circ i_X ) \omega = \eta_{X \omega} \ L_X \omega.
\end{eqnarray}
For p = 0, it takes the form $(dA)(X) = \eta_{XA} \  X(A)$.

Taking, in Eq.(20), $\omega$ a homogeneous p-form and contracting
both sides with homogeneous derivations $X_1,..,X_{p}$   gives the
quantity $(d \omega)(X,X_1,..,X_p)$ in terms of evaluations of
exterior derivatives of lower degree forms. This determines $ d
\omega $ recursively giving
\begin{eqnarray}
(d \omega)(X_0,X_1,..,X_p)
= \sum_{i=0}^{p} (-1)^{i+ a_i} X_i [ \omega(X_0,..,\hat{X}_{i},..,X_p)]
\nonumber \\
+ \sum_{0 \leq i < j \leq p}(-1)^{j+ b_{ij}}
\omega (X_0,.., X_{i-1}, [X_i,X_j],X_{i+1}, ..,\hat{X}_j,..,X_p)
\end{eqnarray}
where the hat indicates omission and
\begin{eqnarray*}
a_i = \epsilon_{X_i} (\epsilon_{\omega} + \sum_{k=0}^{i-1}
\epsilon_{X_k}); \ b_{ij} = \epsilon_{X_j}\sum_{k=i+1}^{j-1} \epsilon_{X_k}.
\end{eqnarray*}
Some important properties of the exterior derivative are
\begin{eqnarray}
d^2 (= d \circ d) = 0 ; \ \ \ d \circ L_Y = L_Y \circ d
\end{eqnarray}
and
\begin{eqnarray}
d (\alpha \wedge \beta) = (d \alpha) \wedge \beta +
(-1)^p \alpha \wedge (d \beta)
\end{eqnarray}
where $\alpha$ is a p-cochain. The first of equations (22) shows
that the pair, (C(\sdera, \sca), d) constitutes a cochain complex.
We shall call a cochain $\alpha$ \emph{closed} if $ d \alpha = 0$
and \emph{exact} if $ \alpha = d \beta $ for some cochain $\beta$.

The cochain complex obtained above is a special case of the
Chevalley-Eilenberg cochain complex. For later use, we collect here
some results relating to the Chevalley-Eilenberg cohomology
(employing ordinary Lie algebras and vector spaces; no gradings are
involved).

Let $\mathcal{G}$ be a Lie algebra over the field K (which may be R
or C) and V a $\mathcal{G}$-module which means it is a vector space
over K having defined on it a $\mathcal{G}$-action  associating a
linear mapping $\Psi(\xi)$ on V with every element $\xi$ of
$\mathcal{G}$ such that
\[ \Psi(0) = id_V \ \ \textnormal{and} \ \ \Psi([\xi,\eta]) =
\Psi(\xi) \circ \Psi(\eta) - \Psi(\eta) \circ \Psi(\xi). \] The
analogue of Eq.(21) for a V-valued p-cochain $\lambda^{(p)}$ of
$\mathcal{G}$ is
\begin{eqnarray} (d\lambda^{(p)})(\xi_0,\xi_1,..,\xi_p)  =
\sum_{i=0}^{p} (-1)^i
\Psi(\xi)[\lambda^{(p)}(\xi_0,..,\hat{\xi}_i,.., \xi_p)] + \nonumber
\\
\sum_{0 \leq i<j \leq p}(-1)^j \lambda^{(p)}(\xi_0,..,\xi_{i-1},
[\xi_i,\xi_j],\xi_{i+1},..,\hat{\xi}_j,..,\xi_p) \end{eqnarray} for
 $\xi_0,..,\xi_p \in \mathcal{G}$. The subspaces of the vector
space $C^p(\mathcal{G},V)$ consisting of closed cochains (cocycles)
and exact cochains (coboundaries) are denoted as
$Z^p(\mathcal{G},V)$ and $B^p(\scg,V)$ respectively; the quotient
space $H^p(\scg,V) \equiv Z^p(\scg,V)/B^p(\scg,V)$ is called the
p-th cohomology group of \scg \ with coefficients in V.

For the special case of the trivial action of \scg \ on V [i.e.
$\Psi(\xi) = 0 \ \forall \xi \in \scg$], a subscript zero is
attached to these spaces [$C^p_0(\scg,V)$ etc].  In this case, for p
= 1 and p = 2, Eq.(24) takes the form
\begin{eqnarray} d \lambda^{(1)}(\xi_0,\xi_1) & = & -
\lambda^{(1)}([\xi_0,\xi_1]) \nonumber \\
d \lambda^{(2)} (\xi_0,\xi_1,\xi_2)  & = & -
[\lambda^{(2)}([\xi_0,\xi_1], \xi_2) + \textnormal{cyclic terms in}
\ \xi_0, \xi_1, \xi_2]. \end{eqnarray}

\vspace{.12in} \noindent \textbf{3.3. Differential forms}

Taking clue from (Dubois-Violette 1995, 1999)  [where the subcomplex
of $Z(\sca)$-linear cochains (Z(\sca) being, in his notation, the
center of the algebra \sca) was adopted as the space of differential
forms], we consider the subset $\Omega(\sca)$ of $C(\sdera, \sca)$
consisting of $Z_0(\sca)$-linear cochains. Eq.(2) ensures that this
subset is closed under the action of d and, therefore, a subcomplex.
We shall take this space to be the space of differential forms in
subsequent geometrical work. We have, of course,
\begin{eqnarray*}
\Omega(\sca) = \oplus_{p \geq 0}\Omega^p(\sca)
\end{eqnarray*}
with $\Omega^0(\sca) = \sca$ and $\Omega^p(\sca) = \Omega^{p,0}(\sca)
\oplus \Omega^{p,1}(\sca)$ for all $ p \geq 0.$

\vspace{.12in} \noindent \textbf{3.4. Induced mappings on
differential forms}

\vspace{.12in} A superalgebra
*-isomorphism $ \Phi : \sca \rightarrow \scb $ induces, besides the Lie
superalgebra-isomorphism $ \phst :\sdera \rightarrow SDer(\scb),$ a
mapping \[ \phstup : C^{p,s}(SDer (\scb), \scb) \rightarrow
C^{p,s}(\sdera,\sca)\] given, in obvious notation, by
\begin{eqnarray}
(\phstup \omega)(X_1,..,X_p) = \Phi^{-1} [ \omega ( \phst X_1,..,
\phst X_p)].
\end{eqnarray}
The mapping $ \Phi$ preserves (bijectively) all the algebraic
relations. Eq.(3) shows that \phst \ preserves $ Z_0(\sca)$-linear
combinations of the superderivations. It follows that \phstup \ maps
differential forms onto differential forms.

These mappings are analogues (and generalizations) of the pull-back mappings
on differential forms (on manifolds) induced by diffeomorphisms. They satisfy
the expected relations [with $\Psi : \scb \rightarrow \mathcal{C}$]
\begin{eqnarray}
(\Psi \circ \Phi)^* = \Phi^* \circ \Psi^*
\end{eqnarray}
\begin{eqnarray}
\phstup (\alpha \wedge \beta) = (\phstup \alpha) \wedge (\phstup \beta)
\end{eqnarray}
\begin{eqnarray}
\phstup (d \alpha ) = d ( \phstup \alpha ).
\end{eqnarray}
\emph{Outlines of proofs of Eqs.(27-29)} :

\vspace{.1in} \noindent Eq.(27) : For $ \omega \in
C^{p,s}(SDer(\mathcal{C}), \mathcal{C})$ and $ X_1,..,X_p \in
SDer(\sca)$,
\begin{eqnarray*}
[(\Psi \circ \Phi)^* \omega] (X_1,..,X_p)
& = & (\Phi^{-1} \circ \Psi^{-1})[\omega(\Psi_*(\Phi_*X_1),..,
       \Psi_*(\Phi_*X_p))] \\
& = & \Phi^{-1} [(\Psi^* \omega)(\Phi_*X_1,..,\Phi_*X_p)] \\
& = & [\Phi^*(\Psi^* \omega)] (X_1,..,X_p). \ \Box
\end{eqnarray*}

\noindent Eq.(28) : For $\alpha \in C^{p,r}(SDer(\scb), \scb), \beta
\in C^{q,s}(SDer(\scb),\scb)$ and $ X_1,..,X_{p+q} \in \sdera$,
\begin{eqnarray*}
[\Phi^*(\alpha \wedge \beta)](X_1,.., X_{p+q})
& = & \Phi^{-1}[ (\alpha \wedge \beta) (\phst X_1,..,\phst X_{p+q})].
\end{eqnarray*}
Expanding the wedge product and noting that \[ \Phi^{-1}[\alpha
(..)\beta(..)] = \Phi^{-1}[\alpha(..)]. \Phi^{-1}[\beta(..)],
\] the right hand side is easily seen to be equal to $ [(\phstup
\alpha)\wedge (\phstup \beta)](X_1,..,X_{p+q}). \ \Box$

\vspace{.1in} \noindent Eq.(29) : We have
\begin{eqnarray*}
[\phstup (d \alpha)](X_0,..,X_p)
= \Phi^{-1} [(d \alpha)(\phst X_0,..,\phst X_p)].
\end{eqnarray*}
Using Eq.(21) for $d \alpha$ and noting that
\begin{eqnarray*}
\Phi^{-1}[(\phst X_i)(\alpha (\phst X_0,..))
& = & \Phi^{-1} [\Phi (X_i [\Phi^{-1}(\alpha(\phst X_0,..))]]  \\
& = & X_i [(\phstup \alpha) (X_0,..)]
\end{eqnarray*}
and making similar (in fact, simpler) manipulations with the second
term in the expression for $d \alpha$, it is easily seen that the
left hand side of Eq.(29), evaluated at $(X_0,..,X_p)$, equals $[(d
(\phstup \alpha)](X_0,..,X_p)$.\   $\Box$

Now, let $ \Phi_t : \sca \rightarrow \sca $ be a
one-parameter family  of transformations (i.e. superalgebra isomorphisms)
given, for small t, by
\begin{eqnarray}
\Phi_t (A) \simeq A + t g(A)
\end{eqnarray}
where g is some (linear, even) mapping of \sca \ into itself. The condition
$ \Phi_t (AB) = \Phi_t (A) \Phi_t (B) $ gives $ g(AB) = g(A) B +
A g(B)$ implying that $ g(A) = Y(A) $ for some even superderivation Y
of \sca \ ( to be called the \emph{infinitesimal generator} of
$ \Phi_t $). From Eq.(3), we have, for small t,
\begin{eqnarray}
(\Phi_t)_* X \simeq X + t [Y, X] = X + t L_Y X.
\end{eqnarray}
Similarly, for any p-form $ \omega, $ we have
\begin{eqnarray}
\Phi_t^* \omega \simeq \omega - t L_Y \omega.
\end{eqnarray}
\emph{Proof} :  We have
\begin{eqnarray*}
(\phstup_t \omega)(X_1,..,X_p)
& = & \Phi_t^{-1} [\omega\left( (\Phi_t)_* X_1,..(\Phi_t)_* X_p \right)] \\
& \simeq & \omega(X_1,..,X_p) - tY\omega(X_1,..,X_p)  \\
    & \ & + t \sum _{i=1}^p \omega(X_1,.., [Y,X_i],.., X_p) \\
& = & [\omega - t L_Y \omega] ](X_1,..,X_p).  \hspace{.2in} \Box
\end{eqnarray*}
Note that (as in the commutative geometry), the Lie
derivative term appears
with a plus sign in Eq.(31) and a minus sign in Eq.(32). This is because, in
the pull-back action, the effective mapping is $ \Phi_t^{-1}. $

\vspace{.12in} \noindent \textbf{3.5. A generalization of the DVNCG
scheme}(Dass 1993,2002)

In the formula (21) for $d \omega$, the superderivations $X_j$
appear on the right either singly or as supercommutators. It should,
therefore, be possible to restrict them to a Lie sub-superalgebra
\scx \ of \sdera \ and develop the whole formalism with the pair
$(\sca, \scx)$ obtaining thereby a useful generalization of the
formalism developed in the previous three subsections. Working with
such a pair is the analogue of restricting oneself to a leaf of a
foliated manifold as the first example below indicates.

\vspace{.1in} \noindent \emph{Examples} : (i) $ \sca =
C^{\infty}(R^3)$; \scx = the Lie subalgebra of the Lie algebra
$\scx(R^3)$ of vector fields on $R^3$ generated by the Lie
differential operators $ L_j = \epsilon_{jkl}x_k\partial_l$ for the
SO(3)-action on $R^3$. These differential operators, when expressed
in terms of the polar coordinates $r, \theta, \phi$, contain only
the partial derivatives with respect to $\theta$ and $\phi$; they,
therefore, act on the 2-dimensional spheres that constitute the
leaves of the foliation $R^3 - \{(0,0,0) \} \simeq S^2 \times R$.
The restriction [of the pair $(\sca, \scx(R^3))$] to (\sca,\scx)
amounts to restricting oneself to a leaf ($S^2$) in the present
case.

\vspace{.1in} \noindent (ii) $ \sca = M_4(C)$, the algebra of
complex $4 \times 4$ matrices. The vector space $C^4$ on which these
matrices act serves as the carrier space of the spin $s =
\frac{3}{2}$ projective irreducible representation of the rotation
group SO(3). Denoting by $S_j$ (j=1,2,3) the $4 \times 4$ matrices
representing the generators of the Lie algebra so(3), let \scx  \ be
the real Lie algebra generated by the inner  derivations $D_{S_j}$.
The pair (\sca, \scx) is relevant for the treatment of spin dynamics
for $s= \frac{3}{2}$.

\vspace{.1in} In the generalized formalism, one obtains the cochains
$C^{p,s}(\scx, \sca)$ for which the formulas of  sections  3.2 and
3.3 are valid (with the $X_js$ restricted to \scx). The differential
forms $\Omega^{p,s}(\sca)$ will now be replaced by the objects $
\Omega^{p,s} (\scx, \sca)$ obtained by restricting the cochains to
the $Z_0(\sca)$-linear ones. [In the new notation, the objects
$\Omega^{p,s}(\sca)$ will be called $\Omega^{p,s}(\sdera,\sca)$.]

\vspace{.1in} \noindent \emph{Note}. In (Dass 2002), the notation
$\Omega^p(\sca, \scx)$ was used for the space of differential
p-forms [which appeared natural in view of the notation (\sca, \scx)
for the pairs called algebraic differential systems there]. In the
present work, we have changed it to $\Omega^p(\scx, \sca)$ to bring
it in tune with the notation for Lie algebra cochains in the
mathematics literature.

\vspace{.1in} To define the induced mappings \phst \ and \phstup \
in the present context, one must employ a \emph{pair-isomorphism}
$\Phi : (\sca,\scx) \rightarrow (\scb, \scy)$ which consists of  a
superalgebra *- isomorphism $ \Phi : \sca \rightarrow \scb$ such
that the induced mapping $\Phi_* : \sdera \rightarrow SDer(\scb)$
restricts to an isomorphism of \scx \  onto \scy. Various properties
of the induced mappings hold as before.

Given a one-parameter family of transformations $\Phi_t : (\sca,
\scx) \rightarrow (\sca, \scx)$, the condition $(\Phi_t)_* \scx
\subset \scx $ implies that the infinitesimal generator Y of
$\Phi_t$ must satisfy the condition $[Y,X] \in \scx $ for all $X \in
\scx$. In practical applications one will generally have $Y \in
\scx$ which trivially satisfies this condition.

This generalization will be used in sections 4.5 and 7.5 below.

\vspace{.15in} \noindent \textbf{3.6. Symplectic structures}

\emph{Note}. The sign conventions about various quantities adopted
below are parallel to those of Woodhouse (1980). This results in a
(super-) Poisson bracket which, when applied to classical
Hamiltonian mechanics, gives one differing from the Poisson bracket
in most current books on mechanics by a minus sign. [See Eq.(46).]
The main virtue of the adopted conventions is that Eq.(38) below has
no unpleasant minus sign.

\vspace{.12in} A \emph{symplectic structure} on a superalgebra \sca
\ is  a 2- form $\omega$ (the \emph{symplectic form}) which is even,
closed and \emph{non-degenerate} in the sense that, for every $A \in
\sca$, there exists a unique superderivation $Y_A$ in \sdera \ [the
\emph{(globally) Hamiltonian superderivation} corresponding to A]
such that
\begin{eqnarray}
i_{Y_A} \omega = - dA.
\end{eqnarray}
The pair $(\sca,\omega)$ will be called a \emph{symplectic
superalgebra}. A symplectic structure is said to be \emph{exact} if
the symplectic form is exact ( $ \omega = d \theta$  for some 1-form
$ \theta$ called the \emph{symplectic potential}).

A \emph{symplectic mapping} from a symplectic superalgebra
$(\sca,\alpha)$ to another one $(\scb,\beta)$ is a superalgebra
isomorphism $\Phi : \sca \rightarrow \scb$ such that $\Phi^*\beta =
\alpha.$ (If the symplectic structures involved are exact, one
requires a symplectic mapping to preserve the symplectic potential
under the pull-back action; Eq.(29) then guarantees the preservation
of the symplectic form.) A symplectic mapping from a symplectic
superalgebra onto itself will be called a \emph{canonical/symplectic
transformation}. The symplectic form and its exterior powers are
invariant under canonical transformations.

If $\Phi_t$ is a one-parameter family of canonical transformations
generated by $X\in \sdera, $ the condition $ \Phi_t^* \omega =
\omega$ implies, with Eq.(32),
\begin{eqnarray}
L_X \omega = 0.
\end{eqnarray}
The argument just presented gives Eq.(34) with X an even
superderivation. More generally, a superderivation X (even or odd or
inhomogeneous) satisfying Eq.(34) will be called a \emph{locally
Hamiltonian} superderivation. Eq.(20) and the condition $ d \omega
=0 $  imply that Eq.(34) is equivalent to the condition
\begin{eqnarray} d (i_X \omega) = 0.
\end{eqnarray}
 The (globally) Hamiltonian superderivations defined by Eq(33)
constitute a subclass of locally Hamiltonian superderivations for
which $i_X \omega$ is exact. Note from Eq(33) that $\epsilon(Y_A) =
\epsilon(A)$. In analogy with the commutative case,  the
supercommutator of two locally Hamiltonian superderivations is a
globally Hamiltonian superderivation. Indeed, given two locally
Hamiltonian superderivations X and Y, we have, recalling Equations
(19) and (20),
\begin{eqnarray*}
\eta_{X \omega} i_{[X,Y]} \omega & = & (L_Y \circ i_X - i_X \circ
L_Y) \omega
                                         \nonumber \\
                 & = & \eta_{Y \omega} (i_Y \circ d +
                       d \circ i_Y)(i_X \omega)
                       \nonumber \\
                 & = & \eta_{Y \omega} d (i_Y i_X \omega)
\end{eqnarray*}
which is exact. It follows that the locally Hamiltonian
superderivations constitute a Lie superalgebra in which the globally
Hamiltonian superderivations constitute an ideal.

The \emph{Poisson bracket} (PB) of two elements A and B of \sca \ is
defined as
\begin{eqnarray}
\{ A, B \} = \omega (Y_A, Y_B) = Y_A (B) = - \eta_{AB} Y_B (A).
\end{eqnarray}
It obeys the superanalogue of the Leibnitz rule :
\begin{eqnarray}
\{A, BC \} = Y_A (BC) & = & Y_A(B) C + \eta_{AB} B Y_A (C) \nonumber \\
                      & = & \{ A, B \} C + \eta_{AB} B \{ A, C \}.
\end{eqnarray}
As in the classical case, we have the relation
\begin{eqnarray}
[Y_A, Y_B] = Y_{ \{A,B \} }.
\end{eqnarray}
Eqn.(38) follows by using the equation for $ i_{[X,Y]} \omega$ above
with $X = Y_A$ and $ Y =Y_B$ and equations (16), (36) and (33),
remembering that Eq.(33) determines $Y_A$ uniquely. The super-Jacobi
identity
\begin{eqnarray}
0 & =  & \frac{1}{2} (d \omega)(Y_A, Y_B, Y_C)  \nonumber \\
  & =  & \{ A, \{ B, C \} \} + (-1)^{\epsilon_A(\epsilon_B + \epsilon_C)}
            \{ B,  \{ C, A \} \}  \nonumber \\
  & \, &    + (-1)^{\epsilon_C (\epsilon_A  + \epsilon_B)}
  \{ C, \{ A, B \} \}
\end{eqnarray}
is obtained by using Eq.(21) and noting that
\begin{eqnarray*}
Y_A[\omega(Y_B,Y_C)] & = & \{ A, \{B, C \} \} \\
\omega ([Y_A,Y_B],Y_C) & = &  \omega (Y_{\{A,B \}},Y_C) = \{ \{A,B
\}, C \}.
\end{eqnarray*}
Clearly, the pair (\sca, \{ ,  \}) is a Lie superalgebra. Eq.(38)
shows that the linear mapping $ A \mapsto Y_A $ is a
Lie-superalgebra homomorphism.

An element A of \sca \ can act, via $ Y_A $, as the infinitesimal
generator of a one-parameter family of canonical transformations.
The change in $ B \in \sca$ \ due to such an infinitesimal
transformation is
\begin{eqnarray}
\delta B = \epsilon Y_A(B) = \epsilon \{ A, B \}.
\end{eqnarray}
In particular, if $ \delta B = \epsilon I$ (infinitesimal
`translation' in B), we have
\begin{eqnarray}
\{ A, B \} = I
\end{eqnarray}
which is the noncommutative analogue of the classical PB relation $
\{ p,q \} = 1$. A pair (A,B) of elements of \sca \ satisfying the
condition (41) will be called a \emph{canonical pair}.

\vspace{.15in} \noindent \textbf{3.7. Reality properties of the
symplectic form and the Poisson bracket}

\vspace{.12in} For classical superdynamical systems, conventions
about reality properties of the symplectic form are based on the
fact that the Lagrangian is a real, even object (Berezin and marinov
1977; Dass 1993b). The matrix of the symplectic form is then real-
antisymmetric in the `bosonic sector' and imaginary-symmetric in the
`fermionic sector' (which means anti-Hermitian in both sectors).
Keeping this in view, it appears appropriate to impose, in supmech,
the following reality condition on the symplectic form $ \omega$:
\begin{eqnarray}
\omega^* (X,Y) = - \eta_{XY}\omega (Y,X) \hspace{.2in} \mbox{for
all} \ X, Y \in \sdera;
\end{eqnarray}
but this means,by Eq.(5), that $ \omega^* = \omega$ (i.e. $\omega$
is real) which is the most natural assumption to make about
$\omega$. Eq.(42) is equivalent to the condition
\begin{eqnarray}
\omega(X^*,Y^*) = - \eta_{XY} [\omega (Y,X)]^*.
\end{eqnarray}
Now, for arbitrary $ A,B \in \sca,$ we have
\begin{eqnarray*}
\{ A,B \}^* & = & Y_A (B)^* = Y_A^*(B^*) = \eta_{AB} dB^*(Y_A^*) \\
            & = & -\eta_{AB} \omega(Y_{B^*}, Y_A^*) =
                  [\omega(Y_A,Y_{B^*}^*)]^* \\
        & = & - [dA(Y_{B^*}^*)]^* = - \eta_{AB}[Y_{B^*}^*(A)]^* \\
        & = & - \eta_{AB}Y_{B^*}(A^*)
\end{eqnarray*}
giving finally
\begin{eqnarray}\{ A,B \}^* = - \eta_{AB}\{B^*, A^* \}.
\end{eqnarray}
Eq.(44) is consistent with the  reality properties of the classical
and quantum Poisson brackets.[See equations (46) and (54) below.]

\vspace{.15in} \noindent \textbf{3.8. The algebra $\sca_{cl}$ and
the classical Poisson bracket}

\vspace{.12in} Classical symplectic structure, traditionally treated
in the framework of a symplectic manifold $(M, \omega_{cl})$ [where
M is a differentiable manifold of even dimension (say, 2n) and
$\omega_{cl}$ , the classical symplectic form, a nondegenerate
differential 2-form on M], can be realized as a special case of the
symplectic structure treated above by taking $\sca = C^{\infty}(M,C)
\equiv \sca_{cl}$, the commutative algebra of smooth complex-valued
functions on M [with pointwise product: (fg)(x) = f(x)g(x)]. The
star operation in this case is the complex conjugation : $f^*(x) =
\overline{f(x)}$. The derivations of $\sca_{cl}$ are the smooth
complex vector fields. The differential forms of subsection C above
are easily seen, for $\sca = \sca_{cl}$, to be the traditional
differential forms on M. The symplectic structure is defined in
terms of the classical differential form (in canonical coordinates)
\begin{eqnarray} \omega_{cl} = \sum_{j=1}^{n} dp_j \wedge dq_j;
\end{eqnarray} in terms of the general local coordinates $\xi^a
(a=1,..,2n)$, we have $\omega_{cl} = (\omega_{cl})_{ab} d\xi^a
\wedge d\xi^b$. The hamiltonian derivation corresponding to $f \in
\sca_{cl}$ is the Hamiltonian vector field
\[ X_f = \omega_{cl}^{ab} \frac{\partial f}{\partial
\xi^a}\frac{\partial}{\partial \xi^b} \] where $(\omega_{cl}^{ab})$
is the inverse of the matrix $((\omega_{cl})_{ab}).$ The classical
PB is given by
\begin{eqnarray*}
\{  f,g \}_{cl} = X_f (g) = \omega_{cl}^{ab} \frac{\partial
f}{\partial \xi^a} \frac{\partial g}{\partial \xi^b};
\end{eqnarray*}
in canonical coordinates, it takes the traditional form
\begin{eqnarray} \{ f,g \}_{cl} = \sum_{j=1}^{n}\left(\frac{\partial f}
{\partial p_j}\frac{\partial g}{\partial q_j}- \frac{\partial
f}{\partial q_j} \frac{\partial g}{\partial p_j}\right).
\end{eqnarray}
Noting that the form $\omega_{cl}$ and the coordinates $\xi^a$ are
real, the reality properties embodied in the equations (42) and (44)
are obvious in the present case.

\vspace{.12in} \noindent \textbf{3.9. Special algebras; the
canonical symplectic form}

\begin{sloppypar}
In this subsection, we shall consider a distinguished class of
superalgebras (Dass 1993a, 2002; Dubois-Violette 1995, 1999) which
have a canonical symplectic structure associated with them. As we
shall see in the next subsection and in section 7, these
superalgebras play an important role in Quantum mechanics.
\end{sloppypar}

A complex, associative, non-supercommutative  *-superalgebra will be
called \emph{special} if all its superderivations are inner. The
differential 2-form $\omega_c$ defined on such a superalgebra \sca \
by
\begin{eqnarray}
\omega_c (D_A, D_B) = [A,B]
\end{eqnarray}
is said to be the \emph{canonical form} on \sca. It is easily seen
to be closed [the equation $(d \omega_c)(D_A,D_B,D_C) = 0$ is
nothing but the Jacobi identity for the supercommutator], imaginary
(i.e. $\omega_c^* = - \omega_c$) and dimensionless. For any $A \in
\sca$, the equation \begin{eqnarray*} \omega_c(Y_A, D_B) = -
(dA)(D_B) = [A,B]
\end{eqnarray*}
has the unique solution $ Y_A = D_A$. (To see this, note that, since
all derivations are inner, $Y_A = D_{C}$ for some $C \in \sca$; the
condition $\omega_c (D_C, D_B) = [C,B] = [A,B]$ for all $B \in \sca$
\ implies that $(C - A) \in Z(\sca)$. But then $D_C = D_A. \Box$)
This gives
\begin{eqnarray}
i_{D_A} \omega_c = - dA.
\end{eqnarray}
The closed and non-degenerate form $\omega_c$ defines, on \sca, the
\emph{canonical symplectic structure}. It gives, as Poisson bracket,
the supercommutator :
\begin{eqnarray}
\{ A,B \} = Y_A(B) = D_A(B) = [A,B].
\end{eqnarray}
Using equations (48), (20) and (22), it is easily seen that the form
$\omega_c$ is \emph{invariant} in the sense that $ L_X \omega_c = 0$
for all $X \in \sdera$. The invariant symplectic structure on the
algebra $M_n(C)$ of complex $n \times n$ matrices obtained by
Dubois-Violette and coworkers (1994) is a special case of the
canonical symplectic structure on special algebras described above.

If, on a special superalgebra \sca, instead of $\omega_c$, we take
$\omega = b \omega_c$ as the symplectic form (where $b$ is a nonzero
complex number), we have
\begin{eqnarray}
Y_A = b^{-1} D_A ,  \hspace{.4in} \{A, B \} = b^{-1}[A,B].
\end{eqnarray}
We shall make use of such a symplectic structure (with $b = -i
\hbar$)  in the following subsection and in the treatment of quantum
mechanics in section 7. (Note that b must be imaginary to make
$\omega$ real.) Such a symplectic structure with general nonzero b
will be referred to as the \emph{quantum symplectic structure with
parameter b}.

\vspace{.15in} \noindent \textbf{3.10. The quantum symplectic form}

\vspace{.12in} In the traditional QM of a non-relativistic spinless
particle, one generally operates in the Schr$\ddot{o}$dinger
representation which employs the  Hilbert space $\sch = L^2(R^3,dx)$
of complex square-integrable functions on $R^3$ and a suitable dense
domain \scd \ in \sch \ which is generally taken to be the space
$\mathcal{S}(R^3)$ of Shwartz functions.. The fundamental
observables of such a particle are the Cartesian components $X_j,
P_j (j=1,2,3)$ of position and momentum vectors which are
self-adjoint linear operators represented as
\begin{eqnarray}
(X_j \phi)(x) = x_j \phi(x); \ \ (P_j \phi)(x) = -i \hbar
\frac{\partial \phi}{\partial x_j} \ \ \textnormal{for all} \ \phi
\in \mathcal{D}.
\end{eqnarray} These operators satisfy the canonical commutation
relations (CCR)
\begin{eqnarray} [X_j, X_k] = 0 = [P_j, P_k]; \ \ \   [X_j, P_k] =
i \hbar I  \ \ (j,k = 1,2,3) \end{eqnarray} where I is the unit
operator. Other operators appearing in QM of the particle belong to
the algebra \sca \ generated by the operators $X_j, P_j $ (j= 1,2,3)
and I [subject to the CCR (52)]. The space $ \scd =
\mathcal{S}(R^3)$ is clearly an invariant domain for all elements of
\sca. Defining a
*-operation on \sca \ by $ A^* = A^{\dagger}|\scd$ (where $A^{\dagger}$
is the Hilbert space adjoint of A), the Hermitian elements of \sca \
represent the general observables of the particle.

  The algebra \sca \ obtained above is `special' in the sense
defined in the previous subsection (Dubois-Violette 1995,1999); one
has, therefore, a canonical form $\omega_c$ defined on it. The
\emph{quantum symplectic structure} is defined
 on \sca \ by employing the \emph{quantum symplectic form}
 \begin{eqnarray} \omega_Q = -i \hbar \omega_c. \end{eqnarray}
 Note that the factor i serves to make $\omega $ real and $\hbar$
 to give it the dimension of action (which is the correct dimension
 of a symplectic form in mechanics). The minus sign is a matter of
 convention. Eq.(50) now gives the \emph{ quantum Poisson bracket}
 \begin{eqnarray} \{ A, B \}_Q = (-i \hbar)^{-1} [A,B].
 \end{eqnarray}
The CCR (52) can now be expressed as the quantum Poisson brackets
involving the canonical pairs ($X_j, P_j$) :
\begin{eqnarray}
\{ X_j, X_k \}_Q = 0 = \{ P_j, P_k \}_Q; \ \ \{ P_j, X_k \}_Q =
\delta_{jk} I \quad (j,k = 1,2,3).
\end{eqnarray}

 \vspace{.15in} \noindent \textbf{4. THE FORMALISM OF SUPMECH}

We shall now develop an algebraic scheme of mechanics employing
noncommutative differential forms and symplectic structures of the
previous section. Most developments are parallel to those in
classical Hamiltonian mechanics. States are defined as in
traditional algebraic approaches; a condition of `compatible
completeness' between observables and pure states is an important
extra input. The treatment of symmetries and conservation laws is
along the lines dictated by the formalism. Lie group actions on a
symplectic superalgebra are treated in some detail. Augmented
symplectics including time are treated which includes noncommutative
generalizations of the Poincar$\acute{e}$-Cartan form and the
symplectic version of Noether's theorem. Systems with configuration
space are treated and observables corresponding to localization are
introduced.

\vspace{.12in} \noindent \textbf{4.1. The system algebra and states}

\vspace{.1in} Supmech associates, with every physical system, a
symplectic superalgebra $(\sca, \omega)$ of the type considered in
the previous section. Here we shall treat the term `physical system'
informally as is traditionally done; some formalities in this
connection will be taken care of in section 9 where the axioms are
stated. The even Hermitian elements of \sca \ represent
\emph{observables} of the system. The collection of all observables
in \sca \ will be denoted as $\mathcal{O}(\sca)$.

To take care of limit processes and continuity of mappings, we must
employ topological algebras. The choice of the admissible class of
topological algebras must meet the following reasonable
requirements:

\vspace{.1in} \noindent (i) It should be closed under the formation
of (a) topological completions and (b) tensor products. [Both are
nontrivial requirements (Dubin and Hennings 1990).]

\noindent (ii) It should include

\noindent (a) the Op$^{*}$-algebras (Horuzhy 1990) based on the
pairs $(\sch, \scd)$ where \sch \ is a separable Hilbert space and
\scd \ a dense linear subset of \sch. [Recall that such an algebra
is an algebra of operators which, along with their adjoints, map
\scd \ into itself. The *-operation on the algebra is defined as the
restriction of the Hilbert space adjoint on $\mathcal{D}$ as in
section 3.10. These are the algebras of operators (not necessarily
bounded) appearing in the traditional Hilbert space QM; for example,
the algebra \sca \ in section 3.10 belongs to this class.];

\noindent (b) Algebras of smooth functions on manifolds (to
accommodate classical dynamics).

\noindent (iii) The GNS representations of the system algebra (in
the non-supercommutative case) induced by various states must have
\emph{separable} Hilbert spaces as the representation spaces.

\vspace{.1in} The right choice appears to be the
$\hat{\otimes}$-(star-)algebras of Helemskii (1989) (i.e. locally
convex *-algebras which are complete and Hausdorff with a jointly
continuous product) satisfying the additional condition of being
separable. [\emph{Note}. The condition of separability may have to
be dropped in applications to quantum field theory.] Henceforth all
(super-)algebras employed will be assumed to belong to this class.
For easy reference, unital *-algebras of this class will be called
\emph{supmech-admissible}. Mappings between topological spaces
should henceforth be understood as continuous.

 A \emph{state} on a (unital) *-algebra \sca \  is a linear
functional $\phi$ on \sca \ which is (i) positive [which means $
\phi(A^*A) \geq 0$ for all $A \in \sca$) and (ii) normalized [i.e.
$\phi (I) =1$]. Given a state $\phi$, the quantity $\phi(A)$ for any
observable A is real (this can be seen by considering, for example,
the quantity$\phi[(I + A)^*(I + A)]$) and is to be interpreted as
the expectation value of
 A in the state $\phi$. Following general usage in literature, we
shall call observables of the form $A^*A$ or sum of such terms
\emph{positive} (strictly speaking, the term `non-negative' would be
more appropriate); states assign non-negative expectation values to
such observables. The family of all states on \sca \ will be denoted
as $\mathcal{S}(\sca)$. It is easily seen to be closed under convex
combinations: given $ \phi_i \in \mathcal{S}(\sca)$ , i = 1,..,n and
$p_i \geq 0 $ with $p_1 + ..+ p_n = 1$, we have $ \phi =
\sum_{i=1}^n p_i \phi_i $ also in $\mathcal{S}(\sca)$.[More
generally, the integral of an $\mathcal{S}(\sca)$-valued function on
a probability space integrated  with respect to  the probability
measure is an element of $\mathcal{S}(\sca)$.] States which cannot
be expressed as nontrivial convex combinations of other states will
be called \emph{pure} states and others \emph{mixed} states or
\emph{mixtures}. The family of pure states of \sca \  will be
denoted as \sone(\sca). The triple $(\sca, \sone(\sca), \omega)$
will be referred to as a \emph{symplectic triple}.

\vspace{.1in} \noindent \emph{Note}. In physics literature, it is
sometimes found convenient to include, among states, those for which
the magnitudes of expectation values of some observables are
infinite. For example, in practical work in QM, one employs wave
functions which give infinite values for the expectation values of
some unbounded observables like position, momentum or energy. It
needs to be made clear, however, that \emph{physical states} (or
\emph{physically realizable states}) must be restricted to those for
which the expectation values of \emph{all} observables are finite.
In practical quantum mechanical work this would mean that only wave
functions lying in a common invariant dense domain of all
observables must be treated as representing (pure) physical states.
Our formal definition of state, in fact, allows only physical
states. (This is because the domain of definition of the functionals
defining states is the whole algebra \sca.)

\vspace{.1in} In a sensible physical theory, the collection of pure
states must be rich enough to distinguish between two different
observables. (Mixtures represent averaging over ignorances over and
above those implied by the irreducible probabilistic aspect of the
theory; they, therefore, are not the proper objects for a statement
of the above sort.) Similarly, there should be enough observables to
distinguish between different pure states. These requirements are
taken care of in supmech  by stipulating that the pair (\oa,
\sone(\sca)) be \emph{compatibly complete} in the sense that

\noindent (i) given $A,B \in \oa, A \neq B, $ there should be a
state $\phi \in \sone(\sca)$ such that $ \phi(A) \neq \phi(B)$;

\noindent
(ii) given two different states $\phi_1$ and $\phi_2$ in \sone(\sca), there
should be an $A \in \oa$ \ such that $ \phi_1(A) \neq \phi_2(A)$.

\noindent We shall refer to this condition as the `CC condition' for
the pair $ (\oa, \sone(\sca))$.

Expectation values of all even elements of \sca \ can be expressed
in terms of those of the observables (by considering the breakup of
such an element into its Hermitian and anti-Hermitian part). This
leaves out the odd elements of \sca. It appears reasonable to demand
that the expectation values $\phi(A)$ of all odd elements $ A \in
\sca$ must vanish for all  pure states (and, therefore, for all
states).

Next, we  consider the  relation between states and traditional
probability measures. We shall introduce classical probabilities in
the formalism through a straightforward formalization of a
measurement situation. To this end, we consider a measurable space
$(\Omega, \mathcal{F})$ and associate, with  every measurable set $E
\in \mathcal{F}$, a positive observable $\nu(E)$ such that
\[ (i) \ \nu(\emptyset)=0, \ (ii) \  \nu(\Omega) = I, \
(iii)\  \nu(\cup_i E_i) = \sum_i \nu(E_i)\  \textnormal{(for
disjoint unions)}. \] (The last equation means that, in the relevant
topological algebra, the possibly infinite sum on the right hand
side is well defined and equals the left hand side.) Then, given a
state $\phi$, we have a probability measure $p_{\phi}$
 on $(\Omega, \mathcal{F})$ given by
\begin{eqnarray}
p_{\phi}(E) = \phi(\nu(E)) \ \ \forall E \in \mathcal{F}.
\end{eqnarray}
The family $ \{ \nu (E), E \in \mathcal{F} \}$ will be called a
\emph{positive observable-valued measure} (PObVM) on $(\Omega,
\mathcal{F})$. It is the abstract counterpart of the `positive
operator-valued measure' (POVM) employed in Hilbert space QM (Holevo
1982; Busch et al. 1995). The objects $\nu (E)$ may be called
\emph{supmech events} (representing possible outcomes in a
measurement situation); a state assigns probabilities to these
events. Eq.(56) represents the desired relationship between the
supmech expectation values and classical probabilities.

Denoting the algebraic dual of the superalgebra \sca \ by $\sca^*$,
an automorphism $\Phi : \sca \rightarrow \sca$ induces the transpose
mapping $\tilde{\Phi} : \sca^* \rightarrow \sca^*$ such that
\begin{eqnarray}
\tilde{\Phi}(\phi)(A) = \phi(\Phi(A)) \ \mbox{or}
\ < \tilde{\Phi}(\phi), A> = <\phi, \Phi(A)>
\end{eqnarray}
where the second alternative has employed the dual space pairing
$<,>$. The mapping $\tilde{\Phi}$ (which is easily seen to be linear
and bijective) maps states (which form a subset of $\sca^*$) onto
states. To see this, note that

\vspace{.1in} \noindent (i) $ \tilde{\Phi}(\phi)(A^*A) =
\phi(\Phi(A^*A)) = \phi(\Phi(A)^*\Phi(A)) \geq 0;$

\vspace{.1in} \noindent (ii)$ [\tilde{\Phi}(\phi)](I) =
\phi(\Phi(I)) =\phi(I) = 1. $

\vspace{.1in} \noindent The linearity of $\tilde{\Phi}$ (as a
mapping on  $\sca^*$  ) ensures that it preserves convex
combinations of states. In particular, it maps pure states onto pure
states. We have, therefore, a bijective mapping $\tilde{\Phi} :
\sone(\sca) \rightarrow \sone(\sca).$

When $\Phi$ is a canonical transformation, the condition  $\phstup \omega =
\omega$ gives, for $X,Y \in \sdera$,
\begin{eqnarray*}
\omega(X,Y) = (\phstup \omega)(X,Y) = \Phi^{-1}[\omega(\Phi_*X, \phst Y)]
\end{eqnarray*}
which gives
\begin{eqnarray}
\Phi[\omega(X,Y)] = \omega (\phst X, \phst Y).
\end{eqnarray}
Taking expectation value of both sides of this equation in a state $\phi$,
we get
\begin{eqnarray}
(\tilde{\Phi}\phi)[\omega(X,Y)] = \phi[\omega(\phst X, \phst Y)].
\end{eqnarray}
The dependence on X,Y in this equation can be gotten rid of by defining
$\omega_{\Phi}$ by
\begin{eqnarray}
\omega_{\Phi}(X,Y) = \omega(\phst X, \phst Y).
\end{eqnarray}
Eq(59)  can now be written as
\begin{eqnarray}
(\tilde{\Phi}\phi)[\omega(.,.)] = \phi[\omega_{\Phi}(.,.)].
\end{eqnarray}
It is generally simpler to use Eq.(59). When $\Phi$ is an
infinitesimal canonical transformation generated by $G \in \sca$, we
have
\begin{eqnarray}
\tilde{\Phi}(\phi) (A) = \phi (\Phi(A)) \simeq \phi (A + \epsilon \{
G,A \}).
\end{eqnarray}
Putting $\tilde{\Phi}(\phi) = \phi + \delta \phi$, we have
\begin{eqnarray}
(\delta \phi)(A) = \epsilon \phi (\{ G,A \}).
\end{eqnarray}

\vspace{.12in} \noindent \textbf{4.2. Dynamics}

\vspace{.1in} Dynamics is described by specifying an observable H,
called the \emph{Hamiltonian};  the  evolution of the system is
given in terms of the one-parameter family $\Phi_t$ of canonical
transformations generated by H. (The parameter t is supposed to be
an evolution parameter which need not always be the conventional
time.) Writing $\Phi_t(A) = A(t)$ and recalling Eq.(40), we have the
\emph{Hamilton's equation} of supmech :
\begin{eqnarray}
\frac{dA(t)}{dt} = Y_H[A(t)] = \{ H, A(t) \}.
\end{eqnarray}
The triple $(\sca, \omega, H)$ [or, more appropriately, the
quadruple $(\sca, \sone (\sca), \omega, H)$] will be called a
\emph{supmech Hamiltonian system}; it is the analogue of a classical
Hamiltonian system $ (M, \omega_{cl}, H_{cl})$ [where (M, \omcl) is
a symplectic manifold and $H_{cl}$, the classical Hamiltonian (a
smooth real-valued function on M); note that the specification of
the symplectic manifold M serves to define both  observables and
pure states in classical mechanics]. As far as the evolution is
concerned, the Hamiltonian is, as in the classical case, arbitrary
up to the addition of a constant multiple of the identity element.
We shall assume that H is bounded below in the sense that its
expectation values in all pure states (hence in all states) must be
bounded below.

This is the analogue of the Heisenberg picture in traditional QM. An
equivalent description is obtained by transferring the time
dependence to states through the relation [see Eq.(57)]
\[ < \phi(t), A> = <\phi, A(t)> \]
where $\phi(t) = \tilde{\Phi}_t(\phi)$. The mapping $\tilde{\Phi}_t$
satisfies the condition (61) which [with $\Phi = \Phi_t$] may be
said to represent the canonicality of the evolution of states. With
$ \Phi = \Phi_t$ and G = H, Eq.(63) gives the \emph{Liouville
equation} of supmech:
\begin{eqnarray}
\frac{d \phi (t)}{dt}(A) = \phi (t)( \{ H,A \}) \hspace{.12in}
\mbox{or} \hspace{.12in} \frac{d \phi(t)}{dt} (.) = \phi (t)(\{ H,.
\}).
\end{eqnarray}
This is the analogue of the Schr$\ddot{o}$dinger picture in
traditional QM.

\vspace{.15in} \noindent \textbf{4.3. Equivalent descriptions;
Symmetries and conservation laws}

\vspace{.12in}
 By a `description' of a system, we shall mean specification of its
 triple $(\sca, \scs(\sca), \omega)$. Two descriptions are said to
 be \emph{equivalent} if they are
related through a pair of isomorphisms $\Phi_1 : \sca \rightarrow
\sca$  and $\Phi_2 : \scs(\sca) \rightarrow \scs(\sca)$ such that
the symplectic form and the expectation values are preserved :
\begin{eqnarray}
\Phi_1^* \omega = \omega; \hspace{.12in}
\Phi_2 (\phi)[\Phi_1(A)] = \phi(A)
\end{eqnarray}
for all $ A \in \sca $ and $\phi \in \scs(\sca)$. The second
equation above and Eq(57) imply tht we must have $\Phi_2 =
(\tilde{\Phi}_1)^{-1}$. Two equivalent descriptions are, therefore,
related through a canonical transformation on \sca \ and the
corresponding inverse transpose transformation on the states. An
infinitesimal transformation of this type generated by $ G \in \sca$
takes the form
\begin{eqnarray}
\delta A = \epsilon \{ G, A \}, \hspace{.2in} (\delta \phi)(A) =
- \epsilon \phi (\{ G,A \})
\end{eqnarray}
for all $A \in \sca$ and $ \phi \in \scs (\sca).$

These transformations may be called symmetries of the formalism;
they are the analogues of simultaneous unitary transformations on
operators and state vectors in a Hilbert space preserving
expectation values of operators. Symmetries of dynamics are the
subclass of these which leave the Hamiltonian invariant:
\begin{eqnarray}
\Phi_1 (H) = H.
\end{eqnarray}
For an infinitesimal transformation generated by $ G \in \sca $, this
equation gives
\begin{eqnarray}
\{ G,H \} = 0.
\end{eqnarray}
It now follows from the Hamilton equation (64) that (in the
`Heisenberg picture' evolution) G is a constant of motion. This is
the situation familiar from classical and quantum mechanics:
generators of symmetries of the Hamiltonian are conserved quantities
and vice-versa.

\noindent \emph{Note}. Noting that a symmetry operation is uniquely
defined by any one of the two mappings $\Phi_1$ and $\Phi_2$, we can
be flexible in the implementation of symmetry operations. It is
often useful to implement them such that the symmetry operations
act, in a single implementation, \emph{either} on states \emph{or}
on observables, and the two actions are related as the Heisenberg
and Schr$\ddot{o}$dinger picture evolutions above [see Eq.(57)]; we
shall refer to this type of implementation as \emph{unimodal}. In
such an implementation, the second equation in (67) will not have a
minus sign on the right.

For future reference, we define equivalence of supmech Hamiltonian
systems. Two supmech Hamiltonian systems \[ (\sca, \sone (\sca),
\omega, H) \ \ \textnormal{and} \ \  (\sca^{\prime}, \sone
(\sca^{\prime}), \omega^{\prime}, H^{\prime}) \] are said to be
equivalent if they are related through a pair $\Phi = (\Phi_1,
\Phi_2)$ of bijective mappings such that $\Phi_1 : (\sca, \omega)
\rightarrow (\sca^{\prime}, \omega^{\prime})$ is a symplectic
mapping connecting the Hamiltonians [i.e. $ \Phi_1^* \omega^{\prime}
= \omega$ and $\Phi_1 (H) = H^{\prime}]$ and $\Phi_2: \sone(\sca)
\rightarrow \sone(\sca^{\prime})$ such that $<\Phi_2(\phi),
\Phi_1(A)> = <\phi, A>.$

\vspace{.12in} \noindent \textbf{4.4. Symplectic actions of Lie
groups}

In this subsection and the next section, we shall generally employ bosonic
objects. The square brackets will, therefore, be commutators; the subscript --
(minus) for the latter will be omitted.

The study of symplectic actions of  Lie groups in supmech proceeds
generally parallel to the classical case (Sudarshan and Mukunda
1974; Arnold 1978; Guillemin and Sternberg 1984; Woodhouse 1980) and
promises to be quite rich and rewarding. Here we shall present the
essential developments mainly to provide background material for the
next section.

Let G be a connected Lie group with Lie algebra \scg. Elements of G,
\scg \ and \gstar \  (the dual space of \scg) will be denoted,
respectively, as g,h,.., $\xi, \eta,..$ and $\lambda, \mu,..$. The
pairing between \gstar \ and \scg \ will be denoted as $<.,.>$.
Choosing a basis $ \{\xi_a; a = 1,..,r\}$ in \scg, we have the
commutation relations $ [ \xi_a, \xi_b] = C_{ab}^c \xi_c.$ The dual
basis in \gstar \ is denoted as $ \{ \lambda^a \}$ (so that $ <
\lambda^a, \xi_b> = \delta^a_b$). The action of G on \scg \ (adjoint
representation) will be denoted as $ Ad_g : \scg \rightarrow \scg$
and that on \gstar \ (the coadjoint representation) by $Cad_g :
\gstar \rightarrow \gstar;$ the two are related as $ < Cad_g
\lambda, \xi> = <\lambda, Ad_{g^{-1}}\xi>.$ With the bases chosen as
above, the matrices in the two representations are related as $
(Cad_g)_{ab} = (Ad_{g^{-1}})_{ba}.$

Recalling the mappings $\Phi_1$ and $\Phi_2$ of the previous
subsection, a \emph{symplectic action} of of G on a symplectic
superalgebra $(\sca, \omega)$ is given by the assignment, to each $
g \in G,$ a symplectic mapping  (canonical transformation)
$\Phi_1(g) : \sca \rightarrow \sca$ which is a group action [which
means that $\Phi_1(g) \Phi_1(h) = \Phi_1(gh)$ and $\Phi_1(e)=
id_{\sca}$ in obvious notation]. The action on the states is given
by the mappings $ \Phi_2(g) = [\tilde{\Phi}_1(g)]^{-1}$.

A one-parameter subgroup g(t) of G generated by $\xi \in \scg$
induces a locally Hamiltonian derivation $Z_{\xi} \in SDer(\sca)$ as
the generator of the one-parameter family $\Phi_1(g(t)^{-1})$ of
canonical transformations of \sca : For small t
\begin{eqnarray*} \Phi_1(g(t)^{-1})(A) \simeq A + t Z_{\xi}
(A). \end{eqnarray*}

\noindent \emph{Note.} We employed $\Phi_1(g(t)^{-1})$ (and not $
\Phi_1(g(t))$) for defining $Z_{\xi}$ above because the former
correspond to the right action of G on \sca.

 The correspondence $ \xi \rightarrow Z_{\xi}$ is a Lie algebra
homomorphism : \begin{eqnarray*}  Z_{[\xi,\eta]} = [Z_{\xi},
Z_{\eta}]. \end{eqnarray*} \emph{Proof.} The objects $Z_{\xi}$ are
linear in $\xi$. To see this, note that, for infinitesimal group
parameters $\epsilon^a$, we have \[ \Phi_1(exp(\epsilon^a
\xi_a)^{-1})(A) \simeq A + \epsilon^a Z_{\xi_a}(A). \] With $\xi =
\alpha^a \xi_a$ and $g(t) = exp(t\xi)$, we have $\epsilon^a = t
\alpha^a$; this gives $t Z_{\xi}(A) = t \alpha^a Z_{\xi_a}(A)$ for
all $A \in \sca$, hence $Z_{\xi} = \alpha^a Z_{\xi_a}$.

Defining the group actions $L_g, R_g$ and $A_g$ by $ L_g(h) = gh,
R_g(h) = hg, A_g(h) =g^{-1}hg$ and recalling that $\xi \in
\mathcal{G}$ is a left-invariant vector field on G (i.e. $(L_g)_*
\xi = \xi \ \forall \ \xi \in \mathcal{G}$), we note that, for any
$h \in G, \xi^{\prime} \equiv (R_h)_* \xi = (A_h)_* \xi$ generates
the one-parameter subgroup $g^{\prime}(t) = h^{-1}g(t)h$. Now,
recalling Eq.(3), we have, for small t,
\begin{eqnarray*} \Phi_1(g^{\prime}(t)^{-1})(A) & = &
\Phi_1(h^{-1}g( t)^{-1}h)(A) \\ & = & \Phi_1(h^{-1})
\Phi_1(g(t)^{-1}) \Phi_1(h) (A) \\ & \simeq & A + t \Phi_1(h^{-1})
Z_{\xi}[ \Phi_1(h)(A)] \\ & = & A + t [\Phi_1(h^{-1})_*Z_{\xi}](A) \\
& = & A + t Z_{\xi^{\prime}} (A) \end{eqnarray*} which gives
\begin{eqnarray*} Z_{\xi^{\prime}} = \Phi_1(h^{-1})_* Z_{\xi}.
\end{eqnarray*}
Replacing h above by the general element h(s) of the one-parameter
subgroup generated by $\eta \in \mathcal{G}$, we have
\begin{eqnarray*} \xi^{\prime}(s) \equiv (R_{h(s)})_* \xi \simeq \xi
+ s L_{\eta} (\xi) = \xi + s [\eta, \xi]. \hspace{.9in}  (*)
\end{eqnarray*} Moreover
\begin{eqnarray*} Z_{\xi^{\prime}(s)} & = & \Phi_1(h(s)^{-1})_* Z_{\xi}
\\ & \simeq & Z_{\xi} + s L_{Z_{\eta}}(Z_{\xi}) \\ & = & Z_{\xi} +
s[Z_{\eta}, Z_{\xi}].  \hspace{.9in} (**) \end{eqnarray*} Recalling
the linearity of $Z_{\xi^{\prime}(s)}$ in $\xi^{\prime}(s)$,
equations (*) and (**) give the desired result. $\Box$

The G-action is said to be \emph{hamiltonian} if the derivations
$Z_{\xi}$ are Hamiltonian, i.e. for each $\xi \in \mathcal{G},
Z_{\xi} = Y_{h_{\xi}}$  for some $h_{\xi} \in \sca$ (called the
\emph{hamiltonian} corresponding to $\xi$). These hamiltonians are
arbitrary up to  addition of  multiples of the unit element. This
arbitrariness can be somewhat reduced by insisting that $h_{\xi}$ be
linear in $\xi$.(This can be achieved by first defining the
hamiltonians for the members of a basis in $\mathcal{G}$ and then
defining them for general elements as appropriate linear
combinations of these.) We shall always assume this linearity.

A hamiltonian G-action satisfying the additional condition
\begin{eqnarray}
\{ h_{\xi}, h_{\eta} \} = h_{[\xi,\eta]} \ \mbox{for all \ } \xi, \eta \in
\mathcal{G}
\end{eqnarray}
is called a \emph{Poisson action}. The hamiltonians of a Poisson action
have the following equivariance property :
\begin{eqnarray}
\Phi_1(g) (h_{\xi}) = h_{Ad_g(\xi)}.
\end{eqnarray}
Since G is connected, it is adequate to verify this relation for
infinitesimal group actions. Denoting by g(t) the one-parameter
group generated by $\eta \in \mathcal{G}$, we have, for small t,
\begin{eqnarray*}
\Phi_1(g(t))(h_{\xi}) \hspace{.1in}   \simeq \hspace{.1in}  h_{\xi} +
t \{ h_{\eta}, h_{\xi} \} \hspace{.1in}
                         = \hspace{.1in}  h_{\xi} + t h_{[\eta, \xi]}
         \hspace{.1in}  = \hspace{.1in}  h_{\xi + t [\eta, \xi]}
           \hspace{.1in} \simeq \hspace{.1in}  h_{Ad_{g(t)}\xi}
 \end{eqnarray*}
completing the verification.

A Poisson action is not always admissible.
The obstruction to such an  action is determined by the objects
\begin{eqnarray}
\alpha (\xi, \eta) = \{h_{\xi}, h_{\eta} \} - h_{[\xi, \eta]}
\end{eqnarray}
which are easily seen to have vanishing Hamiltonian derivations :
\begin{eqnarray*} Y_{\alpha(\xi,\eta)} & = & [Y_{h_{\xi}},
Y_{h_{\eta }}] - Y_{h_{[\xi, \eta]}} \nonumber \\ & = & [Z_{\xi},
Z_{\eta}] - Z_{[\xi,\eta]} = 0 \end{eqnarray*} and hence vanishing
Poisson brackets with all elements of \sca. [This last condition
defines the so-called  \emph{neutral elements} (Sudarshan and
Mukunda 1974) of the Lie algebra$(\sca, \{, \})$. They clearly form
a complex vector space which will be denoted as $\mathcal{N}$.] We
also have
\begin{eqnarray*}
\alpha([\xi, \eta], \zeta) + \alpha([\eta, \zeta], \xi) + \alpha
([\zeta,\xi], \eta) = 0.
\end{eqnarray*}
The derivation (Woodhouse 1980) of this result in classical
mechanics employs only the standard properties of PBs and remains
valid in supmech. Comparing this equation with Eq.(25), we see that
$ \alpha(.,.) \in Z^2_0(\scg,\mathcal{N})$. A redefinition of the
hamiltonians $ h_{\xi} \rightarrow h^{\prime}_{\xi} = h_{\xi} +
k_{\xi}I$ (where the scalars $k_{\xi}$ are linear in $\xi$) changes
$\alpha$ by a coboundary term:
\begin{eqnarray*}
\alpha^{\prime} (\xi,\eta) = \alpha(\xi, \eta) - k_{[\xi,\eta]}I
\end{eqnarray*}
showing that the obstruction is characterized by a cohomology class
of $\mathcal{G}$ [i.e. an element of $H^2_0(\scg,\mathcal{N})$]. A
necessary and sufficient condition for the admissibility of Poisson
action of G on \sca \ is that it should be possible to transform
away all the obstruction 2-cocycles by redefining the hamiltonians,
or, equivalently, $H_0^2(\mathcal{G},\mathcal{N}) = 0.$

We restrict ourselves to the special case, relevant for application
in section 5, in which \scg \ is a finite dimensional real Lie
algebra and the cocycles $\alpha$ are multiples of the unit element
: \[ \alpha(\xi, \eta) = \underline{\alpha}(\xi, \eta) I; \] here
the quantities $\underline{\alpha}(\xi, \eta)$ must be real numbers
because the set of observables is closed under Poisson brackets.
This implies $\mathcal{N} = R$, the set of real numbers. In this
case, the relevant cohomology group $H^2_0(\scg,R)$ is a real finite
dimensional vector space; we shall take it to be $R^m$. In this
case, as in classical symplectic mechanics (Sudarshan and Mukunda
1974; Cari$\tilde{n}$ena and Santander 1975;  Alonso 1979),
Hamiltonian group actions (more generally, Lie algebra actions) with
nontrivial neutral elements can be treated as Poisson actions of a
(Lie group with a) larger Lie algebra $\hat{\scg}$ obtained as
follows: Let $ \eta_r(.,.)(r = 1,..,m)$ be a set of representatives
in $Z^2_0(\scg,R)$ of a basis in $H^2_0(\scg,R)$. We add extra
generators $M_r $ to the basis $\{ \xi_a \}$ of \scg \ and take the
commutation relations of the larger Lie algebra $\hat{\scg}$ as
\begin{eqnarray}
[\xi_a, \xi_b] = C_{ab}^c \xi_c + \sum_{r=1}^{m}\eta_r(\xi_a, \xi_b) M_r; \ \
  [\xi_a, M_r] \ = \ 0 \ = \ [M_r, M_s].
\end{eqnarray}
The simply connected Lie group $\hat{G}$ with the Lie algebra
$\hat{\scg}$ is called the \emph{projective group} (Alonso 1979) of
G; it is generally a central extension of the universal covering
group $\tilde{G}$ of G.

The hamiltonian action of G with the cocycle $\alpha$ now becomes a
Poisson action of $\hat{G}$ with the Poisson bracket relations
(writing $h_{M_r}= h_r$)
\begin{eqnarray} \{ h_a, h_b \} = C_{ab}^c h_c + \sum _{r=1}^{m}
\eta_r(\xi_a, \xi_b) h_r \nonumber \\
\{h_a, h_r\} = 0 = \{h_r, h_s\}. \end{eqnarray}

\vspace{.12in}\begin{sloppypar} \noindent \emph{Momentum map}. In
classical mechanics, given a Poisson action of a connected Lie group
G on a symplectic manifold  $(M, \omega_{cl})$ [with
hamiltonians/comoments $ h^{(cl)}_{\xi} \in C^{\infty}(M)$], a
useful construction is the so-called \emph{momentum map}(Souriau
1997; Arnold 1978; Guillemin and Sternberg 1984)  $ P: M \rightarrow
\mathcal{G}^*$ given by
\end{sloppypar}
\begin{eqnarray}
< P(x), \xi>\mbox{\  = \ }h^{(cl)}_{\xi}(x) \ \ \forall x \in M
\mbox{\ and \ } \xi \in \mathcal{G}.
\end{eqnarray}
This map relates the symplectic action $\Phi_g$ of G on M ($\Phi_g :
M \rightarrow M, \Phi_g^* \omega_{cl} = \omega_{cl}\  \forall g \in
G$) and the transposed adjoint action on $\mathcal{G}^*$ through the
equivariance property
\begin{eqnarray}
P(\Phi_g(x)) = Ad_g^*(P(x)) \ \forall x \in M \mbox{\ and \ } g \in G.
\end{eqnarray}

Noting that points of M are pure states of the algebra $ \sca_{cl} =
C^{\infty}(M)$, the map P may be considered as the restriction to M of the
dual/transpose $\tilde{h}^{(cl)}: \sca_{cl}^* \rightarrow \mathcal{G}^*$ of the
linear map $h^{(cl)}: \mathcal{G}
\rightarrow  \sca_{cl}$ [given by $ h^{(cl)}(\xi) = h^{(cl)}_{\xi}$]:
\begin{eqnarray*}
< \tilde{h}^{(cl)}(u), \xi> = <u, h^{(cl)}(\xi)> \
\forall \ u \in \mathcal{A}_{cl}^* \mbox{\ and \ } \xi \in \mathcal{G}.
\end{eqnarray*}
The analogue of M in supmech is $\sone = \sone(\sca)$. Defining
$ h : \mathcal{G} \rightarrow
\sca $ by $ h(\xi) = h_{\xi},$ the analogue of the momentum map in supmech is
the mapping $\tilde{h}: \sone \rightarrow \mathcal{G}^* $ (considered as the
restriction to \sone \ of the mapping $\tilde{h} : \sca^* \rightarrow
\mathcal{G}^*$) given by
\begin{eqnarray}
<\tilde{h}(\phi), \xi> = < \phi, h(\xi) > = < \phi, h_{\xi}>.
\end{eqnarray}
Recalling the symplectic  mappings $\Phi_1$ and $\Phi_2$ and
Eq.(71), we have
\begin{eqnarray*}
< \tilde{h}(\Phi_2(g) \phi ), \xi> & = &  <\Phi_2(g) \phi, h_{\xi}>
                         =  < \phi, \Phi_1 (g^{-1})(h_{\xi})>
                         =  < \phi, h_{Ad_{g^{-1}}(\xi)} >  \\
                       & = & <\phi, h(Ad_{g^{-1}}(\xi))>
                   = < Cad_g (\tilde{h}(\phi)), \xi>
\end{eqnarray*}
giving finally
\begin{eqnarray}
\tilde{h}(\Phi_2(g) \phi) = Cad_g(\tilde{h}(\phi))
\end{eqnarray}
which is the supmech analogue of Eq.(75). [\emph{Note}. In
Eq.(78),the co-adjoint (and not the transposed adjoint) action
appears on the right because $\Phi_2(g)$ is inverse transpose (and
not transpose) of $\Phi_1(g)$.]

\vspace{.12in} \noindent \textbf{4.5. Generalized symplectic
structures and Hamiltonian systems}

\vspace{.12in} The generalization of the DVNCG scheme introduced in
section 3.5 can be employed to obtain the corresponding
generalization of the supmech formalism. One picks up a
distinguished Lie sub-superalgebra \scx \ of \sdera \ and restricts
the superderivations of \sca \ in all definitions and constructions
to those in \scx. Thus, a symplectic superalgebra $(\sca, \omega)$
is now replaced by a \emph{generalized symplectic superalgebra}
$(\sca, \scx, \omega)$ and a symplectic mappings $\Phi : (\sca,
\scx, \alpha) \rightarrow (\scb, \scy, \beta)$ is restricted to a
superalgebra-isomorphism  $\Phi : \sca \rightarrow \scb$  such that
$\phst : \scx \rightarrow \scy$ is a Lie-superalgebra- isomorphism
and $\phstup \beta = \alpha$. A supmech Hamiltonian system $(\sca,
\sone(\sca), \omega, H)$ is now replace by a \emph{generalized
supmech Hamiltonian sytem} $(\sca, \sone(\sca), \scx, \omega, H)$.
In section 7.5, we shall employ the pairs (\sca, \scx) with \scx =
ISDer(\sca) to define quantum symplectic structure on superalgebras
admitting outer as well as inner superderivations.

\vspace{.1in} \noindent \emph{Note}. In (Dass 2002), the formalism
in section IV (of that paper) was developed right from the beginning
in terms of the pairs (\sca, \scx) (called `algebraic differential
systems' there; we have chosen to dispense with this nomenclature to
avoid confusion of the term `differential system' with its use
elsewhere in mathematics literature). This has the advantage of
extra generality; however, noting that this generality is needed
only at very few places and its use everywhere would  make the
formalism unnecessarily more complicated, the author has not opted
for it in the present work.

\vspace{.12in} \noindent \textbf{4.6. Augmented symplectics
including time; the generalized Poincar$\acute{e}$-Cartan form}

We shall now augment the kinematic framework of supmech by including
time and obtain the non-commutative analogues of the
Poincar$\acute{e}$-Cartan form and the symplectic version of
Noether's theorem (Souriau 1997).

For a system S with associated symplectic superalgebra $ (\sca,
\omega)$ we construct the \emph{extended system algebra} $\sca^e =
C^{\infty}(R) \otimes \sca $ (where the real line R is the carrier
space of  the `time' t) whose elements are finite sums $\sum_i  f_i
\otimes A_i $ (with $f_i \in C^{\infty}(R) \equiv \sca_0$)  which
may be written as $\sum_i f_i A_i $. This algebra is the analogue of
the algebra of functions on the evolution space of Souriau  (the
Cartesian product of the time axis and the phase space --- often
referred to as the phase bundle).The superscript e in $\sca^e$, may,
therefore, also be taken to refer  to `evolution'.

  Derivations on $\sca_0$ are of the form $g(t)\frac{d}{dt}$ and
one-forms of the form h(t)dt where g and h are smooth functions;
there are no nonzero higher order forms. We have, of course, $ dt
(\frac{d}{dt}) = 1$.

A (super-)derivation $D_1$ on $\sca_0$ and $D_2$ on \sca \ extend
trivially to (super-)derivations on $\sca^e$ as $D_1 \otimes
id_{\sca}$ and $id_{\sca_0} \otimes D_2$ respectively (where
$id_{\sca}$ is the identity mapping on \sca); these trivial
extensions may be informally denoted as $D_1$ and $D_2$. With $f
\otimes A$ written as fA, we can write $D_1 (fA) = (D_1f)A$ and
$D_2(fA) = f (D_2A)$.

  The mapping $ \Xi : \sca \rightarrow \sca^e$ given by $ \Xi (A) = 1
\otimes A (=A)$ is an isomorphism of the algebra \sca \ onto the
subalgebra $\tilde{\sca} \equiv 1 \otimes \sca $ of $\sca^e$ and can
be employed to pull back the differential forms on \sca \ to those
on $\tilde{\sca}$. We write, for a p-form $\alpha$ on \sca, $
(\Xi^{-1})^* (\alpha) = \tilde{\alpha}$ and extend this form on
$\tilde{\sca}$ to one on $\sca^e$ by defining
$\tilde{\alpha}(\frac{d}{dt}, ...) = 0.$ We shall generally skip the
tilde. Similarly, we may extend the one-form dt on $\sca_0$ to one
on $\sca^e$ by defining (dt)(X) =0 for all $X \in \sdera$.

  The symplectic structure $\omega$ on \sca \ induces, on $\sca^e$,  a
generalized symplectic structure (of the type introduced in section
4.5) with the distinguished Lie sub-superalgebra \scx \ of
$Sder(\sca^e)$ taken to be the one consisting of the objects
$\{id_{\sca_0} \otimes D; D\in \sdera \}$ which constitute a Lie
sub-superalgebra of $SDer(\sca^e)$ isomorphic to \sdera, thus giving
a generalized symplectic superalgebra $(\sca^e, \scx,
\tilde{\omega})$. The corresponding PBs on $\sca^e$ are trivial
extensions of those on \sca \ obtained by treating the `time' t as
an external parameter; this amounts to extending the C-linearity of
PBs on \sca \ to what is essentially $\sca_0$-linearity :
\begin{eqnarray*}
\{ fA + gB, hC \}_{\sca^e} = fh \{A,C \}_{\sca} + gh \{ B,C
\}_{\sca} \end{eqnarray*} where, for clarity, we have put subscripts
on the PBs to indicate the underlying superalgebras. We shall often
drop these subscripts; the underlying (super-)algebra will be clear
from the context.

To describe dynamics in $\sca^e$, we extend the one-parameter family
$\Phi_t$ of canonical transformations on \sca \ generated by a
Hamiltonian $H \in \sca$ to a one-parameter family $\Phi_t^e$ of
transformations on $\sca^e$ (which are `canonical' in a certain
sense, as we shall see below)  given by
\begin{eqnarray*} \Phi_t^e
(fA) \equiv (fA)(t) = f(t) A(t) \equiv (\Phi_t^{(0)} f) \Phi_t(A)
\end{eqnarray*} where $\Phi_t^{(0)}$ is the one-parameter group of
translations on $\sca_0$ generated by the derivation $\frac{d}{dt}$.
An infinitesimal transformation under the evolution $\Phi_t^e$ is
given by
\begin{eqnarray*} \delta(fA)(t) & \equiv & (fA)(t + \delta t) -
(fA)(t) \\
& = & [\frac{df}{dt} A + f \{H, A \}_{\sca}] \delta t \equiv
\hat{Y}_H (fA) \delta t \end{eqnarray*} where \begin{eqnarray}
\hat{Y}_H = \frac{\partial} {\partial t} + \tilde{Y}_H.
\end{eqnarray} Here $\frac{\partial} {\partial t}$ is the derivation
on $\sca^e$ corresponding to the derivation $\frac{d}{dt}$ on
$\sca_0$ and
\begin{eqnarray*} \tilde{Y}_H = \{ H, . \}_{\sca^e}. \hspace{3.5in}  (*)
\end{eqnarray*} Note that

\noindent (i) $ dt(\hat{Y}_H ) = 1$;

\noindent (ii) the right hand side of the equation (*) remains well
defined if $H \in \sca^e$ (`time dependent' Hamiltonian).
Henceforth, in various formulas in this subsection, H will be
understood to be an element of $\sca^e$.

The obvious generalization of the supmech Hamilton equation (64) to
$\sca^e$ is the equation
\begin{eqnarray}
\frac{d F(t)}{dt} = \hat{Y}_H F(t) = \frac{\partial F(t)}{\partial
t} + \{ H(t), F(t) \}.
\end{eqnarray}

We next consider an object in $\sca^e$ which contains complete
information about the symplectic structure \emph{and} dynamics [i.e.
about $\tilde{\omega}$ and H (up to an additive constant multiple of
I)] and is canonically determined by these objects. It is the 2-form
\begin{eqnarray}
\Omega =  \tilde{\omega} +dt  \wedge dH
\end{eqnarray}
which is `obviously' closed. [To have a formal proof, apply Eq.(133)
below with $ \Omega = 1 \otimes \omega + d_1t \otimes d_2H $.] If
the symplectic structure on \sca \ is exact (with $ \omega = d
\theta$), we have (`obviously') $ \Omega = d \Theta $ where
\begin{eqnarray}
\Theta = \tilde{\theta} - H dt
\end{eqnarray}
is the supmech avatar of the Poincar$\acute{e}$-Cartan form in
classical mechanics. [Again, for a formal derivation, use Eq.(133)
with $\Theta = 1 \otimes \theta - dt \otimes H$.]

The closed form $\Omega$ is generally not non-degenerate. It defines
what may be called a \emph{presymplectic structure} (Souriau 1997)
on $\sca^e$. In fact, we have here the noncommutative analogue of a
special type of presymplectic structure called \emph{contact
structure} (Abraham and Marsden 1978; Berndt 2001); it may be called
the Poincar$\acute{e}$- Cartan contact structure. We shall, however,
not attempt a formal development of noncommutative contact
structures here.

A \emph{symplectic action} of a Lie group G on the presymplectic
space $(\sca^e, \Omega)$ is the assignment, to every $g \in G$, an
automorphism $\Phi(g)$ of the superalgebra $\sca^e$ having the usual
group action properties and such that $\Phi(g)^* \Omega = \Omega$.
This implies, as in section 4.4, that, to every element $\xi$ of the
Lie algebra \scg \ of G, corresponds a derivation $Z_{\xi}$ such
that $L_{Z_{\xi}} \Omega = 0$ which, in view of the condition $d
\Omega = 0$, is equivalent to the condition
\begin{eqnarray}
d(i_{Z_{\xi}} \Omega) = 0.
\end{eqnarray}

We shall now verify that the one-parameter family $\Phi_t^{(e)}$ of
transformations on $\sca^e$ is symplectic/canonical. For this, it is
adequate to verify that Eq.(83) holds with $ Z_{\xi} = \hat{Y}_H$.
We have, in fact, the stronger relation
\begin{eqnarray}
  i_{\hat{Y}_H} \Omega = 0.
\end{eqnarray}
Indeed
\begin{eqnarray*}
i_{\hat{Y}_H} \Omega & = & i_{\partial/\partial t} \Omega +
                           i_{\tilde{Y}_H} \Omega \\
& = & i_{\partial/\partial t}(dt \wedge dH) +
                     i_{\tilde{Y}_H} \tilde{\omega} + i_{\tilde{Y}_H} (dt \wedge dH) \\
                       & = & dH - dH -i_{\tilde{Y}_H}(dH) dt \\
               & = & [i_{\tilde{Y}_H}(i_{\tilde{Y}_H} \tilde{\omega})]dt = 0.
\end{eqnarray*}

  The equation in note (i) above and Eq.(84) are analogous to the
properties of the `characteristic vector field' of a contact
structure. The derivation $\hat{Y}_H$ may, therefore, be called the
\emph{characteristic derivation} of the Poincar$\acute{e}$-Cartan
contact structure.

A symplectic G-action (in the present context) is said to be
\emph{hamiltonian} if the 1-forms $i_{Z_{\xi}} \Omega$ are exact,
i.e. to each $\xi \in \scg$, corresponds a `hamiltonian'
$\hat{h}_{\xi} \in \sca^e$ (unique up to an additive constant
multiple of the unit element) such that
\begin{eqnarray}
i_{Z_{\xi}} \Omega = -d \hat{h}_{\xi}.
\end{eqnarray}
These `hamiltonians' (\emph{Noether invariants}) are constants of motion :
\begin{eqnarray}
\frac{d \hat{h}_{\xi}(t)}{dt} & = & \hat{Y}_H(\hat{h}_{\xi}(t))
                                      = (d \hat{h}_{\xi})(\hat{Y}_H)(t)
                         \nonumber \\
                  & = & - (i_{Z_{\xi}} \Omega)(\hat{Y}_H)(t)
                        = 0
\end{eqnarray}
where, in the last step, Eq.(84) has been used. This is the supmech
analogue of the symplectic version of Noether's theorem. For some
concrete examples of Noether invariants, see section 5.2.

\vspace{.12in} \noindent \textbf{4.7. Systems with configuration
space; lacalizability}

\vspace{.12in} We shall now consider the class of systems each of
which has a configuration space (say, M) associated with it and it
is meaningful to ask questions about the localization of the system
in subsets of M. To start with, we shall take M to be a topological
space and take the permitted domains of localization to belong to
B(M), the family of Borel subsets of M.

Some good references containing detailed treatment of localization
in conventional approaches are (Newton and Wigner 1949; Wightman
1962; Varadarajan 1985; Bacry 1988). We shall follow a relatively
more economical path exploiting some of the constructions described
above.

We shall say that a system S [with associated symplectic
superalgebra $(\sca,\omega)$] is \emph{localizable} in M if we have
a positive observable-valued measure (as defined in section 4.1) on
the measurable space (M,B(M)), which means that, corresponding to
every subset $ D \in B(M)$, there is a positive  observable P(D) in
\sca \ satisfying  the three conditions

\noindent
(i) $ P(\emptyset) = 0;$ \hspace{.5in} (ii) P(M) = I;

\noindent
(iii)for any countable family of mutually disjoint sets $D_i \in B(M)$,
\begin{eqnarray}
P (\cup_iD_i) = \sum_i P(D_i).
\end{eqnarray}
For such a system, we can associate, with any state $\phi$, a
probability measure $\mu_{\phi}$ on the measurable space $(M,B(M))$
defined by [see Eq.(56)]
\begin{eqnarray}
\mu_{\phi}(D) = \phi (P(D)),
\end{eqnarray}
making the triple $(M, B(M), \mu_{\phi})$ a probability space. The
quantity $\mu_{\phi}(D)$ is to be interpreted as the probability of
the system, given  in the state $\phi$, being found (on
observation/measurement) in the domain D.

Generally it is of  interest to consider localizations having
suitable invariance properties under a transformation group G.
Typically G is a topological group  with continuous action on M
assigning, to each $g \in G$, a bijection $T_g : M \rightarrow M$
such that, in obvious notation, $T_g T_{g^{\prime}} =
T_{gg^{\prime}}$ and $T_e = id_M$; it also has a symplectic action
on \sca \ and $\scs(\sca)$ given by  the mappings $\Phi_1(g)$ and
$\Phi_2(g)$ introduced in section 4.4. The localization on M
described above will be called \emph{G-covariant} (or, loosely,
G-invariant) if \begin{eqnarray} \Phi_1(g)(P(D)) = P(T_g(D)) \ \
\forall g \in G \ \textnormal{and} \ D \in B(M). \end{eqnarray}
Equation (89) translates into the following useful relation for the
probabilities (88):
\begin{eqnarray} \mu_{\Phi_2(g)(\phi)} (P(D)) = \mu_{\phi}(T_g(D)).
\end{eqnarray} \emph{Proof} \begin{eqnarray*}
\mu_{\Phi_2(g)(\phi)}(D) & = & <\Phi_2(g)(\phi), P(D)> = <\phi,
\Phi_1(g)(P(D))> \\ & = & <\phi, P(T_g(D))> = \mu_{\phi}(T_g(D)). \
\ \Box \end{eqnarray*} In most practical applications, M is a
manifold and G is a Lie group with smooth action on M and a Poisson
action on the symplectic superalgebra $(\sca, \omega)$.

In Hilbert space QM, the problem of G-covariant localization is
traditionally formulated in terms of the so-called `systems of
imprimitivity' (Mackey 1949; Varadarajan 1985; Wightman 1962). We
are operating in the more general algebraic setting trying to
exploit the machinery of noncommutative symplectics developed above.
Clearly, there is considerable scope for mathematical developments
in this context parallel to those relating to systems of
imprimitivity. We shall, however, restrict ourselves to some
essential developments relevant to the treatment of localizable
elementary systems (massive particles) later.

We shall be mostly concerned with $ M= R^n$ (equipped with the
Euclidean metric). In this case, one can consider averages of the
form (denoting the natural coordinates on $R^n$ by $x_j$)
\begin{eqnarray*}
\int_{R^n} x_j d \mu_{\phi}(x), \ \ j= 1,...,n. \ \ \ (***)
\end{eqnarray*}
It is natural to introduce \emph{position/configuration observables}
$X_j$ such that the quantity (***) is $\phi(X_j)$. Let $E_n$ denote
the (identity component of) Euclidean group in n dimensions and let
$p_j, m_{jk} (= - m_{kj})$ be its generators satisfying the
commutation relations \[ [p_j,p_k] = 0, \ \ [m_{jk}, p_l] =
\delta_{jl}p_k - \delta_{kl}p_j \] \[ [m_{jk}, m_{pq}] =
\delta_{jp}m_{kq} - \delta_{kp}m_{jq} - \delta_{jq}m_{kp} +
\delta_{kq}m_{jp}. \] We shall say that a system S with
configuration space $R^n$ has \emph{concrete Euclidean-covariant
localization} if it is localizable as above in $ R^n$ and

\noindent (i) it has position observables $X_j \in \sca$ \ such that
the above mentioned relation holds :
\begin{eqnarray}
\phi(X_j) & = & \int_{R^n} x_j d \mu_{\phi}(x);
\end{eqnarray}
(The term `concrete' is understood to imply this condition.)

\noindent (ii) the group $E_n$ has a Poisson action on \sca \ so
that we have the hamiltonians $P_j, M_{jk}$ associated with $p_j,
m_{jk}$ such that
\begin{eqnarray}
\{ P_j, P_k \} = 0, \hspace{.12in}  \{ M_{jk}, P_l \} = \delta_{jl}P_k -
\delta_{kl} P_j \nonumber \\
\{ M_{jk}, M_{pq} \} = \delta_{jp} M_{kq} - \delta_{kp}M_{jq} -
\delta_{jq} M_{kp} + \delta_{kq} M_{jp};
\end{eqnarray}

\noindent (iii) the covariance condition (89) holds with the
Euclidean group action on $R^n$ given by
\begin{eqnarray} T_{(R,a)} x = Rx + a, \ \ R \in SO(n), \ a \in R^n.
\end{eqnarray}

\noindent Equations (91) and (89) for infinitesimal Euclidean
transformations and Eq.(67) then give the analogues of the classical
canonical PBs of $X_j$s with the Euclidean generators :

\begin{eqnarray}
\{ P_j, X_k \} = \delta_{jk} I , \hspace{.12in} \{M_{jk}, X_l \} =
\delta_{jl} X_k - \delta_{kl} X_j.
\end{eqnarray}
\emph{Proof}. Using Eq.(91) with $\phi$ replaced by $\phi^{\prime} =
\Phi_2(g)(\phi)$, we have
\begin{eqnarray}
\phi^{\prime}(X_j) = \int x_j d\mu_{\phi^{\prime}}(x) = \int x_j d
\mu_{\phi}(x^{\prime}) = \int (x_j^{\prime} - \delta x_j) d
\mu_{\phi}(x^{\prime}) \end{eqnarray} where $x^{\prime} \equiv
T_g(x) = x + \delta x$ and we have used Eq.(90) to write $d
\mu_{\phi^{\prime}}(x) = d \mu_{\phi}(x^{\prime})$. Writing
$\phi^{\prime} = \phi + \delta \phi$ and taking $T_g$ to be a
general infinitesimal transformation generated by $ \epsilon \xi =
\epsilon^a \xi_a$, we have
\begin{eqnarray} \epsilon \phi ( \{ h_{\xi}, X_j \}) = \int_{R^n}
\delta x_j d \mu_{\phi}(x). \end{eqnarray} For translations, with $
\xi = p_k, h_{p_k}= P_k, \delta x_j = \epsilon \delta_{jk}$, Eq.(96)
gives \begin{eqnarray*} \phi(\{ P_k, X_j \}) = \delta_{jk} =
\delta_{jk} \phi(I). \end{eqnarray*} Since this holds for all $\phi
\in \scs (\sca)$, we have the first of the equations (94). The
second equation is similarly obtained by taking, in obvious
notation, $ \epsilon \xi = \frac{1}{2}\epsilon_{jk} m_{jk}$ and
\begin{eqnarray*} \delta x_l = \epsilon _{lk}x_k = \epsilon_{jk}
\delta_{ jl} x_k = \frac{1}{2} \epsilon_{jk} (\delta_{jl}x_k -
\delta_{kl} x_j). \  \Box \end{eqnarray*}

 The hamiltonians $ P_{j} $ and $ M_{jk} $ will be referred to as the
\emph{momentum} and \emph{angular momentum} observables of the
system S. It should be noted that the PBs obtained above do not
include the expected relations $ \{ X_j, X_k \} =0$; these
relations, as we shall see in the following section, come from the
relativity group. [Recall that, in the treatments of localalization
based on systems of imprimitivity, the commutators $[X_j,X_k] = 0$
appear because there the analogues of the objects P(D) are assumed
to be projection operators satisfying the relation
$P(D)P(D^{\prime}) = P(D \cap D^{\prime})  (= P(D^{\prime})P(D))$.
In our more general approach, we have no basis for making such
assumptions.]

\vspace{.15in} \noindent \textbf{5. RELATIVITY GROUPS, ELEMENTARY
SYSTEMS AND FUNDAMENTAL OBSERVABLES}

\vspace{.12in}

Having presented the general formalism of supmech, we now proceed to
treat concrete systems. We start with the simplest ones : particles.
In this section, we take up the question of the definition of a
particle and the fundamental observables relating to the
characterization/labelling and kinematics of a particle. Relativity
group will be seen to play a very important role in this context.

\vspace{.12in} \noindent \textbf{5.1. General considerations about
relativity groups and elementary systems}

\vspace{.12in} A particle is basically an irreducible entity (in the
sense that it cannot be represented as the composite of more than
one identifiable entities) localized in what we traditionally call
`space' and the description of its dynamics involves `time'. We
must, therefore, introduce the concepts of space and time or, more
generally, space-time before we talk about particles.

In the following developments, \emph{space-time} will be understood
to be a (3+1)- dimensional differentiable manifold equipped with a
suitable metric to define spatial distances and time-intervals.[The
breakup (3+1) means that one of the four dimensions is in some way
distinguished from the other three. Details about the metric will be
given only when needed.] A \emph{reference frame} is an atlas
providing a coordinatization of the space-time points.
\emph{Observers} are supposedly intelligent beings employing
reference frames for doing concrete physics; they will be understood
to be in one-to-one correspondence with reference frames.

Observables of systems localized in space are generally
observer-dependent. This observer dependence is systematically taken
into consideration by adopting a \emph{relativity scheme} which
incorporates (i) specification of the geometry of space-time, (ii)
selection of a class of reference frames to to be treated as
distinguished (all members of the chosen class to be treated as
physically equivalent) and (iii) transformation laws between
coordinatizations of different members of the chosen class (these
transformations constitute a group called the \emph{relativity
group} of the scheme).

Assuming a fixed background space-time M, we shall assume the
relativity group to be a connected Lie group $ G_0 $ (with Lie
algebra $\scg_0$) acting as a transformation group on M. For
concrete applications, we shall take $G_0$ to be the Galilean group
and the Poincar$\acute{e}$ group (the inhomogeneous Lorentz group)
in the schemes of Galilean relativity and special relativity
respectively. Both these groups have the one-parameter group
$\mathcal{T}$ of time translations as a subgroup.

Treatment of kinematics and dynamics of a system in accordance with
a relativity scheme involves the action of $G_0$ on the symplectic
algebra $(\sca, \omega)$. To exploit the availability of a
symplectic framework, we would like this to be a hamiltonian action
so that we can associate observables with the infinitesimal
generators of $G_0$. We may formally state, in the sub-domain of
supmech covering theories admitting a background space-time, the
\emph{principle of relativity} as follows :

\vspace{.1in} \noindent (i) There is a preferred class of reference
frames whose space-time coordinatisations are related through the
action of a connected Lie group $G_0$. \\
(ii) For a system with the system algebra \sca, the  relativity
group $G_0$ has a hamiltonian action on the symplectic algebra
$(\sca, \omega)$ [or the generalized symplectic superalgebra
$(\sca, \scx, \omega)$ in appropriate situations]. \\
(iii) All reference frames in the chosen class are physically
equivalent in the sense that the fundamental equations of the theory
are covariant with respect to the $G_0$-transformations of the
relevant variables.

\vspace{.1in} We shall call such a scheme \emph{$G_0$-relativity}
and systems covered by it \emph{$G_0$-relativistic}.

Heisenberg and Schr$\ddot{o}$dinger pictures of dynamics
corresponding to two observers O and O$^{\prime}$ may be related
through the symplectic action of $G_0$ by following the strategy of
Sudarshan and Mukunda (1974) (referred to as SM below) exploiting
the fact that $G_0$ has $\mathcal{T}$ as a subgroup. Showing the
observer dependence of the algebra elements explicitly, the two
Heisenberg picture descriptions A(O,t) and
A(O$^{\prime}$,t$^{\prime}$)of an element A of \sca \ can be related
through the sequence (assuming a common zero of time for the two
observers)
\[ A(O,t) \longrightarrow A(O,0) \longrightarrow A(O^{\prime},0)
\longrightarrow A(O^{\prime},t^{\prime}) \] where the first and the
last steps involve the operations of time translations in the two
frames.We shall be concerned only with the symplectic action of
$G_0$ involved in the middle step.  A similar strategy can be
adopted for the Schr$\ddot{o}$dinger picture. Detailed treatments of
the relativistic Heisenberg and Schr$\ddot{o}$dinger pictures in the
classical hamiltonian formalism may be found in SM.

Construction of Noether invariants, on the other hand, involves
explicit consideration of the transformation of the time variable.
The formalism of section 4.6 has obvious limitations in this regard
because time was treated as an external parameter in the Poisson
brackets employed there. We shall, therefore, construct the Noether
invariants only for the Galilean group where the only admitted
transformations of the time variable are translations.

To formalize the notion of a (relativistic, quantum) particle as an
irreducible entity, Wigner (1939)introduced the concept of an
`elementary system' as a quantum system whose Hilbert space carries
a projective unitary irreducible representation of the
Poincar$\acute{e}$ group. The basic idea is that the state space of
an elementary system should not admit a decomposition into more than
one invariant (under the action of the relevant relativity group)
subspaces. Following this idea, elementary systems in classical
mechanics (SM; Alonso 1979) have been defined in terms of a
transitive action of the relativity group on the phase space of the
system. Alonso (1979) gave a unified treatment of classical and
quantum elementary systems by treating them as special cases of
(irreducible/transitive) kinematical action of the relativity groups
(called `invariance groups' in that work) on the state space of a
dynamical system.

In this section, we shall treat elementary systems in the framework
of supmech. Traditional classical and quantum elementary systems
will be seen as special cases of these. This treatment goes a step
further than that of Alonso in that the unification is achieved in a
single \emph{symplectic} framework.

A system S having associated with it the symplectic triple $(\sca,
\sone, \omega)$ will be called an \emph{elementary system} in
$G_0$-relativity  if it is a $G_0$-relativistic system such that the
action of $G_0$ on the space \sone \ of its pure states is
transitive. Formally, an elementary system may be represented as a
collection $ \mathcal{E} = ({G}_0, \sca, \sone,\omega, \Phi)$ where
$ \Phi = (\Phi_1, \Phi_2)$ are mappings as in section 4.4
implementing the $G_0$-actions --- $\Phi_1$ describing a hamiltonian
action on $(\sca, \omega)$ and $\Phi_2$ a transitive action on
\sone.

\vspace{.1in} \noindent The transitive action of $G_0$ on \sone \
implies that the expectation value of a $G_0$-invariant observable
is the same in every pure state (hence in every state). To see this,
let Q be such an observable and $\phi_1, \phi_2$ two pure states
such that $\Phi_2(g) (\phi_1) = \phi_2$ for some $ g \in G_0$. We
have
\[ <\phi_2, Q> = <\Phi_2(g)(\phi_1), Q> = <\phi_1, \Phi_1(g^{-1})(Q)>
= <\phi_1, Q> \] as desired. Denoting this common expectation value
of Q by q (we shall call it the value of Q for the system), we have,
by the CC condition, Q = qI.

This has the important implication that, for an elementary system, a
Poisson action (of $G_0$ or of its projective group $\hat{G}_0$) is
always available; this is because, if $G_0$ does not admit Poisson
action, the cocycle $\alpha$ of section 4.4 is a multiple of the
unit element and the hamiltonian action of $G_0$ can be extended to
a Poisson action of $\hat{G}_0$.

Let $\xi_a$ (a = 1,..,r) be a basis in the Lie algebra
$\hat{\scg}_0$ of $\hat{G}_0$ satisfying the commutation relations
as in section 4.4. The admissibility of Poisson action of
$\hat{G}_0$ on \sca \ implies that, corresponding to the generators
$\xi_a$, we have the hamiltonians $h_a \equiv h_{\xi_a}$ in \sca \
satisfying the  PB relations
\begin{eqnarray}
\{ h_a, h_b \} = C_{ab}^c \ h_c.
\end{eqnarray}
Recalling Eq.(78), the condition of transitive action on \sone \
implies that the $\tilde{h}$-images of pure states of an elementary
system are coadjoint orbits in $\hat{\scg}_0^*$.

In classical mechanics, one has an isomorphism between the
symplectic structures on the symplectic manifolds of elementary
systems and those on the coadjoint orbits. In our case, the state
spaces of elementary systems and coadjoint orbits of relativity
groups are generally spaces of different types and the question of
an isomorphism does not arise. We can, however, use Eq.(78) to
obtain useful information about the transformation properties of the
quantities $h_a$ under the $\hat{G}_0$- action. Recalling the
notations in section 4.4, writing $Cad_g[\tilde{h}(\phi)] = u_a(g)
\lambda^a$, we have
\begin{eqnarray}
u_a(g) = < Cad_g [\tilde{h}(\phi)], \xi_a> =
<\tilde{h}[\Phi_2(g)(\phi)], \xi_a> = <\phi, \Phi_1(g^{-1})h_a>
\end{eqnarray}
showing that the transformation properties of hamiltonians $h_a$ are
directly related to those of the corresponding coordinates (with
respect to the dual basis) of points on the relevant co-adjoint
orbit. This is adequate to enable us to to use the descriptions of
the relevant co-adjoint actions in (Alonso 1979) and draw parallel
conclusions.

We shall adopt the following strategy :

\vspace{.1in} \noindent (i) Given a relativity scheme, use the
Poisson action of $G_0$ or  $\hat{G}_0$ on the symplectic
superalgebra of an elementary system to obtain the corresponding
hamiltonians and their PBs [Eq.(97)]. These PBs are clearly the same
for all elementary systems of the group $G_0$.

\vspace{.1in} \noindent (ii) Use these PBs to identify some
\emph{fundamental observables} [i.e. those which cannot be obtained
from other observables (through algebraic relations or PBs)]. These
include observables (like mass) that Poisson-commute with all $h_a$s
and the momentum observables (the group of space translations being
a subgroup of both the relativity groups we consider).

\vspace{.1in} \noindent (iii) Determine the transformation laws of
$h_a$s under finite transformations of $G_0$ following the relevant
developments in (SM; Alonso 1979). Use these transformation laws to
identify the $G_0$-invariants and some other fundamental observables
(the latter will be configuration and spin observables). The values
of the invariant observables characterize (or serve to label) an
elementary system.

\vspace{.1in} \noindent (iv) The system algebra \sca \ for an
elementary system is to be taken as the (topological completion of)
the one generated by the fundamental observables and the identity
element.

\vspace{.1in} \noindent (v) Obtain (to the extent possible) the
general form of the Hamiltonian as a function of the fundamental
observables as dictated by the PB relations (97).

\noindent (vi) (For the Galilean group) use the formalism of section
4.6 to consider the action of $G_0$ on the presymplectic space
$(\sca^e, \Omega)$ and, noting that this action satisfies Eq.(85),
identify the appropriate Noether invariants.

We shall now obtain an equation that will be useful for this last
job. Let $\xi \in \scg_0$ generate an infinitesimal transformation
giving $\delta t = \epsilon f(t)$ (and possibly some changes in
other quantities). [In view of the limitations of the formalism of
section 4.6 mentioned above, arguments other than t for the function
f have been excluded.] The relation between the induced derivations
$Z_{\xi}$ on \sca \ and $\hat{Z}_{\xi}$ on $\sca^e$ is given by
\begin{eqnarray}
\hat{Z}_{\xi} = Z_{\xi} + f(t)\frac{\partial}{\partial t}.
\end{eqnarray}
We have $ Z_{\xi} = Y_{h_{\xi}}$ (see section 4.4). We look for the
quantity $\hat{h}_{\xi}$ (the prospective Noether invariant) such
that Eq.(85) holds. (Finding such a quantity will establish
invariance of $\Omega$ under the relevant group action and also
determine the corresponding Noether invariant.) Equations (99) and
(81) now give the desired relation
\begin{eqnarray}
i_{\hat{Z}_{\xi}}\Omega & = & i_{Z_{\xi}} \tilde{\omega} -
i_{Z_{\xi}}(dH) dt + f(t) dH \nonumber \\
              & = & -dh_{\xi} - Y_{h_{\xi}}(H) dt + f(t)dH.
\end{eqnarray}

Most of the equations in the following two subsections have the same
mathematical form as some of the equations in (Alonso 1977, 1979)
and/or SM. The following couple of remarks should serve to clarify
the situation.

\noindent (a) Classical elementary systems are defined in terms of a
transitive canonical action of the relevant relativity group on
symplectic manifolds. These are obviously special cases of supmech
elementary systems corresponding to  commutative system algebras of
the type treated in section 3.8. Those results in the treatment of
classical elementary systems whose derivation does not use the
commutativity of the algebra $\sca_{cl}$ are expected to be valid
for general supmech elementary systems.

\noindent (b) Quantum elementary systems are defined in terms of
projective irreducible unitary representations of the relevant
relativity group on separable Hilbert spaces. Keeping the
developments in section 3.10 in view, these are seen as special
cases of supmech elementary systems when the system algebra \sca \
is a member of a triple ($\sch, \mathcal{D}, \sca$) [a `quantum
triple'; see section 7.2] where \sch \ is a separable Hilbert space,
$\mathcal{D}$ a dense linear subset of \sch \ and \sca \ an
Op$^*$-algebra based on ($\sch,\mathcal{D})$. According to theorem
(3.2) in (Cari$\tilde{n}$ena and Santander 1975), every projective
unitary representation of a relativity group $G_0$ can be lifted to
a unitary representation of the corresponding projective group
$\hat{G}_0$ (called the `projective covering group' of $G_0$ in that
work). The infinitesimal generators of $\hat{G}_0$ arising in such a
(continuous) unitary representation serve as hamiltonians of the
corresponding supmech elementary system. Once the infinitesimal
generators have been obtained, the Hilbert space goes into the
background; the rest of the work is algebraic. All the results
obtained, in the traditional treatments of quantum elementary
systems, by algebraic manipulations (involving the above mentioned
infinitesimal generators) and use of the quantum Poisson brackets
(54) are expected to be valid in the treatments of the corresponding
elementary systems in supmech.

\vspace{.12in} \noindent \textbf{5.2. Nonrelativistic elementary
systems}

\vspace{.12in} In this and the following subsection, we shall keep
close to the notational conventions of Alonso (1979). Our PBs,
however, follow the conventions of Woodhouse  !980) and differ from
those of (SM; Alonso 1979) by a sign; moreover, our Galilean
generator \textbf{K} differs from that of Alonso by a sign.

The relativity group $G_0$ of the nonrelativistic domain of supmech
is (the identity component of) the Galilean group of transformations
of the Newtonian space-time $R^3 \times R$ given by
\begin{eqnarray}
x^{\prime} = Rx + tv + a, \hspace{.2in} t^{\prime} = t + b
\end{eqnarray}
where $R \in SO(3),\  v \in R^3, \ a \in R^3$ and $b \in R$. This
group does not admit Poisson action. After a careful consideration
of the freedom to modify the hamiltonians by additive terms, the
hamiltonians $ J_i, K_i, P_i, H $  corresponding to the ten
generators $\mathcal{J}_i, \mathcal{K}_i, \mathcal{P}_i (i=1,2,3),
\sch$ of $G_0$  [so that $h_{\mathcal{P}_i} = P_i$ etc] can be shown
to satisfy the Poisson bracket relations (SM)
\begin{eqnarray}
\{J_i, J_j \} = - \epsilon_{ijk}J_k, \ \ \{J_i, K_j \} = -
\epsilon_{ijk} K_k,
\ \  \{ J_i, P_j \} = - \epsilon_{ijk} P_k  \nonumber \\
 \{ K_i, H \} = - P_i, \ \ \{ K_i, P_j \} = - \delta_{ij} M,
\end{eqnarray}
where M is a neutral element; all other PBs vanish. By the argument
presented above, we must have M= mI, $m \in R$. We shall identify m
as the mass of the elementary system. The condition $m \geq 0$ will
follow later from an appropriate physical requirement. The objects
$P_i$ and $J_i$, being generators of the Euclidean subgroup $ E_3 $
of $G_0$, are the momentum and angular momentum observables of
section 4.7.

Following the procedure outlined in section 4.4, we augment the Lie
algebra $\mathcal{G}_0$ of $G_0$ to a larger Lie algebra
$\hat{\mathcal{G}}_0$ by including an additional generator
$\mathcal{M}$ corresponding to M (which now becomes the Hamiltonian
corresponding to $\mathcal{M} \in \hat{\mathcal{G}}_0$); it commutes
with all other generators and appears only in the commutator
\begin{eqnarray} [\mathcal{K}_j, \mathcal{P}_k] = -\delta_{jk}
\mathcal{M}. \end{eqnarray}  The remaining commutation relations of
$\hat{\mathcal{G}}_0$ are those of $\mathcal{G}_0$ ($ [J_j, J_k] = -
\epsilon_{jkl} J_l$ etc.). The projective group $\hat{G}_0$ of the
Galilean group $G_0$ is the connected and simply connected Lie group
with the Lie algebra $\hat{\mathcal{G}}_0$.

Representing a general group element of $\hat{G}_0$ in the form
\begin{eqnarray} g & = & (A,v,b,a,\tau) \nonumber \\
                   & = & exp(-\tau \mathcal{M})exp(-a.\mathcal{P})
                   exp(-b \mathcal{H}) exp(-v.\mathcal{K})A
                    \end{eqnarray}
where $A \in SU(2)$ and $\tau \in R$, the group law of $\hat{G}_0$
is obtained, after a straightforward calculation, as
\begin{eqnarray} g^{\prime}g = (A^{\prime}A,v^{\prime} +
R(A^{\prime})v, b^{\prime} + b, a^{\prime} + b v^{\prime}+
R(A^{\prime})a,\tau^{\prime} + \tau + R(A^{\prime})_{jk}v_j^{\prime}
a_k). \end{eqnarray}

The transformation laws of the hamiltonians of $\hat{G}_0$ under its
adjoint action may be found following the procedure of either SM or
Alonso (1979). These transformation laws give the following three
independent invariants :
\begin{eqnarray}
M, \hspace{.12in} C_1 \equiv 2MH - \mathbf{P}^2, \hspace{.12in} C_2
\equiv (M \mathbf{J} - \mathbf{K} \times \mathbf{P})^2.
\end{eqnarray}
Of these, the first one is obvious; the vanishing of PBs of $C_1$
with all the hamiltonians is also easily checked. Writing $C_2 =
B_jB_j$ where
\begin{eqnarray*} B_j = MJ_j - \epsilon_{jkl}K_kP_l, \end{eqnarray*}
it is easily verified that
\begin{eqnarray*} \{J_j, B_k \} = - \epsilon_{jkl} B_l, \ \
\{K_j, B_k \} = \{ P_j, B_k \} = \{H, B_k \} = 0 \end{eqnarray*}
which finally leads to the vanishing of PBs of $C_2$ with all the
hamiltonians. By the argument given above for M, the last two
invariants also should be scalar multiples of the unit element in
\sca. The values of these three invariants characterize a Galilean
elementary system in supmech.

We henceforth restrict ourselves to elementary systems with $ m \neq
0.$ Defining $X_i = m^{-1} K_i$, we have
\begin{eqnarray}
\{ X_j, X_k \} = 0, \hspace{.12in} \{ P_j, X_k \} = \delta_{jk} I, \
\ \hspace{.12in} \{ J_j, X_k \} = - \epsilon_{jkl} X_l.
\end{eqnarray}
 Comparing the last two equations above with the equations (94)(for n=3),
we identify $X_j$  with the position observables of section 4.7.
Note that the fact that the $X_j$s mutually Poisson-commute comes
from the relativity group.

Writing $ \mathbf{S} = \mathbf{J} - \mathbf{X} \times \mathbf{P}, $
we have  $C_2$ = $m^2 \mathbf{S}^2$. We have the PB relations
\begin{eqnarray}
\{ S_i, S_j \} = - \epsilon_{ijk} S_k, \ \  \{ S_i, X_j \} = 0 = \{
S_i, P_j \}.
\end{eqnarray}
We identify $\mathbf{S}$ with the internal angular momentum or spin of the
elementary system.

The invariant quantity
\begin{eqnarray}
U \equiv \frac{C_1}{2m} = H - \frac{\mathbf{P}^2}{2m}
\end{eqnarray}
is interpreted as the \emph{internal energy} of the elementary
system. It is the appearance of this quantity (which plays no role
in Newtonian mechanics) which is responsible for energy being
defined in Newtonian mechanics only up to an additive constant.

Writing $\mathbf{S}^2 = \sigma I$ and U = u I, we see that Galilean
elementary systems with $m \neq 0$ can be taken to be
characterized/labelled  by the parameters m, $\sigma$ and u. The
fundamental kinematical observables are $X_j, P_j$ and $S_j$
(j=1,2,3). Other observables are assumed in supmech to be functions
of the fundamental observables.

Henceforth  we shall take u = 0 (a natural assumption to make if the
elementary systems to be treated are particles). Eq.(109) now gives
\begin{eqnarray}
H = \frac{\mathbf{P}^2}{2m}
\end{eqnarray}
which is the Hamiltonian for a free Galilean particle in supmech.

\vspace{.1in} \noindent Note. (i) Full Galilean invariance (more
generally, full invariance under a relativity group) applies only to
an isolated system. Interactions/(external influences) are usually
described with (explicit or implicit)reference to a fixed reference
frame or a restricted class of frames. For example, the interaction
described by a central potential implicitly assumes that the center
of force is at the origin of axes of the chosen reference frame.

\noindent (ii) In the presence of external influences, translational
invariance is lost and the PB $ \{H, P_i \} = 0 $ must be dropped.
For a spinless particle, the Hamiltonian (assumed to be a function
of the fundamental observables \textbf{X} and \textbf{P}) then has
the general form
\begin{eqnarray}
H = \frac{\mathbf{P}^2}{2m} + V(\mathbf{X}, \mathbf{P}).
\end{eqnarray}
In most practical situations, V is a function of $ \mathbf{X}$ only.

\vspace{.1in} We can now rule out the case $m < 0$ on physical
grounds because, by Eq.(110), this will allow arbitrarily large
negative values for energy. (Expectation values of the observable $
\mathbf{P}^2$ are expected to have no upper bound.)

Lastly, we consider the action of $G_0$ on the augmented algebra
$\sca^e$ for a free massive spinless particle. As noted above, it is
adequate, for each $\xi$ in the chosen basis of $\mathcal{G}_0$, to
find a `hamiltonian' $\hat{h}_{\xi}$ such that Eq.(85) holds; for
this we must show the exactness of the form on the right hand side
of Eq.(100). We have

\vspace{.1in} \noindent (i)for rotations ($ \xi = \mathcal{J}_i,
h_{\xi} = J_i $) f(t) =0, $ Y_{\xi}(H) = 0$,  giving $ \hat{h}_{\xi}
= h_{\xi} = J_i; $

\vspace{.1in} \noindent (ii) for Galilean boosts ($ \xi =
\mathcal{K}_i, h_{\xi} = K_i = m X_i $) \hspace{.12in} f(t) = 0,
\hspace{.12in} $vY_{\xi}(H) = v \{K_i, H \} = - v P_i$ giving
$\hat{h}_{\xi} = m X_i - P_i t$;

\vspace{.1in} \noindent (iii)for space translations ( $ \xi =
\mathcal{P}_i, h_{\xi} = P_i $) f(t) =0, $ Y_{\xi}(H) = 0$, giving
$\hat{h}_{\xi} = h_{\xi} = P_i;$

\vspace{.1in} \noindent (iv) for time translations ($ \xi =
\mathcal{H}, h_{\xi} = H $) f(t)=1, $Y_{\xi} =0$, giving
$\hat{h}_{\xi}= H$.

\vspace{.1in}
\noindent
Finally, the Noether invariants of the Galilean group are
\begin{eqnarray}
\mathbf{J}, \ \ m \mathbf{X} - \mathbf{P}t , \ \ \mathbf{P}, \ \ H
\end{eqnarray}
which are (up-to signs) the supmech avatars of those in (Souriau
1997).

\vspace{.12in} \noindent \textbf{5.3. Relativistic elementary
systems}

\vspace{.12in} In the scheme of special relativity, the relativity
group $G_0$ is the (identity component of) Poincar$\acute{e}$ group
of transformations on the Minkowski space-time [($R^4, \eta_{\mu
\nu})$ where $\mu, \nu$ =0,1,2,3 and $\eta_{\mu \nu}$ =
diag(-1,1,1,1)]:
\begin{eqnarray}
x^{\prime} = \Lambda x + a, \hspace{.12in} \Lambda \in SO(3,1) \ \
(\textnormal{with} \Lambda^0 \ _0 \geq 1), \ \ a \in R^4.
\end{eqnarray}Sudars
This group admits Poisson actions (SM; Alonso 1979; Guillemin and
Sternberg 1984). Since the general method (of treating elementary
systems in supmech) has been illustrated in the previous subsection,
we shall be  more brief here. Some more details may be found in (
Alonso 1977,1979; SM). We shall generally keep close to the
developments in (Alonso 1977,1979).

The generators of $G_0$ are $ ( \mathcal{M}_{\mu \nu} = -
\mathcal{M}_{\nu \mu}, \mathcal{P}_{\mu}) $ or, equivalently, $ (
\mathcal{J}_i, \mathcal{K}_i, \mathcal{P}_i, \mathcal{H}) $ where $
\mathcal{P}^0= \sch,  \mathcal{M}_{0i} = \mathcal{K}_i$ and $
\mathcal{M}_{ij} = \epsilon_{ijk}\mathcal{J}_k $. The hamiltonians
($J_i, K_i, P_i, H $) arising from the Poisson action of $G_0$ on
the symplectic superalgebra $(\sca, \omega)$ of an elementary system
satisfy the PB relations
\begin{eqnarray}
\{ J_i, J_j \} = - \epsilon_{ijk} J_k,\ \  \{ J_i, K_j, \} = -
\epsilon_{ijk} K_k, \ \
\{J_i, P_j \} = - \epsilon_{ijk} P_k, \nonumber \\
\{K_i, K_j \} =  \epsilon_{ijk} J_k, \ \ \{ K_i, P_j \} =
-\delta_{ij} H, \ \ \{ K_i, H \} = - P_i;
\end{eqnarray}
all other PBs vanish. The PBs for the manifestly Lorentz-covariant
hamiltonians $M_{\mu \nu}, P_{\mu}$ are those of Eq.(92) with
j,k,l,p,q replaced by $\mu, \nu, \lambda, \rho, \sigma$ and
$\delta_{jl}$ by $\eta_{\mu \lambda}$ etc.

Defining $ W^{\mu} = \frac{1}{2} \epsilon^{\mu \nu \lambda \rho}
M_{\nu \lambda}P_{\rho}$ ( the Pauli-Lubanski vector), the two
independent invariants of the $G_0$-action are $P^2 \equiv
P^{\mu}P_{\mu}$ and $W^2$; this can be directly checked from the
covariant PBs mentioned above. For elementary systems, they take
values $p^2 I$ and $w^2 I$; the real-valued quantities $p^2$ and
$w^2$ characterize the elementary systems. It is useful to note that
\begin{eqnarray}
W^0 = - \mathbf{J}.\mathbf{P}, \hspace{.12in}
\mathbf{W} = - H \mathbf{J} + \mathbf{P} \times \mathbf{K}.
\end{eqnarray}

We shall restrict ourselves to the cases with $ p^2 \leq 0$ (i.e.
$p^{\mu}$ non-spacelike) and write $ p^2 = - m^2 $ (with $ m \geq
0$). For situations with $ m > 0$ and H and (H + mI) invertible, one
can define the position and spin observables as follows :
\begin{eqnarray}
\mathbf{X} = -\frac{1}{2} [\mathbf{K},H^{-1}]_+ +
[m H (H + mI)]^{-1} \mathbf{P} \times \mathbf{W}  \\
\mathbf{S} = - m^{-1} \mathbf{W} + [m H (H + mI)]^{-1}
\mathbf{W}.\mathbf{P} \mathbf{P}.
\end{eqnarray}
The expected PB relations hold :
\begin{eqnarray}
\{ X_i, X_j \} = 0, \hspace{.12in} \{P_i, X_j \} = \delta_{ij}I, \nonumber \\
\{X_i, S_j \} = 0 = \{P_i, S_j \} \nonumber \\
\{ J_i, X_j \} = - \epsilon_{ijk} X_k, \hspace{.1in} \{ J_i, S_j \}
= \{ S_i, S_j \} = - \epsilon_{ijk} S_{k}.
\end{eqnarray}
We have $W^2 = m^2 \mathbf{S}^2$ and the relations
\begin{eqnarray}
H^2 = \mathbf{P}^2 + m^2 I, \hspace{.12in} \mathbf{J} =
\mathbf{X} \times \mathbf{P} + \mathbf{S}  \\
\mathbf{K} = -\frac{1}{2} [\mathbf{X}, H]_+ + (H + m I)^{-1}
\mathbf{S} \times \mathbf{P}.
\end{eqnarray}
Writing $\mathbf{S}^2 = \sigma I $, the invariant quantities m
(mass) and $\sigma$ (spin) characterize an elementary system and the
fundamental kinematical observables are again $X_j, P_j, S_j$ (j=
1,2,3).

\vspace{.15in} \noindent \textbf{6. COUPLED SYSTEMS IN SUPMECH}

\vspace{.12in}  We shall now consider the interaction of two systems
$S_1$ and $S_2$ described individually as the supmech Hamiltonian
systems  $ (\sca^{(i)}, \omega^{(i)}, H^{(i)})$ (i=1,2) and treat
the coupled system $S_1 + S_2$ also as a supmech Hamiltonian system.
To facilitate this, we must obtain the relevant mathematical objects
for the coupled system.

\vspace{.12in} \noindent \textbf{6.1. The symplectic form and
Poisson bracket on the tensor product of two superalgebras}

\vspace{.12in} The superalgebra corresponding to the coupled system
($ S_1 + S_2 $ ) will be taken as the (skew) tensor product $\sca =
\sca^{(1)} \otimes \sca^{(2)}$; its  elements are finite sums of
tensored pairs :
\begin{eqnarray*}
\sum_{j=1}^{m} A_j \otimes B_j \ \ \ \  A_j \in \sca^{(1)}, \ \ B_j
\in \sca^{(2)}
\end{eqnarray*}
with the multiplication rule
\begin{eqnarray}
(\sum_{j=1}^{m} A_j \otimes B_j)(\sum_{k=1}^{n} A_k \otimes B_k) =
\sum_{j,k} \eta_{B_jA_k} (A_jA_k) \otimes (B_jB_k).
\end{eqnarray}

The superalgebra $\sca^{(1)}$ (resp. $\sca^{(2)}$) has, in \sca, an
isomorphic copy consisting of the elements ($A \otimes I_2, A \in
\sca^{(1)}$) (resp. $I_1 \otimes B, B \in \sca^{(2)}$) to be denoted
as $\tilde{\sca}^{(1)}$ (resp. $\tilde{\sca}^{(2)}$). Here $I_1$ and
$I_2$ are the unit elements of $\sca^{(1)}$ and $\sca^{(2)}$
respectively. We shall also use the notations $\tilde{A}^{(1)} = A
\otimes I_2$ and $ \tilde{B}^{(2)} = I_1 \otimes B$.

Derivations and differential forms on $\sca^{(i)}$ and
$\tilde{\sca}^{(i)}$ (i = 1,2) are formally related through the
induced mappings corresponding to the isomorphisms $ \Xi^{(i)} :
\sca^{(i)} \rightarrow \tilde{\sca}^{(i)}$ given by $\Xi^{(1)}(A) =
A \otimes I_2$ and $\Xi^{(2)}(B) = I_1 \otimes B$. For example,
corresponding to $X \in SDer(\sca^{(1)}),$ we have $ \tilde{X}^{(1)}
= \Xi^{(1)}_*(X)$ in SDer($\tilde{\sca}^{(1)}$) given by [see
Eq.(3)]
\begin{eqnarray}
\tilde{X}^{(1)}(\tilde{A}^{(1)}) = \Xi^{(1)}_*(X)(\tilde{A}^{(1)}) =
\Xi^{(1)}[X(A)] = X(A) \otimes I_2.
\end{eqnarray}
Similarly, corresponding to $Y \in SDer(\sca^{(2)})$, we have
$\tilde{Y}^{(2)} \in \mbox{SDer}(\tilde{\sca}^{(2)})$ given by
$\tilde{Y}^{(2)}(\tilde{B}^{2}) = I_1 \otimes Y(B)$. For the 1-forms
$\alpha \in \Omega^1(\sca^{(1)})$ and $\beta \in
\Omega^1(\sca^{(2)})$, we have $\tilde{\alpha}^{(1)} \in
\Omega^1(\tilde{\sca}^{(1)})$ and $\tilde{\beta}^{(2)} \in
\Omega^1(\tilde{\sca}^{(2)})$ given by [see Eq.(26)]
\begin{eqnarray}
\tilde{\alpha}^{(1)}(\tilde{X}^{(1)}) = \Xi^{(1)}[\alpha
(((\Xi^{(1)})^{-1})_* \tilde{X}^{(1)})] = \Xi^{(1)}[\alpha(X)] =
\alpha(X) \otimes I_2
\end{eqnarray}
and $ \tilde{\beta}^{(2)}(\tilde{Y}^{(2)}) = I_1 \otimes \beta(Y)$.
Analogous formulas hold for the higher forms.

We can extend the action of the superderivations $ \tilde{X}^{(1)}
\in SDer(\tilde{\sca}^{(1)})$ and $\tilde{Y}^{(2)} \in
SDer(\tilde{\sca}^{(2)})$ to $\tilde{\sca}^{(2)}$ and
$\tilde{\sca}^{(1)}$ respectively by defining
\begin{eqnarray}
\tilde{X}^{(1)}(\tilde{B}^{(2)}) = 0 , \mbox{\ \ } \tilde{Y}^{(2)}
(\tilde{A}^{(1)}) = 0 \mbox{\ \ for all \ } A \in \sca^{(1)} \mbox{\
and \ } B \in \sca^{(2)}.
\end{eqnarray}
Note that an $X \in \sdera$ is determined completely by its action
on the subalgebras $\tilde{\sca}^{(1)}$ and $\tilde{\sca}^{(2)}$ :
\[ X(A \otimes B) = X(\tilde{A}^{(1)} \tilde{B}^{(2)}) =
(X\tilde{A}^{(1)})\tilde{B}^{(2)} + \eta_{XA} \tilde{A}^{(1)}
X(\tilde{B}^{(2)}). \]  With the extensions described above, we have
available to us superderivations belonging to the span of terms of
the form [see Eq.(122)]
\begin{eqnarray}
X = X^{(1)} \otimes I_2 + I_1 \otimes X^{(2)}.
\end{eqnarray}
Replacing $I_2$ and $I_1$ in Eq.(125) by elements of $Z(\sca^{(2)})$
and $Z(\sca^{(1)})$ respectively, we again obtain superderivations
of \sca. We, therefore, have  the space of superderivations
\begin{eqnarray}
[SDer(\sca^{(1)}) \otimes Z(\sca^{(2)})] \oplus [Z(\sca^{(1)})
\otimes SDer(\sca^{(2)})].
\end{eqnarray}
This space, however, is generally only a Lie sub-superalgebra of
SDer(\sca). For example, for $\sca^{(1)} = M_m(C)$ and $\sca^{(2)} =
M_n(C)$, recalling that all the derivations of these matrix algebras
are inner and that their centers consist of scalar multiples of the
respective unit matrices
, we have the (complex) dimensions of
SDer$(\sca^{(1)})$, and SDer$(\sca^{(2)})$ respectively, $(m^2 -1)$
and $(n^2-1)$ [so that the dimension of the space (126) is $m^2 +
n^2 -2$ ] whereas that of \sdera \ is $(m^2n^2-1)$.

We shall need to employ a class of superderivations more general
than (126). To this end, it is instructive to obtain explicit
representation(s) for a general derivation  of the matrix algebra $
\sca = M_m(C) \otimes M_n(C)$. We have
\begin{eqnarray*}
[A \otimes B, C \otimes D]_{ir,js} = A_{ik} B_{rt} C_{kj} D_{ts} -
C_{ik} D_{rt} A_{kj} B_{ts}
\end{eqnarray*}
which gives
\begin{eqnarray}
[A \otimes B, C \otimes D]_- & = & AC \otimes BD - CA \otimes DB \\
                             & = & [A,C]_- \otimes \frac{1}{2} [B,D]_+ +
                       \frac{1}{2}[A,C]_+ \otimes [B,D]_-.
\end{eqnarray}
This gives, in obvious notation,
\begin{eqnarray}
D_{A \otimes B} \equiv [A \otimes B,.]_-
& = & A.(.) \otimes B_.(.) - (.).A \otimes (.).B \\
& = & D_A \otimes J_B + J_A \otimes D_B
\end{eqnarray}
where $J_B$ is the linear mapping on $\sca^{(2)}$ given by
\mbox{$J_B (D) = \frac{1}{2}[B,D]_+$} and a similar expression for
$J_A$ as a linear mapping on $\sca^{(1)}$.  Eq.(129) shows that a
derivation of the algebra $ \sca = \sca^{(1)} \otimes \sca^{(2)} $
need not explicitly contain those of $\sca^{(i)}$. We shall,
however, not get involved in the search for the most general
expression for a derivation of the tensor product algebra \sca
(although such an expression would be very useful). The expression
(130) is more useful for us; it is a special case of the more
general form
\begin{eqnarray}
X = X_1 \otimes \Psi_2 + \Psi_1 \otimes X_2
\end{eqnarray}
where $X_i \in SDer(\sca^{(i)})$ (i=1,2) and $ \Psi_i : \sca^{(i)}
\rightarrow \sca^{(i)}$ (i =1,2) are linear mappings. Our
constructions below will lead us to the form (131). It is important
to note, however, that an expression of the form (131) (which
represents a linear mapping of \sca \ into itself) need not always
be a derivation as can be easily checked. We shall impose the
condition (1) on such an expression  to obtain a derivation.

To obtain the differential forms and the exterior product on \sca,
the most straightforward procedure is to obtain the graded
differential space $(\Omega(\sca),d)$  as the tensor product (Greub
1978) of the graded differential spaces $(\Omega(\sca^{(1)}), d_1)$
and $ (\Omega(\sca^{(2)}), d_2)$. A (homogeneous) differential
k-form on \sca \ is of the form (in obvious notation)
\begin{eqnarray}
\alpha_{kt} \ = \sum_{\begin{array}{c} i+j=k \\
r+s= t \ mod(2) \end{array}} \alpha^{(1)}_{ir} \otimes
\alpha^{(2)}_{js}. \end{eqnarray} The d operation on $\Omega(\sca)$
is given by [here $\alpha \in \Omega^p(\sca^{(1)})$ and $ \beta \in
\Omega(\sca^{(2)})] $
\begin{eqnarray}
d(\alpha \otimes \beta) = (d_1 \alpha) \otimes \beta + (-1)^p \alpha
\otimes d_2 \beta.
\end{eqnarray}

Given the symplectic forms $\omega^{(i)}$ on $\sca^{(i)}$ (i=1,2) we
shall construct  the induced symplectic form $\omega$ on \sca \
satisfying the following conditions :\\
(a) It should not depend on anything other than the objects
$\omega^{(i)}$ and $I_{(i)}$ (i=1,2) [the
`naturality'/`canonicality' assumption for $\omega$. (Note that the
unit elements are the only distinguished elements of the algebras
being considered)]. \\
(b) The restrictions of $\omega$ to $\tilde{\sca}^{(1)}$ and
$\tilde{\sca}^{(2)}$ be, respectively, $\omega^{(1)} \otimes I_2$
and $I_1 \otimes
\omega^{(2)}$. \\
This determines $\omega$ uniquely :
\begin{eqnarray}
\omega =  \omega^{(1)} \otimes I_2 + I_1 \otimes \omega^{(2)}.
\end{eqnarray}
 To verify that it is a symplectic form, we must show
that it is (i) closed and (ii) nondegenerate. Eq.(133) gives
\begin{eqnarray*}
d \omega = (d_1 \omega^{(1)}) \otimes I_2 + \omega^{(1)} \otimes d_2
(I_2) + d_1 (I_1) \otimes \omega^{(2)} + I_1 \otimes d_2
\omega^{(2)} = 0
\end{eqnarray*}
showing that $\omega$ is closed. To show the nondegeneracy of
$\omega$, we must show that, given $A \otimes B \in \sca$, there
exists a unique superderivation $Y = Y_{A\otimes B}$ in \sdera \
such that
\begin{eqnarray}
i_{Y} \omega  =  - d (A\otimes B)
             & = & - (d_1 A ) \otimes B - A \otimes d_2 B  \nonumber \\
             & = & i_{Y_A^{(1)}} \omega^{(1)} \otimes B
               + A \otimes i_{Y_B^{(2)}} \omega^{(2)}.
\end{eqnarray}
where $Y_A^{(1)}$ and $Y_B^{(2)}$ are the Hamiltonian
superderivations associated with $A \in \sca^{(1)}$ and $B \in
\sca^{(2)}$. The structure of Eq.(135) suggests that Y must be of
the form [see Eq.(131)]
\begin{eqnarray}
Y = Y_A^{(1)} \otimes \Psi_{B}^{(2)} + \Psi_A^{(1)} \otimes
Y_B^{(2)}
\end{eqnarray}
where the linear mappings $\Psi_A^{(1)}$ and $\Psi_B^{(2)}$ satisfy
the conditions $\Psi_A^{(1)}(I_1) = A$ and $ \Psi_B^{(2)}(I_2) = B$.
Recalling the discussion after Eq.(131) and Eq.(1) [and denoting the
multiplication operators in $\sca^{(1)}, \sca^{(2)}$ and \sca \ by
$\mu_1, \mu_2$ and $\mu$ respectively], the condition for Y to be a
superderivation may be written as
\begin{eqnarray}
Y \circ \mu(C \otimes D) - \eta_{Y, C \otimes D} \mu(C \otimes D)
\circ Y = \mu(Y(C \otimes D)). \end{eqnarray} Noting that $ \mu (C
\otimes D) = \mu_1(C) \otimes \mu_2(D)$ (the skew tensor product
causes no problems here), Eq.(137) with Y of Eq.(136) gives
\begin{eqnarray} \eta_{BC} \{ [Y_A^{(1)} \circ \mu_1(C)] \otimes
[\Psi_B^{(2)} \circ \mu_2(D)] + [\Psi_A^{(1)} \circ \mu_1(C)]
\otimes
Y_B^{(2)} \circ \mu_2(D)] \}  \nonumber \\
- (-1)^{\epsilon} \{ [\mu_1(C) \circ Y_A^{(1)}] \otimes [\mu_2(D)
\circ \Psi_B^{(2)}] + [\mu_1(C) \circ \Psi_A^{(1)}] \otimes
[\mu_2(D) \circ Y_B^{(2)}]
\} \nonumber \\
= \eta_{BC} [\mu_1( \{ A,C \}_1) \otimes \mu_2(\Psi_B^{(2)}(D)) +
\mu_1(\Psi_A^{(1)}(C)) \otimes \mu_2(\{B,D \}_2)] \ \ \
\end{eqnarray} where $ \epsilon \equiv \epsilon_A \epsilon_C +
\epsilon_B \epsilon_D + \epsilon_B \epsilon_C$ and we have used the
relations $Y_A^{(1)}(C) = \{ A, C \}_1$ and $ Y_B^{(2)}(D) = \{ B,D
\}_2$.

The objects  $Y_A^{(1)}$ and $Y_B^{(2)}$, being superderivations,
satisfy relations of the form (1) :
\begin{eqnarray} Y_A^{(1)} \circ \mu_1(C) - \eta_{AC} \mu_1(C) \circ
Y_A^{(1)} = \mu_1(Y_A^{(1)}(C)) = \mu_1(\{ A, C \}_1) \nonumber  \\
Y_B^{(2)} \circ \mu_2(D) - \eta_{BD} \mu_2(D) \circ Y_B^{(2)} =
\mu_2( \{ B,D \}_2). \end{eqnarray} Putting $D = I_2$ in Eq.(138),
we have [noting that $\mu_2(D) = \mu_2(I_2) = id_2,$ the identity
mapping on $\sca^{(2)}$ and that $ \{B, I_2 \}_2 = Y_B^{(2)}(I_2) =
0$]
\begin{eqnarray}
  [Y_A^{(1)} \circ  \mu_1(C)] \otimes \Psi_B^{(2)} & + &
       [\Psi_A^{(1)} \circ \mu_1(C)] \otimes Y_B^{(2)} \nonumber \\
 & \ &     - \eta_{AC} \{ [\mu_1(C) \circ Y_A^{(1)}] \otimes \Psi_B^{(2)} +
      [\mu_1(C) \circ \Psi_A^{(1)}] \otimes Y_B^{(2)} \} \nonumber \\
 & \  & = \mu_1(\{ A, C \}_1) \otimes \mu_2(B) \end{eqnarray}
 which, along with equations (139), gives
 \begin{eqnarray} \mu_1( \{ A,C \}_1) \otimes
[\Psi_B^{(2)} - \mu_2(B)] & = & \nonumber \\
- [\Psi_A^{(1)} \circ \mu_1(C) & - & \eta_{AC} \mu_1(C) \circ
\Psi_A^{(1)}] \otimes Y_B^{(2)}. \end{eqnarray} Similarly, putting $
C = I_1$ in Eq.(138), we get
\begin{eqnarray} [\Psi_A^{(1)} - \mu_1(A)] \otimes \mu_2( \{ B,D
\}_2) & = & \nonumber \\
 - Y_A^{(1)} \otimes [\Psi_B^{(2)} \circ \mu_2(D) & - & \eta_{BD}
\mu_2(D) \circ \Psi_B^{(2)}]. \end{eqnarray}

Now, equations (142) and (141) give the relations
\begin{eqnarray}
\Psi_A^{(1)} - \mu_1(A) = \lambda_1 Y_A^{(1)} \end{eqnarray}
\begin{eqnarray} \Psi_B^{(2)} \circ \mu_2(D) - \eta_{BD} \mu_2(D)
\circ \Psi_B^{(2)} = - \lambda_1 \mu_2( \{ B,D \}_2)
\end{eqnarray} \begin{eqnarray} \Psi_B^{(2)} - \mu_2(B) = \lambda_2
Y_B^{(2)} \end{eqnarray} \begin{eqnarray} \Psi_A^{(1)} \circ
\mu_1(C) - \eta_{AC} \mu_1(C) \circ \Psi_A^{(1)} = - \lambda_2
\mu_1( \{A,C \}_1) \end{eqnarray} where $\lambda_1$ and $\lambda_2$
are complex numbers.

Equations (136), (143) and (145) now give
\begin{eqnarray}
Y & = & Y_A^{(1)} \otimes [\mu_2(B) + \lambda_2
        Y_B^{(2)}] + [\mu_1(A) + \lambda_1 Y_A^{(1)}] \otimes Y_B^{(2)}
        \nonumber \\
  & = & Y_A^{(1)} \otimes \mu_2(B) + \mu_1(A) \otimes Y_B^{(2)} +
(\lambda_1 + \lambda_2) Y_A^{(1)} \otimes Y_B^{(2)}.
\end{eqnarray}
Note that only the combination $(\lambda_1 + \lambda_2) \equiv
\lambda$ appears in Eq.(147). To have a unique Y, we must obtain an
equation fixing $\lambda$ in terms of given quantities.

Substituting for $\Psi_A^{(1)}$ and $\Psi_B^{(2)}$ from equations
(143) and (145) into equations (144) and (146) and using equations
(139), we obtain the equations
\begin{eqnarray}
\lambda \mu_1( \{ A,C \}_1) & = & - \mu_1([A,C]) \ \ \textnormal{for
all} \ \ A,C \in \sca^{(1)} \\
\lambda \mu_2( \{ B,D \}_2) & = & - \mu_2 ([B,D]) \ \
\textnormal{for all} \ \ B,D \in \sca^{(2)}. \end{eqnarray} We have
not one but two equations of the type we have been looking for. This
is a signal for the emergence of nontrivial conditions (for the
desired symplectic structure on the tensor product superalgebra to
exist).

Let us consider the equations (148,149) for the various possible
situations (corresponding to whether or not one or both the
superalgebras are super-commutative) :

\vspace{.12in} \noindent (i) Let $\sca^{(1)}$ be supercommutative.
Assuming that the PB $\{, \}_1$ is nontrivial, Eq.(148) implies that
$\lambda = 0.$ Eq.(149) then implies that $\sca^{(2)}$ must also be
super-commutative. It follows that \\
(a) when both the superalgebras $\sca^{(1)}$ and $\sca^{(2)}$ are
super-commutative, the unique Y is given by Eq.(147) with $\lambda =
0$; \\
(b) a `natural'/`canonical' symplectic structure does not exist on
the tensor product of a super-commutative and a non-supercommutative
superalgebra.

\vspace{.12in} \noindent (ii) Let the superalgebra $ \sca^{(1)}$ be
non-supercommutative. Eq.(148) then implies that $ \lambda \neq 0,$
which, along with Eq.(149) implies that the superalgebra
$\sca^{(2)}$ is also non-supercommutative [which is also expected
from (b) above]. Equations (148,149) now give
\begin{eqnarray} \{ A,C \}_1 = - \lambda^{-1} [A,C], \ \
\{ B, D \}_2 = - \lambda^{-1} [B,D] \end{eqnarray} which shows that,
when both the superalgebras are non-supercommutative, a
`natural'/`canonical' symplectic structure on their (skew) tensor
product exists if and only if each superalgebra has a quantum
symplectic structure with the \emph{same} parameter $(- \lambda)$,
i.e.
\begin{eqnarray} \omega^{(1)} = - \lambda \omega^{(1)}_c, \ \
\omega^{(2)} = - \lambda \omega^{(2)}_c \end{eqnarray} where
$\omega^{(i)}_c$ (i=1,2) are the canonical symplectic forms on the
two superalgebras. The traditional quantum symplectic structure is
obtained with $ \lambda = i \hbar$.

\vspace{.1in} \noindent \emph{Note}. The two forms
$\omega^{(i)}$(i=1,2) of Eq.(151) represent bonafide symplectic
structures only if the superalgebras $\sca^{(i)}$ (i=1,2) have only
inner superderivations (see section 3.9). More generally, we can
have generalized symplectic superalgebras $(\sca^{(i)}, \scx^{(i)},
\omega^{(i)})$ (i=1,2) where $\scx^{(i)} = ISDer(\sca^{(i)})$.

\vspace{.1in} In all the permitted cases, the PB on the superalgebra
$ \sca = \sca^{(1)} \otimes \sca^{(2)}$ is given by
\begin{eqnarray} \{ A \otimes B, C \otimes D \} = Y_{A \otimes B}(C
\otimes D)
 & =  & \eta_{BC}
[\{ A, C \}_1 \otimes BD + AC \otimes \{ B, D \}_2  \nonumber \\
 & \ & + \lambda \{ A, C \}_1 \otimes \{ B,D  \}_2 ] \end{eqnarray}
 where the parameter $ \lambda$ vanishes in
 the super-commutative case; in the non-supercommutative case, it is
 the universal parameter appearing in the symplectic forms (151).

Noting that, in the non-supercommutative case,
\begin{eqnarray} \lambda \{A,C \}_1 \otimes \{ B, D \}_2 =
-[A,C] \otimes \{B,D \}_2 = - \{ A,C \}_1 \otimes [B,D] \nonumber \\
=-\frac{1}{2} [A,C] \otimes \{B,D \}_2 - \frac{1}{2} \{A,C \}_1
\otimes [B,D],  \end{eqnarray} the PB of Eq.(152) can be written in
the more symmetric form
\begin{equation} \begin{array}{l}
\{ A \otimes B, C \otimes D \}   \\
= \eta_{BC}[ \{ A,C \}_1 \otimes \frac{BD + \eta_{BD}DB}{2} +
\frac{AC + \eta_{AC} CA}{2} \times  \{ B,D \}_2 ]. \end{array}
\end{equation}

Recalling that, for the matrix algebra $M_n(C) (n \geq 2),$ the
Poisson bracket (with the canonical symplectic form) is a
commutator, Eq.(128) is a special case of  Eq.(154). In fact, had we
employed supermatrices, we would have got exactly Eq.(154) as can be
easily verified using the equation preceding Eq.(127). As shown
below, a direct calculation for the tensor product of two classical
algebras of observables also gives results consistent with Eq.(154).

\vspace{.12in} \noindent \emph{Example [Both algebras commutative]}
\begin{eqnarray*}
\sca^{(1)} = C^{\infty}(R^m), \mbox{\ \ } \sca^{(2)} =
C^{\infty}(R^n); \hspace{.2in} \mbox{(m,n even)}.
\end{eqnarray*}
Let $x^i$ and $y^r$ be the coordinates on $ R^m $ and $ R^n $
respectively and let the  Poisson brackets on them be
\begin{eqnarray*}
\{ f,g\}_1 = \omega_1^{ij}\frac{\partial f}{\partial x^i}
\frac{\partial g}{\partial x^j};  \ \ \{u, v\}_2 =
\omega_2^{rs}\frac{\partial u}{\partial y^r} \frac{\partial
v}{\partial y^s}.
\end{eqnarray*}
Let $z^a = (x^i, y^r)$ be the coordinates on $ R^m \times R^n =
R^{m+n}$. The PB on $ \sca^{(1)} \otimes \sca^{(2)} $ is [putting
$F(z) \equiv (f\otimes u)(x,y) = f(x) u(y)$ and G(z) = g(x)v(y) and
choosing the symplectic form on $R^{m+n}$ in accordance with
Eq.(134)]
\begin{eqnarray}
\{F,G \} & = & \omega^{ab}\frac{\partial F}{\partial z^a}
               \frac{\partial G}{\partial z^b} \nonumber \\
     & = & \omega_1^{ij}\frac{\partial f}{\partial x^i}
            \frac{\partial g}{\partial x^j} uv +
        \omega_2^{rs} \frac{\partial u}{\partial y^r}
        \frac{\partial v}{\partial y^s} fg \nonumber \\
     &  = & \{f, g \}_1 uv + \{u, v \}_2 fg,
\end{eqnarray}
which is consistent with Eq.(154).

\vspace{.12in} \noindent \textbf{6.2. Dynamics of coupled systems}

Given the individual systems $S_1$ and $S_2$ as the supmech
Hamiltonian systems $(\sca^{(i)}, \omega^{(i)},H^{(i)})$ (i = 1,2),
the coupled system ($S_1 + S_2$) is a supmech Hamiltonian system
with the system algebra and symplectic form as discussed above and
the Hamiltonian H given by
\begin{eqnarray}
H = H^{(1)} \otimes I_2 + I_1 \otimes H^{(2)} + H_{int}
\end{eqnarray}
where the interaction Hamiltonian is generally of the form
\begin{eqnarray}
H_{int} = \sum_{i=1}^{n} F_i \otimes G_i.
\end{eqnarray}
The evolution (in the Heisenberg type picture) of a typical
obsevable $A(t)\otimes B(t)$ is governed by the supmech Hamilton's
equation
\begin{eqnarray}
\frac{d}{dt} [ A(t) \otimes B(t)] & = & \{ H, A(t) \otimes B(t)\} \nonumber \\
   & = & \{H^{(1)},A(t)\}_1 \otimes B(t)
                     + A(t) \otimes \{H^{(2)},B(t) \}_2
                     \nonumber \\
                    & \ &  + \{H_{int}, A(t) \otimes B(t) \}.
\end{eqnarray}
The last Poisson bracket in this equation can be evaluated using
Eq.(152) or (154).

In the Schr$\ddot{o}$dinger type picture, the time evolution of
states of the coupled system is given by the supmech Liouville
equation (65) with the Hamiltonian of Eq.(156).

In favorable situations, the supmech Heisenberg or Liouville
equations may be written for finite time intervals by using
appropriate exponentiations of operators. We shall do this in
section 8 below in which  a concrete application of the formalism of
this section to measurements in quantum mechanics will be described.

The main lesson from this section is that \emph{all} systems in
nature whose interaction with other systems can be talked about must
belong to one of the two `worlds' : the `commutative world' in which
all system superalgebras are super-commutative and the
`noncommutative world' in which all system superalgebras are
non-supercommutative with a \emph{universal} quantum symplectic
structure. (No restriction is implied by the above requirement on
the type of symplectic structure on system superalgebras in the
commutative world.) In view of the familiar inadequacy of the
commutative world, the `real' world must clearly be the
noncommutative (hence quantum) world; its systems will be called
quantum systems. (This is formalized as axiom \textbf{A7} in section
9.) The classical systems with commutative system algebras and
traditional symplectic structures will appear only in the
appropriately defined classical limit (or, more generally, in the
classical approximation) of quantum systems.

This brings us on the threshold of an autonomous development of QM.

\vspace{.15in} \noindent \textbf{7. QUANTUM SYSTEMS,
(SUPER-)CLASSICAL SYSTEMS AND QUANTUM-CLASSICAL CORRESPONDENCE}

\vspace{.12in}

We start by  describing what we call `standard quantum systems'
(eventually to be seen as quantum systems without superselection
rules) in purely algebraic terms.

\vspace{.12in} \noindent \textbf{7.1. Standard quantum systems}

\vspace{.12in} By a \emph{standard quantum system} (SQS) we shall
mean a supmech Hamiltonian system $(\sca, \sone, \omega, H)$  in
which the system algebra \sca \ is special (in the sense of section
3.9) and has a trivial graded center and $ \omega $ is the
\emph{quantum symplectic form} $ \omega_Q$ given by [see Eq.(53)]
\begin{eqnarray}
\omega_Q = -i \hbar \omega_c.
\end{eqnarray}
(We have, in the terminology of section 3.9, the quantum symplectic
structure with parameter $ b = -i \hbar$.) This is the only place
where we  put the Planck constant `by hand' (the most natural place
to do it --- such a parameter is \emph{needed} here); its appearance
at all conventional places (canonical commutation relations,
Heisenberg and Schr$\ddot{o}$dinger equations, etc) will be
automatic.

The \emph{quantum Poisson bracket} implied by the quantum symplectic
form is [see Eq.(54)]
\begin{eqnarray}
\{ A, B \} = (-i\hbar)^{-1}[A, B].
\end{eqnarray}
Recall that the bracket [,] represents a supercommutator; it follows
that the bracket on the right in Eq.(160) is an anticommutator when
both A and B are odd/fermionic and a commutator in all other
situations with homogeneous A,B.

A \emph{quantum canonical transformation} is an automorphism $\Phi$ of the
system algebra \sca \ such that $\phstup \omega_Q = \omega_Q$. Now
\begin{eqnarray}
(\phstup \omega_Q )(X_1, X_2) = \Phi^{-1}[ \omega_Q (\phst X_1, \phst X_2)]
\end{eqnarray}
where $ X_1, X_2$ are inner superderivations, say, $D_A$ and $D_B$.
We have
\begin{eqnarray*}
(\phst D_A)(B) = \Phi [D_A (\Phi^{-1}(B)] = \Phi ( [A, \Phi^{-1}(B)])
 = [\Phi(A),B] \end{eqnarray*}
which gives
\begin{eqnarray}
\phst D_A = D_{\Phi(A)}.
\end{eqnarray}
Equations (161) and (47) now give
\begin{eqnarray}
\Phi (i [A,B]) = i [\Phi (A), \Phi (B)]
\end{eqnarray}
which shows, quite plausibly, that quantum canonical transformations
are (in the present algebraic setting --- we have not yet come to
the Hilbert space) the automorphisms of the system algebra
preserving the quantum PBs.

The evolution of an SQS in time is governed, in the Heisenberg
picture, by the supmech Hamilton's equation (64) which now becomes
the familiar \emph{Heisenberg equation} of motion
\begin{eqnarray}
\frac{ dA(t)}{dt} = (-i \hbar)^{-1} [H, A(t)].
\end{eqnarray}
In the Schr$\ddot{o}$dinger  picture, the time dependence is carried
by the states and the evolution equation (65) takes the form
\begin{eqnarray}
\frac{d \phi (t)}{dt}(A) = (-i \hbar)^{-1} \phi(t)([H, A])
\end{eqnarray}
which may be called the \emph{generalized von Neumann equation}.

We shall call two SQSs $\Sigma = (\sca, \sone, \omega, H)$ and
$\Sigma^{\prime} = (\sca^{\prime}, \sone^{\prime}, \omega^{\prime},
H^{\prime})$ equivalent if they are equivalent as supmech
Hamiltonian systems. (See section 4.3.)

\noindent \emph{Note}. In the abstract algebraic framework, the CC
condition is to be kept track of. An advantage, as we shall see
below, of the Hilbert space based realizations of quantum systems is
that the CC condition is automatically satisfied in them.

\vspace{.12in} \noindent \textbf{7.2. Hilbert space based
realizations of standard quantum systems}

\vspace{.12in} Recalling the brief treatment of the
Schr$\ddot{\textnormal{o}}$dinger representation in section 3.10, it
is useful to introduce the concept of a \emph{quantum triple}
$(\sch, \mathcal{D}, \sca)$ where \sch \ is a complex Hilbert space,
$\mathcal{D}$ a dense linear subset of \sch \ and \sca \ an
Op-$^*$-algebra of operators based on ($\sch,\mathcal{D}$). We shall
assume that, for a given \sca, $\mathcal{D}$ is maximal, i.e.
largest such domain. When \sca \ is generated by a finite set of
fundamental observables $ F_1,..,F_n,$ then in the notation of
(Dubin and Hennings 1990), $ \mathcal{D} = C^{\infty}(F_1,..,F_n)$
(i.e. intersection of the domains of all polynomials in
$F_1,..,F_n$).

If, in the quantum triple above, we take \sca \ as our system
algebra, then its states are given by the subclass of density
operators $\rho$ on \sch \ for which $ |Tr (\rho A)| < \infty$ for
all observables A in \sca; the quantity $Tr(\rho A) \equiv
\phi_{\rho}(A)$ (where $\phi_{\rho}$ is the state represented by the
density operator $\rho$) is the expectation value of the observable
A in the state $\phi_{\rho}$. Pure states are the subclass of these
states consisting of one-dimensional projection operators. In view
of the maximality of $\mathcal{D}$, the latter are precisely the
one-dimensional projectors $|\psi><\psi|$ where $\psi$ is any
normalized element of $\mathcal{D}$ (which means that pure states
are the unit rays corresponding to the elements of $\mathcal{D}$).
[\emph{Note.} Here $|\psi><\psi|$ is only a formal notation for the
projector $P_{\psi}$ defined by $P_{\psi} \chi = (\psi, \chi) \psi$
for all $\chi \in \sch$; Dirac bra and ket vectors will be
introduced later.]

When the algebra \sca \ of the quantum triple above is special, we
obtain a Hilbert space based SQS by choosing the quantum symplectic
form as above and an even Hermitian element H of \sca \ as the
Hamiltonian. It is clear that, when the choice of Hamiltonian is not
under consideration, a Hilbert space-based SQS is adequately
described as a quantum triple with the algebra \sca \ qualified as
above.

The CC condition  for the pair $(\mathcal{O}(\sca), \sone)$ can be
explicitly verified for a  Hilbert space based SQS\ : \\
(i) Given $A,B \in \mathcal{O}(\sca)$, and $ (\psi, A \psi) = (\psi,
B \psi)$ for all normalized $\psi$ in $\mathcal{D}$ (hence for all
$\psi$ in $\mathcal{D}$), we have $(\phi, A \psi) = (\phi, B \psi)$
for all $\phi, \psi \in \mathcal{D}$, implying A = B. [Hint :
Consider the given equality with the state vectors $(\phi +
\psi)/\sqrt 2$ and $(\phi +i \psi)/\sqrt 2$.] \\
(ii) Given normalized vectors $ \psi_1, \psi_2 $ in $\mathcal{D}$
and $ (\psi_1, A \psi_1) = (\psi_2, A \psi_2)$ for all $ A \in
\mathcal{O}(\sca)$, the equality $ \psi_1 = \psi_2$ (up to a phase)
can be seen by taking, for A, the projection operators corresponding
to members of an orthonormal basis in \sch \ containing $\psi_1$ as
a member.

\vspace{.1in} \noindent \emph{Note}. We have implicitly assumed
above that all elements of $\mathcal{D}$ represent pure states. This
excludes the situations when \sch \ is a direct sum of more than one
coherent subspaces in the presence of superselection rules.

\vspace{.1in} An interesting feature of the Hilbert space-based SQSs
is that we have density operators representing states which, being
Hermitian operators, are also observables. A density operator $\rho$
is the observable corresponding to the property of  the system being
in the state $\phi_{\rho}$. Given two states represented by density
operators $\rho_1$ and $\rho_2$, we have the quantity $ w_{12} =
Tr(\rho_1 \rho_2)$ defined (representing the expectation value of
the observable $\rho_1$ in the state $\rho_2$ and  vice versa) which
has the natural interpretation of transition probability  from one
of the states to the other (the two are equal because $ w_{12} =
w_{21}).$ When $ \rho_i = |\psi_i><\psi_i|$ (i = 1,2) are pure
states, we have $Tr(\rho_1 \rho_2) = |(\psi_1,\psi_2)|^2$ --- the
familiar text book expression for the transition probability between
two pure quantum states.

\vspace{.1in} \noindent [\emph{Note}. It is desirable to represent
the quantities $w_{12}$ as bonafide probabilities in the standard
form (56) employing an appropriate PObVM. It is clearly adequate to
have such a representation for the case of pure states with $\rho_j
= |\psi_j><\psi_j|$ (j = 1,2), say. To achieve this, let $ \phi =
\phi_{\rho_1}$ and $ \{ \chi_r; r = 1,2,...\}$ an orthonormal basis
in \sch \ having $\chi_1 = \psi_2$. The desired PObVM is obtained by
taking, in the notation of section 4.1,
\[ \Omega = \{ \chi_r; r = 1,2,...\}, \ \ \mathcal{F} = \{
\textnormal{All subsets of}\  \Omega \} \] and, for $ E = \{ \chi_r;
r \in J \} \in \mathcal{F}$ where J is a subset of the positive
integers,
\[ \nu (E) = \sum_{r \in J} |\chi_r><\chi_r|. \]
We now have  $w_{12} = |(\psi_1, \psi_2)|^2 = p_{\phi}(E)$ of
Eq.(56) with $\phi = \phi_{\rho_1}$ and $E = |\chi_1><\chi_1| =
|\psi_2><\psi_2|$.]

 \vspace{.1in} We next consider the implementation, in a Hilbert
space based SQS, of the mappings $(\Phi_1, \Phi_2)$ representing a
quantum canonical transformation. These mappings are subject to the
invariance condition \[ <\Phi_2(\phi_{\rho}), \Phi_1(A)> =
<\phi_{\rho}, A> \] for any $A \in \sca$ and state $\phi_{\rho}$.
The mapping $\Phi_2$ maps pure states onto pure states; hence, for
$\rho = |\psi><\psi|$, we have $\Phi_2(\phi_{\rho}) =
\phi_{\rho^{\prime}}$ with $\rho^{\prime} = |\psi^{\prime}><
\psi^{\prime}|$ for some $\psi^{\prime}$. In this case, writing
$\Phi_1(A) = A^{\prime}$, the above mentioned condition takes the
form \[ (\psi^{\prime}, A^{\prime} \psi^{\prime}) = (\psi, A \psi).
\] These mappings will now be shown to be unitarily implemented,
i.e. for any such pair $(\Phi_1, \Phi_2)$, there is a unitary
operator U on \sch \ such that \[ A^{\prime} = UAU^{-1} \ \
\textnormal{and} \ \ \psi^{\prime} = U \psi. \] \emph{Proof.} In
view of the invariance condition (which fixes the mapping $\Phi_2$
in terms of $\Phi_1$), it is adequate to prove the first relation.
Let $\chi_r$ (r=1,2,...) be an orthonormal basis in \sch \ (with all
the $\chi_r$ in $\mathcal{D}$) and $\chi_r^{\prime}$ `the' image of
$\chi_r$ under the $\Phi_2$ action (i.e. $\chi_r^{\prime}$ is a
representative of the unit ray which is the $\Phi_2$-image of the
unit ray corresponding to $\chi_r$). Expanding $\chi_s^{\prime}$ in
the original basis, we have
\[ \chi_s^{\prime} = = \sum_r U_{rs} \chi_r \ \ \textnormal{where} \
\  U_{rs} = (\chi_r, \chi_s^{\prime}). \] Defining an operator U by
$ (\chi_r, U \chi_s) = U_{rs},$ we have $\chi_s^{\prime} = U \chi_s$
(s=1,2,...). Putting $\psi = \chi_s$ in the invariance condition
above, we have \[ (\chi_s, U^{\dagger}A^{\prime}U \chi_s) = (\chi_s,
A \chi_s).  \] Writing similar equations with $\chi_s$ replaced by $
(\chi_r + \chi_s)/\sqrt{2}$ and $(\chi_r + i \chi_s)/\sqrt{2}$, it
is easily seen that \[ (\chi_r, U^{\dagger} A^{\prime} U \chi_s) =
(\chi_r, A \chi_s) \] which gives $U^{\dagger} A^{\prime} U = A.$

Now, for A = I, we must have $ A^{\prime} = I$ ( the mapping
$\Phi_1$ being an automorphism of the unital algebra \sca); this
gives $U^{\dagger} U = I $ or, remembering the invertibility of the
mapping $\Phi_2$, $U^{\dagger} = U^{-1}$. We have, therefore, $
A^{\prime} = UAU^{-1}.$ The condition (163) implies \[ U (i [A,B])
U^{-1} = i [UAU^{-1}, UBU^{-1}] \] which shows that U must be a
unitary operator. $\Box$

 Note that, while unitary transformations (mapping $\mathcal{D}$
onto itself -- only these are permitted to represent quantum
symmetries in our formalism) are genuine symmetry operations, the
antiunitary transformations are not.

The unitarily implemented $\Phi_2$ actions on states leave the
transition probabilities invariant [in fact, they leave transition
amplitudes invariant : $(\psi^{\prime}, \chi^{\prime}) = (\psi,
\chi)$.] Note that, in contrast with the traditional formalism of
QM, invariance of transition probabilities is not postulated but
proved in the present setting.

A symmetry implemented (in the unimodal sense, as defined in section
4.3) by a unitary operator U acts on a state vector $\psi \in
\mathcal{D}$ according to $ \psi \rightarrow \psi^{\prime} = U \psi$
and (when its action is transferred to operators) on an operator $A
\in \sca$ according to $A \rightarrow A^{\prime}$ such that, for all
$\phi \in \mathcal{D}$,
\begin{eqnarray}
(\phi^{\prime}, A \phi^{\prime}) = (\phi, A^{\prime} \phi)
\hspace{.15in} \Rightarrow  A^{\prime} = U^{-1} A U .
\end{eqnarray}
For an infinitesimal unitary transformation $ U \simeq I + i
\epsilon G$ where G is an even, Hermitian element of \sca \ [this
follows from the condition $ (U \phi, U \psi) = (\phi, \psi)$ for
all $\phi, \psi \in \mathcal{D}$]. Considering the transformation $
A \rightarrow A^{\prime}$ in Eq.(166) as a quantum canonical
transformation, generated (through PBs) by an element $ T \in \sca$,
we have
\begin{eqnarray}
\delta A = - i \epsilon [G, A] = \epsilon \{ T, A \}
\end{eqnarray}
giving
\begin{eqnarray}
T = -i (-i\hbar) G = - \hbar G
\end{eqnarray}
and
\begin{eqnarray}
U \simeq I -i \frac{\epsilon}{\hbar}T.
\end{eqnarray}
It is the appearance of $\hbar$ in Eq(169) which is responsible for
its appearance at almost all conventional places in QM.

The quantum canonical transformation representing evolution in time of an
SQS  is implemented on the state vectors by a one-parameter family of
unitary operators  [in the form $ \psi(t) = U(t-s)\psi(s)$]
generated by the Hamiltonian operator H :
$U(\epsilon) \simeq I -i\frac{\epsilon}{\hbar} H.$
This gives, in the Schr$\ddot{o}$dinger picture, the Schr$\ddot{o}$dinger
equation for the evolution of pure states of a Hilbert space based SQS :
\begin{eqnarray}
i \hbar \frac{d \psi(t)}{dt} = H \psi(t).
\end{eqnarray}
In the Heisenberg picture, we have, of course, the Heisenberg
equation of motion (164), which is now an operator equation.

Quantum triples provide a natural setting for a mathematically
rigorous development of the Dirac bra-ket formalism. The essence of
this formalism lies in generalizing the orthogonal expansions in a
Hilbert space to include integrals over `generalized eigenvectors'
(Fourier transforms, for example). This becomes necessary when some
observables of interest have (partly or wholly) continuous spectrum.
The appropriate formalism for this is provided by a `rigged Hilbert
space' [or Gel'fand triple (Gelfandand Vilenkin 1964)]; the latter
appears as a natural development once a pair $(\sch, \mathcal{D})$
consisting of a Hilbert space \sch \ and a dense linear subset
$\mathcal{D}$ in it is  given.

Given a dense domain $\mathcal{D}$ in \sch, one can define the
*-algebra $L^+(\mathcal{D})$ [in the notation of Lassner (1984)]
-- the largest Op$^*$-algebra based on($\sch,\mathcal{D}$). [If,
instead of choosing \sca \ first and then constructing
$\mathcal{D}$, we had chosen $\mathcal{D}$ first and then proceeded
to choose  \sca, then the natural/simplest  choice of \sca \ would
be $L^+( \mathcal{D})$.] On $\mathcal{D}$, a locally convex topology
$t$ is defined by the seminorms
 $ \| . \|_A $ given by
\begin{eqnarray}
\|\psi \|_A = \| A \psi \|, \ \ A \in L^+(\mathcal{D});
\end{eqnarray}
we denote the resulting locally convex topological vector space by
$\mathcal{D}[t]$. Let $\mathcal{D}^{\prime}[t^{\prime}]$ be the dual
space of $\mathcal{D}[t]$ with the strong topology (Kristensen,
Mejlbo and Thue Poulsen 1965) $t^{\prime}$ (it is defined by the
seminorms
\[ p_B(\chi) = sup_{\psi \in B} |\chi(\psi)| \]
for all bounded subsets B of $\mathcal{D}$). Then the Gelfand triple
\begin{eqnarray*}
\mathcal{D}[t] \subset \sch \subset \mathcal{D}^{\prime}[t^{\prime}]
\end{eqnarray*}
constitutes the \emph{canonical rigged Hilbert space}(Lassner 1984)
based on the pair $(\sch, \mathcal{D})$. The space
$\mathcal{D}^{\prime}[t^{\prime}]$ ( the space of continuous linear
functionals or distributions on the test function space
$\mathcal{D}[t]$) is the space of bra vectors of Dirac. The space of
kets is the space $\mathcal{D}^{\times}$ of continuous antilinear
functionals on $\mathcal{D}[t]$. [An element $\chi \in \sch$ defines
a continuous linear functional $F_{\chi}$ and an antilinear
functional $K_{\chi}$ on \sch \ (hence on $\mathcal{D}$) given by $
F_{\chi}(\psi) = (\chi, \psi)$ and $K_{\chi}(\psi) = (\psi, \chi)$;
both the bra and ket spaces, therefore, have \sch \ as a subset.]

A Hermitian operator A (=$A^*$) in $L^+(\mathcal{D})$ which admits a
unique self adjoint extension in \sch \ ( often called `essentially
self adjoint') and is cyclic [i.e. there exists a vector $\psi$ in
$\mathcal{D}$ such that the vectors $P(A)\psi$, where P(A) is a
polynomial in A, are dense in  \sch] has complete sets of
generalized eigenvectors [eigenkets $\{ |\lambda>; \lambda \in
\sigma (A)$, the spectrum of A $ \}$ and eigenbras $ \{ <\lambda |;
\lambda \in \sigma(A) \}]$ :
\[ A |\lambda> = \lambda | \lambda>; \hspace{.2in} <\lambda | A =
\lambda < \lambda |; \] \begin{eqnarray} \int_{\sigma(A)} d
\mu(\lambda) |\lambda>< \lambda| = I \end{eqnarray} where I is the
unit operator in \sch \ and $\mu $ is a unique measure on
$\sigma(A)$. These equations are to be understood in the sense that,
for all $\phi, \psi \in \mathcal{D}$,
\[ <\phi| A | \lambda> = \lambda <\phi| \lambda>; \hspace{.2in}
<\lambda | A | \phi> = \lambda <\lambda| \phi>; \] \[
\int_{\sigma(A)} d \mu (\lambda)< \phi| \lambda><\lambda| \psi> =
<\phi| \psi>. \] The last equation implies the expansion (in
eigenkets of A)
\begin{eqnarray*}
| \psi> = \int_{\sigma(A)} d \mu (\lambda) |\lambda>< \lambda| \psi>.
\end{eqnarray*}
More generally, one has complete sets of generalized eigenvectors
associated with complete sets of commuting observables. For more
details on the mathematically rigorous development of the bra-ket
formalism, we refer to the literature (Roberts 1966; Antoine 1969;
A. B$\ddot{o}$hm 1978; de la Madrid 2005).

\vspace{.12in} \noindent \textbf{7.3. Inevitability of the Hilbert
space}

\vspace{.12in} Having shown the advantages of a Hilbert space-based
realization of a standard quantum system, we now proceed to consider
the existence and inevitability of such a realization.

Given an (abstract) SQS $\Sigma = (\sca,\sone,\omega,H)$, by a
Hilbert space realization of it we mean an SQS $ \hat{\Sigma} =
(\hat{\sca}, \hat{\sone}, \hat{\omega}, \hat{H})$ of the type
treated in the previous subsection which is equivalent to $\Sigma$
as a supmech Hamiltonian system. This  amounts to constructing a
quantum triple $(\sch, \mathcal{D}, \hat{\sca})$ in which the
algebra $\hat{\sca}$ is isomorphic, as a topological algebra, to the
system algebra \sca \ and choosing $\mathcal{D}$ to be maximal (in
the sense of section 7.2). Elements of $\mathcal{D}$ then provide
pure states such that the observables (i.e. the Hermitian elements
of $\hat{\sca}$) and pure states satisfy the CC condition; moreover,
once  $\hat{\sca}$ has been obtained (as a special algebra
isomorphic to \sca), construction of $\hat{\omega}$ and $\hat{H}$ is
automatic.

From the above definition it is clear that, such a realization, if
it exists, is unique up to equivalence.

Mathematically we have the problem of obtaining a faithful
irreducible *-representation of the *-algebra \sca. Good references
for the treatment of relevant mathematical concepts are Powers
(1971) and Dubin and Hennings (1990). By a *-representation of a
*-algebra \sca \ we mean a triple $(\sch, \mathcal{D}, \pi)$ where
\sch \ is a (separable) Hilbert space, $\mathcal{D}$ a dense linear
subset of \sch \ and $\pi$ a *-homomorphism of \sca \ into the
operator algebra $L^{+}(\mathcal{D})$ (defined in section 7.2)
satisfying the relation
\begin{eqnarray*} (\chi, \pi(A) \psi) = (\pi(A^*) \chi, \psi) \ \
\textnormal{for all} \ \ A \in \sca \ \textnormal{and} \ \  \chi,
\psi \in \mathcal{D}. \end{eqnarray*}

We shall build up our arguments such that no new assumptions will be
involved in going from the abstract algebraic setting to the Hilbert
space setting; emergence of the Hilbert space formalism will be
automatic.

To this end, we shall exploit the fact that the CC condition
guarantees the existence of plenty of (pure) states of the algebra
\sca. Given a state $\phi$ on \sca, a standard way to obtain a
representation of \sca \ is to employ the so-called GNS
construction. Some essential points related to this construction are
given below :

\noindent (i) Noting that the given algebra \sca \ is itself a
complex vector space, one tries to define a scalar product on it
using the state $\phi$, the obvious choice being $(A, B) =
\phi(A^*B)$. This, however, is prevented from being positive
definite if the set
\begin{eqnarray*} L_{\phi} = \{ A \in \sca; \ \phi(A^*A) = 0 \}
\end{eqnarray*}
(which can be shown to be a left ideal of \sca) has nonzero elements
in it. On the quotient space $\mathcal{D}^{(0)}_{\phi} =
\sca/L_{\phi}$, the object
\begin{eqnarray} ([A], [B]) = \phi (A^*B) \end{eqnarray}
is a well defined scalar product. Here $ [A] = A + L_{\phi}$ denotes
the equivalence class of A in $\mathcal{D}^{(0)}_{\phi}$.

\noindent (ii) One then completes the inner product space
$(\mathcal{D}^{(0)}_{\phi}, (,))$ to obtain the Hilbert space
$\sch_{\phi}$; it is guaranteed to be separable by the separability
of the topological algebra \sca.

\noindent (iii) One defines a representation $\pi^{(0)}_{\phi}$ of
\sca \ on the pair $(\sch_{\phi}, \mathcal{D}^{(0)}_{\phi})$ by
\begin{eqnarray} \pi^{(0)}_{\phi}(A)[B] = [AB]; \end{eqnarray}
it can be easily checked to be a well defined *-representation.

\noindent (iv) The operators $\pi^{(0)}_{\phi}(A)$ induce a topology
on $\mathcal{D}^{(0)}_{\phi}$ [see Eq.(171)]; the completion
$\mathcal{D}_{\phi}$ of $\mathcal{D}^{(0)}_{\phi}$ in this topology
acts as the common invariant domain for the operators
$\pi_{\phi}(A)$ (where $\pi_{\phi}$ is the closure of the
representation $\pi^{(0)}_{\phi}$).

\noindent (v) The original state $\phi$ is represented as a vector
state in the representations $\pi^{(0)}_{\phi}$ and $\pi_{\phi}$ by
the vector $\chi_{\phi} = [I]$ (the equivalence class of the unit
element of \sca); indeed, we have, from Eq.(173),
\begin{eqnarray} \phi(A) & = & ([I], [A]) = ([I],
\pi^{(0)}_{\phi}(A)[I]) \nonumber \\
& = & (\chi_{\phi}, \pi^{(0)}_{\phi}(A) \chi_{\phi}) = (\chi_{\phi},
\pi_{\phi}(A) \chi_{\phi}). \end{eqnarray}

\noindent (vi) The triple $( \sch_{\phi}, \mathcal{D}_{\phi},
\pi_{\phi})$ satisfying Eq.(175), referred to as the GNS
representation of \sca \ induced by the state $\phi$ [ some authors
refer to the triple $(\sch_{\phi}, \mathcal{D}^{(0)}_{\phi},
\pi^{(0)}_{\phi})$ as the GNS representation of \sca], is determined
uniquely, up to unitary equivalence, by the state $\phi$.

\noindent (vii) The representation $\pi_{\phi}$ of \sca \ is
irreducible if and only if the state $\phi$ is pure.

This construction, however, does not completely solve our problem
because a GNS representation is generally not faithful; for all $A
\in L_{\phi}$, we have obviously $\pi_{\phi}(A) = 0$. [For example,
a state with zero expectation value for the kinetic energy of a
particle will yield a GNS representation which will represent the
momentum observable of the particle by the zero operator.]

\vspace{.1in} \noindent \emph{Note}. The GNS representation is
faithful if the state $\phi$ is faithful (i.e. if $L_{\phi} = \{ 0
\}$). Such a state, however, is not guaranteed to exist by our
postulates.

\vspace{.1in} A faithful but generally reducible representation of
\sca \ can be obtained by taking the direct sum of the
representations of the above sort corresponding to \emph{all} the
pure states $\phi$. [For the construction of the direct sum of a
possibly uncountable set of Hilbert spaces, see Rudin (1974).] Let
$\mathcal{K}$ be the Cartesian product of the Hilbert spaces $ \{
\sch_{\phi}: \phi \in \sone(\sca) \}$. A general element $\psi$ of
$\mathcal{K}$ is a collection $ \{ \psi_{\phi}: \phi \in \sone(\sca)
\}$; here $\psi_{\phi}$ is called the component of $\psi$ in
$\sch_{\phi}$. The desired Hilbert space \sch \ consists of those
elements $\psi$ in $\mathcal{K}$ which have an at most countable set
of nonzero components $ \psi_{\phi}$ which, moreover, satisfy the
condition
\[ \sum_{\phi} \| \psi_{\phi} \|^2_{\sch_{\phi}} < \infty. \]
The scalar product in \sch \ is given by
\[ (\psi, \psi^{\prime}) = \sum_{\phi} (\psi_{\phi},
\psi^{\prime}_{\phi})_{\sch_{\phi}}. \] The direct sum of the
representations $\{ (\sch_{\phi}, \mathcal{D}_{\phi}, \pi_{\phi});
\phi \in \sone(\sca) \} $ is the representation $(\sch, \mathcal{D},
\pi)$ where \sch \ is as above, $\mathcal{D}$ is the subset of \sch
\ consisting of vectors $\psi$ with $ \psi_{\phi} \in
\mathcal{D}_{\phi}$ for all $\phi \in \sone(\sca)$ and, for any $A
\in \sca$,
\[ \pi(A)\psi = \{ \pi_{\phi}(A)\psi_{\phi}; \phi \in \sone(\sca)
\}. \]

Now, given any two different elements $A_1, A_2$ in \sca, let
$\phi_0$ be a pure state (guaranteed to exist by the CC condition)
such that $\phi_0(A_1) \neq \phi_0(A_2)$. Let $ \psi_0 \in \sch$ be
the vector with the single nonzero component $\chi_{\phi_0}$. For
any $A \in \sca$, we have
\[ (\psi_0, \pi(A) \psi_0) = ( \chi_{\phi_0}, \pi_{\phi_0}(A)
\chi_{\phi_0}) = \phi_0(A). \] This implies
\[ (\psi_0, \pi(A_1) \psi_0) \neq (\psi_0, \pi(A_2) \psi_0), \ \
\textnormal{hence} \ \pi(A_1) \neq \pi(A_2) \] showing that the
representation $(\sch, \mathcal{D}, \pi)$ is faithful.

The Hilbert space \sch \ obtained above may be non-separable (even
if the spaces $\sch_{\phi}$ are separable); this is because the set
$\sone(\sca)$ is generally uncountable. To obtain a faithful
representation of \sca \ on a separable Hilbert space, we shall use
again the separability of \sca \ as a topological algebra. Let $
\sca_0 = \{ A_1, A_2, A_3,...\}$ be a countable dense subset of \sca
\ consisting of nonzero elements. The CC condition guarantees the
existence of pure states $\phi_j$ (j=1,2,...) such that
\begin{eqnarray} \phi_j (A_j^* A_j) \neq 0, \ \ j= 1,2,...
\end{eqnarray} Now consider the GNS representations $(\sch_{\phi_j},
\mathcal{D}_{\phi_j}, \pi_{\phi_j})$ (j=1,2,...). Eq.(176)
guarantees that
\begin{eqnarray} \pi_{\phi_j}(A_j) \neq 0 , \ \ j= 1,2,...
\end{eqnarray} Indeed
\begin{eqnarray*} 0 \neq \phi_j(A_j^*A_j) & = & (\chi_{\phi_j},
\pi_{\phi_j}(A_j^*A_j) \chi_{\phi_j}) \\
& = & (\pi_{\phi_j}(A_j) \chi_{\phi_j},
\pi_{\phi_j}(A_j)\chi_{\phi_j}). \end{eqnarray*} Now consider the
direct sum $(\sch^{\prime}, \mathcal{D}^{\prime}, \pi^{\prime})$ of
these representations. To show that $\pi^{\prime}$ is faithful, we
must show that, for any nonzero element A of \sca, $ \pi^{\prime}(A)
\neq 0.$ This is guaranteed by Eq.(177) because, $\sca_0$ being
dense in \sca, A can be arranged to be as close as we like to some
$A_j$ in $\sca_0$.

The representation $\pi^{\prime}$, however, is, in general,
reducible. To obtain a faithful irreducible representation, we
should try to obtain the relations $\pi(A_j) \neq 0$ (j= 1,2,..) in
a single representation $\pi$. To this end, let $B^{(k)} = A_1
A_2...A_k$ and choose $\phi^{(k)} \in \sone(\sca)$ such that
\[ \phi^{(k)} (B^{(k)*}B^{(k)}) \neq 0. \]
In the GNS representation $(\sch_{\phi^{(k)}},
\mathcal{D}_{\phi^{(k)}}, \pi_{\phi^{(k)}})$, we have
\[ 0 \neq \pi_{\phi^{(k)}}(B^{(k)}) = \pi_{\phi^{(k)}}(A_1)...
\pi_{\phi^{(k)}}(A_k) \] which implies
\begin{eqnarray} \pi_{\phi^{(k)}}(A_j) \neq 0 , \ \ j=1,...,k.
\end{eqnarray}
This argument works for arbitrarily large but finite k. If the $ k
\rightarrow \infty $ limit of the above construction leading to a
limiting GNS representation $(\underline{\sch},
\underline{\mathcal{D}}, \underline{\pi})$ exists, giving
\begin{eqnarray} \underline{\pi}(A_j) \neq 0, \ \ j= 1,2,...,
\end{eqnarray}
then, by an argument similar to that for $\pi^{\prime}$ above, one
must have $\underline{\pi}(A) \neq 0 $ for all non-zero A in \sca \
showing faithfulness of $\underline{\pi}$.

\vspace{.1in} \noindent \emph{Note}. (i) For finitely generated
system algebras (this covers all applications of QM in atomic
physics), a limiting construction is not needed; the validity of
Eq.(178) for sufficiently large k is adequate. [Hint : Take  the
generators of the algebra \sca \ as some of the elements of
$\sca_0$.]

 \noindent (ii) For general algebras, it appears that some extra
condition is needed to arrive at a faithful irreducible
representation.

\noindent (iii) The developments in this subsection did not require
the algebra \sca \ to belong to the restricted class employed for
SQSs in section 7.1; the results obtained are, therefore, valid for
more general quantum systems. We shall use this fact in section 7.5.

\noindent (iv) In fact, even non-commutativity of the algebra \sca \
was not used above. This, however, is not surprising; commutative
algebras, under fairly general conditions, can be realized as
algebras of operators in Hilbert spaces.

\vspace{.1in} A more complete treatment of these matters is intended
to be presented when the treatment of quantum field theory in an
appropriately augmented supmech framework is taken up.

Having shown the existence and desirability of the Hilbert
space-based realizations for finitely generated system algebras, we
now have a formal justification for the direct route to the Hilbert
space taken in the traditional treatment of QM of localizable
elementary systems (massive particles), namely, employment of
projective unitary irreducible representations of the relativity
group $G_0$. This is the simplest way to satisfy the condition of
transitive action of $ G_0$ on the space of pure states and
simultaneously satisfy the CC condition.

We take up the QM of these objects in the next subsection.

\vspace{.12in} \noindent \textbf{7.4. Quantum mechanics of
localizable elementary systems (massive particles)}

\vspace{.12in} A \emph{quantum elementary system} is an SQS which is
also an elementary system. The concept of a quantum elementary
system, therefore, combines the concept of quantum symplectic
structure with that of a relativity scheme. The basic entities
relating to an elementary system are its fundamental observables
which generate the system algebra \sca. For quantum elementary
systems, this algebra \sca \ has the quantum symplectic structure as
described in section 7.1. All the developments in section 5 can now
proceed with the  PBs understood as quantum PBs. We shall employ the
Hilbert space-based realizations of these systems (which are
guaranteed to exist because the algebra \sca \ is finitely
generated).

The  relativity group $G_0$ (or its projective group $\hat{G}_0$)
has a Poisson action on \sca \ and a transitive action on the set
$\sone(\sca)$ of pure states of \sca. We have seen in section 7.2
that, in a Hilbert space based realization of an SQS in terms of a
quantum triple $(\sch, \mathcal{D},\sca)$, a symmetry operation can
be represented as a unitary operator on \sch \ mapping $\mathcal{D}$
onto itself. A symmetry group is then realized as a unitary
representation on \sch \ such that the representative operators map
$\mathcal{D}$ onto itself. For an elementary system the condition of
transitive action on \sone \ implies that this representation must
be irreducible.(There is no contradiction between this requirement
and that of invariance of $\mathcal{D}$ because $\mathcal{D}$ is not
a closed subspace of \sch \ when \sch \ is infinite dimensional.)

By a (quantum) \emph{particle} we shall mean a localizable (quantum)
elementary system. We shall first consider nonrelativistic
particles. The configuration space of a nonrelativistic particle is
the 3-dimensional Euclidean space $R^3$. The fundamental observables
for such a system were identified, in section 5.2, as the mass (m)
and Cartesian components of  position ($X_j$), momentum ($P_j$) and
spin($S_j$) (j = 1,2,3) satisfying the PB relations in equations
(107,108,102). The mass m will be treated, as before, as a positive
parameter. The system algebra \sca \ of the particle is the
*-algebra generated by the fundamental observables (taken as
hermitian) and the unit element. Since it is an ordinary
*-algebra (i.e. one not having any fermionic objects), the
supercommutators reduce to ordinary commutators. The PBs mentioned
above now take the form of the commutation relations
\begin{eqnarray}
  [X_j, X_k] = 0 = [P_j, P_k], \ \  [X_j, P_k] =
i \hbar \delta_{jk} I \hspace{1.2in} (A)\nonumber \\
 {[S_j, S_k]} = i \hbar \epsilon_{jkl} S_l, \ \ [S_j, X_k] = 0 =
[S_j, P_k]. \hspace{1.4in} (B)
\end{eqnarray}

We shall first consider the spinless particles ($\mathbf{S} = 0$).
We, therefore, need to consider only the Heisenberg commutation
relations (180A)[often referred to as the \emph{canonical
commutation relations}(CCR)]. Assuming the existence of a quantum
triple $ (\sch, \mathcal{D}, \hat{\sca})$ corresponding to this SQS,
we shall employ some results obtained in section 7.2 to obtain the
explicit construction. Here $\hat{\sca}$ is the algebra generated by
(representatives of) the fundamental observables $X_j, P_j$ (j =
1,2,3) and the unit element I subject to the commutation relations
(180A) and the pair $(\sch, \mathcal{D})$ carries a faithful
irreducible representation of the system algebra as explained above.
We introduce the bra and ket spaces  as in section 7.2. Let $ x =
(x_1, x_2, x_3), dx = dx_1 dx_2 dx_3 $ and $|x>, <x|$ the
simultaneous eigenkets and eigenbras of the operators $X_j$ (j=
1,2,3):
\begin{eqnarray}
X_j |x> \ = \ x_j |x>, \ \ <x|X_j \ = \ <x| x_j, \ \ x_j \in R, \ \
j=1,2,3;
\end{eqnarray}
they are assumed (with a promise of justification later) to form a
complete set providing a resolution of identity in the form
\begin{eqnarray}
I =  \int_{R^3} |x> dx <x|.
\end{eqnarray}
Given any vector $ |\psi> \in \mathcal{D}$, the corresponding wave
function $ \psi(x) = <x|\psi>$ must satisfy the relation
\begin{eqnarray}
(X_j \psi)(x) = <x|X_j|\psi> = x_j \psi(x).
\end{eqnarray}

 Recalling the discussion of localization in section 4.7, the localization
observable  P(D) corresponding to a Borel set D in $R^3$ is
represented as  the operator
 \begin{eqnarray}
 P(D) = \int_D |x> dx <x|.
 \end{eqnarray}
[The required properties of P(D) are easily verified.]
 Given the particle in the state corresponding to $|\psi> \in \mathcal{D}$, the probability
 that it will be found in the domain D is given by
 \begin{eqnarray}
 <\psi| P(D) |\psi> = \int_D <\psi|x> dx <x| \psi> = \int_D |\psi(x)|^2 dx
 \end{eqnarray}
 giving the traditional Born interpretation of the wave function $\psi$. The
 integral above is meaningful for all Borel sets D only if $\psi$ is square
 integrable over $R^3$ which implies $\sch = L^2(R^3, dx)$.

To determine the  operators $P_j$, we must choose  the unitary
operators $ U(a)$ representing space translations such that the
infinitesimal generators satisfy the last two equations in (180A).
The simplest choice for $U(a)$, namely,
 \begin{eqnarray*}
[U(a) \psi](x) = \psi (x-a)
\end{eqnarray*}
[which is a special case of of the relation [$U(g) \psi](x) =
\psi(T_{g}^{-1}x)$; these operators are unitary when the
transformation $T_g$ of $R^3$ preserves the Lebesgue measure]
happens to be adequate. Recalling Eq.(169), we have, for an
infinitesimal translation,
\begin{eqnarray*}
\delta \psi = - \frac{i}{\hbar} \mathbf{a.P} \psi =
- \mathbf{a.\bigtriangledown} \psi
\end{eqnarray*}
giving the  operators $P_j$  representing momentum components as
 \begin{eqnarray}
 (P_j \psi)(x) = -i \hbar \frac{\partial \psi}{\partial x_j}
 \end{eqnarray}
which satisfy the desired commutation relations.

We now identify the space $\mathcal{D}$ as (Dubin and Hennings 1990)
\[ \mathcal{D} = C^{\infty}(X_j,P_j; j=1,2,3) = \scs (R^3). \]
The operators U(a) clearly map this domain onto itself. With this
choice  of $\mathcal{D}$,  the operators $X_j$ and $P_j$ given by
equations (183) and (186) are essentially  self adjoint. The two
triples of operators $\{X_j \} $ and $ \{ P_j \}$ separately
constitute complete sets of commuting operators. The completeness of
the  $X_j$s can be easily seen by operating in the X-representation
and taking the harmonic oscillator ground state wave function as the
cyclic vector. Similar argument works for the $P_j$s in the momentum
representation. This confirms the legitimacy of the Dirac
constructions employed above.

 The pair  $ (\sch, \mathcal{D}) = (L^2(R^3), \scs(R^3))$ with operators $X_j$ and
 $P_j$ as constructed above is known as the \emph{Schr$\ddot{o}$dinger
 representation} of the CCR (180A).

Closures of the operators $P_j, X_j$ (which are self adjoint and are
denoted by the same symbols) generate the unitary groups of
operators $ U(a) = exp(-ia.P)$ and $V(b) = exp(-ib.X)$ (where $a.P =
\sum_j a_jP_j$ etc. and we have put $\hbar = 1.$) which satisfy the
Weyl commutation relations
\begin{eqnarray}
U(a)U(b) &=& U(b)U(a) = U(a+b), \ V(a)V(b) = V(b)V(a) = V(a+b) \nonumber \\
U(a) V(b) &=& e^{ia.b} V(b) U(a).
\end{eqnarray}
For all $\psi \in \mathcal{D}$, we have
\begin{eqnarray}
(U(a)\psi)(x) = \psi(x-a), \ \ (V(b)\psi)(x) = e^{-ib.x} \psi(x);
\end{eqnarray}
this is referred to as the Schr$\ddot{o}$dinger representation of
the Weyl commutation relations. According to the uniqueness theorem
of von Neumann (Varadarajan 1985), the irreducible representation of
the Weyl commutation relations is, up to unitary equivalence,
uniquely given by the Schr$\ddot{o}$dinger representation (188).

\vspace{.1in} \noindent \emph{Note}. (i) Not every representation of
the CCR (180A) with essentially self adjoint $X_j$ and $P_j$ gives a
representation of the Weyl commutation relation. [For a
counterexample, see Inoue (1998), example (4.3.3).] A necessary and
sufficient condition for the latter to materialize is that the
harmonic oscillator Hamiltonian operator $ H = P^2/(2m) + kX^2/2$ be
essentially self adjoint. In the Schr$\ddot{o}$dinger representation
of the CCR obtained above, this condition is satisfied (Glimm and
Jaffe 1981; Dubin and Hennings 1990).

\vspace{.1in} \noindent (ii) The von Neumann uniqueness theorem
serves to confirm/verify, in the present case, the uniqueness (up to
equivalence) of the Hilbert space realization of an SQS mentioned in
section 7.2.

\vspace{.1in} Quantum dynamics of a free nonrelativistic spinless
particle is governed, in the Schr$\ddot{o}$dinger picture, by the
Schr$\ddot{o}$dinger equation (170) with $\psi \in \mathcal{D} =
\mathcal{S}(R^3)$ and  with the Hamiltonian (110)[where $\mathbf{P}$
is now the operator in Eq.(186)]:
\begin{eqnarray} i \hbar \frac{\partial \psi}{\partial t} = -
\frac{\hbar^2}{2m} \bigtriangledown^2 \psi. \end{eqnarray} Explicit
construction of the projective unitary representation of the
Galilean group $G_0$ in the Hilbert space $ \sch = L^2(R^3,dx)$ and
Galilean covariance of the free particle Schr$\ddot{o}$dinger
equation (189) have been treated in the literature (Bargmann 1954;
Varadarajan 1985; Dass and Sharma 1998).

When external forces are acting, the Hamiltonian operator has the
more general form (111). Restricting V in this equation to a
function of $\mathbf{X}$ only (as is the case in common
applications), and proceeding as above, we obtain the traditional
Schr$\ddot{o}$dinger equation
\begin{eqnarray}
i \hbar \frac{\partial \psi}{\partial t} = [- \frac{\hbar^2}{2m}
 \bigtriangledown^2 + V(\mathbf{X})] \psi.
 \end{eqnarray}

\vspace{.1in} \noindent \emph{Note}. In (Dass 2002), the free
particle Hamiltonian operator of Eq.(127)(there) was arrived at by
direct reasoning from the commutation relations of the projective
Galilean group. (This has some instruction value.) Here we have used
essentially similar reasoning to arrive at the free particle
Hamiltonian of Eq.(110) in the more general context of supmech. In
(Dass 2002), the full Hamiltonian of Eq.(128)(there) has V(X)
[instead of the V(X,P) of Eq.(111) above]. This is due to a mistake
in the last stage of the argument there (which occurred while using
the relation $[H^{\prime},X_k] =0$ which need not be valid when
interactions are present.)

\vspace{.1in} It should be noted that, in the process of obtaining
the Schr$\ddot{o}$dinger equation (190) for a nonrelativistic
spinless particle with the traditional Hamiltonian operator, we did
not use the classical Hamiltonian or Lagrangian for the particle. No
\emph{quantization} algorithm has been employed; the development of
the  quantum mechanical formalism has been autonomous, as promised.

From this point on, the development of QM along the traditional lines can
proceed.

For nonrelativistic particles with $m > 0$ and spin $s \geq 0$, we
have $\sch = L^2(R^3, C^{2s+1})$ and $\mathcal{D} = \scs(R^3,
C^{2s+1})$. The treatment of spin being standard, we skip the
details.

Relativistic elementary quantum systems have been treated
extensively in literature (Wigner 1939; Bargmann and Wigner 1948;
Varadarajan 1985; Alonso 1977,1979). These treatments employ
projective irreducible representations of the Poincar$\acute{e}$
group which can be obtained from the irreducible unitary
representations of its covering group. The justification for this
already has been given in the supmech formalism in the last para of
section 7.3. We shall skip the details.

\vspace{.12in} \noindent \textbf{7.5. Quantum systems with more
general system algebras; Superselection rules}

\vspace{.12in} Now we consider general quantum systems which, as
already defined in section 6.2, are those with (not necessarily
special) non-supercommutative system algebras equipped with quantum
symplectic structure. Standard quantum systems are the subclass of
these in which the system algebra \sca \ has a trivial graded center
and only inner superderivations. We shall now relax these two
conditions. On a noncommutative (super-)algebra \sca \ having a
trivial graded center but having both inner and outer
(super-)derivations, a quantum symplectic structure can be defined
by employing the generalization of the supmech formalism treated in
section 4.5 and operate with the generalized symplectic superalgebra
$(\sca, \scx, \omega_Q)$ where \scx = ISDer(\sca) (this gives,
again, the quantum PBs of section 7.1). For convenience of
reference, we shall call this class of systems \emph{quasi-standard
quantum systems}.

We next consider the generalization involving a nontrivial graded
center. We shall restrict ourselves to considering a nontrivial
center only.  The center $\mathcal{C} \equiv Z_0(\sca)$ of \sca \ is
a commutative locally convex algebra. Keeping in view that a
faithful (not necessarily irreducible) Hilbert space-based
realization of the superalgebra \sca \ is always possible, we shall
operate, in this subsection, in the framework of a quantum triple
$(\sch, \mathcal{D}, \hat{\sca})$ where $\hat{\sca}$ is a faithful
realization of \sca.

A nontrivial center $\mathcal{C}$ implies the presence of
superselection rules and/or external fields. The two are, in fact,
related : values of the external fields define superselection rules.
Before taking up the general case, we consider a couple of
illustrative situations :

\noindent (i) Consider first the situation when $\mathcal{C}$ is
generated by a finite number of self-adjoint operators $Q_s$
(s=1,..,n) each of which has a discrete spectrum (we shall call such
observables \emph{charge type observables}). In this case, we have
$\sch = \oplus_i \sch_i$ where the spaces $\sch_i$ are simultaneous
eigenspaces of the observables $Q_s$. Defining $\mathcal{D}_i =
\mathcal{D}\mid \sch_i$ and $\sca_i = \sca \mid \sch_i$, the
algebras $\sca_i$ have trivial center and the quantum triples
$(\sch_i, \mathcal{D}_i, \sca_i)$ correspond to quasi-standard
quantum systems; the operators $Q_s$ act as superselection operators
and the spaces $\sch_i$ as coherent subspaces labeled by the set of
eigenvalues of the superselection operators.

\noindent (ii) Functions representing (components of) external
fields belong to the commutative algebra of functions on a connected
domain of the space-time manifold; in the relevant situations, this
algebra obviously belongs to the center of the system algebra.

Regarding the general situation (of which, the treatment below is,
admittedly, somewhat heuristic) we note that generalizations of the
famous Gel'fand-Naimark theorem on commutative $C^*$-algebras
(Bratteli and Robinson 1979) to some classes of commutative locally
convex *-algebras have appeared in literature (Iguri and Castagnino
1999; $\acute{A}$lvarez 2004). These generalizations relate (through
isomorphisms or, more generally, homomorphisms) the latter algebras
to those of continuous functions on reasonably `nice' classes of
topological spaces [typically Tychonoff spaces ($\acute{A}$lvarez
2004)]; these topological spaces will be referred to as the spectral
spaces of the respective algebras. In the above-mentioned Hilbert
space-based realization, the center $\mathcal{C} \equiv
Z_0(\hat{\sca})$ will be represented as a commutative algebra of
operators with $\mathcal{D}$ as a common invariant domain.
Generalization of the spectral theorem to commutative algebras of
operators is expected to lead to a representation of \sch \ as a
direct integral
\begin{eqnarray} \sch = \int_{\Sigma}^{\oplus} \sch( \lambda) d
\sigma(\lambda) \end{eqnarray} where $\Sigma$ is the spectral space
of $\mathcal{C}$ and $\sigma$ is a measure uniquely determined by
the algebra $\mathcal{C}$. Defining $\mathcal{D}(\lambda)$ and
$\hat{\sca}(\lambda)$ as the restrictions of $\mathcal{D}$ and
$\hat{\sca}$ to $\sch(\lambda)$, we have the quantum triples
$(\sch(\lambda), \mathcal{D}(\lambda), \hat{\sca}(\lambda))$
representing quasi-standard quantum systems. The Hilbert spaces
$\sch(\lambda)$ are traditionally referred to as coherent subspaces.

Generally the space $\Sigma$ will be disconnected; the integral
(191) will then reduce to a sum of integrals of the same type over
the connected pieces of $\Sigma$. The examples (i) and (ii) above
correspond to the two extreme situations when the space $\Sigma$ is,
respectively, discrete and connected.

An important class of examples corresponds to the situation when
there are one or more mutually commuting operators with continuous
spectra defining superselection rules. The mass operator of a
Galilean particle is a good example. Had we not invoked, in section
5.2, the CC condition to have M = mI and had treated the mass
observable M simply as an element of the center $\mathcal{C}$
(thereby making the latter non-trivial), we could not have treated
the system in question as a standard quantum system. Instead, we
would have the situation in Eq.(191) with $\lambda$ replaced by the
mass parameter m and and $\Sigma$ an interval of the form $[a,
\infty)$ with $a > 0$. Galilean relativity provides no clue to the
value of a; it may tentatively be taken to be the mass of the least
massive among the positive mass particles in nature.

A more systematic treatment  of the general case is expected to be
presented in a future work relating to a supmech-based treatment of
quantum field theory.

Before closing this subsection, it is worth emphasizing that

\vspace{.1in} \noindent (i) in the present formalism, there is a
natural place for superselection rules;

\noindent (ii) the superselection rules arising as described above
are commutative --- a highly desirable feature.

For a somewhat complementary treatment of matters related to this
subsection, we refer to the insightful paper of Jauch and Misra
(1961).

\vspace{.12in} \noindent \textbf{7.6. Classical Systems}

\vspace{.12in} Continuing the treatment, in section 3.8, of
classical symplectic structures as special cases of the symplectic
structures of section 3.6, we note here that a classical Hamiltonian
system $(M, \omega_{cl}, H_{cl})$ is realized in supmech as the
Hamiltonian system $(\sca_{cl}, \omega_{cl},H_{cl})$ where
$\sca_{cl} = C^{\infty}(M)$ and $H_{cl}$ a smooth real-valued
function which is bounded below. The supmech Hamilton equation (64)
is, in the present context, nothing but the traditional Hamilton
equation:
\begin{eqnarray}
\frac{df}{dt} = \{ H_{cl}, f \}_{cl}.
\end{eqnarray}

States in the present context are probability measures on M; in obvious
notation, they are of the form
\begin{eqnarray}
\phi_{\mu}(f) = \int_M f d \mu.
\end{eqnarray}
Pure states are Dirac measures (or, equivalently, points of M) $
\mu_{\xi_0}  (\xi_0 \in M)$ for which $ \phi_{\xi_0}(f) = f
(\xi_0)$. The pair $(\mathcal{O}(\acl), \sone (\acl))$ of classical
observables and pure states is easily seen to satisfy the CC
condition : Given two different real-valued smooth functions on M,
there is a point of M at which they take different values;
conversely, given two different points of M, there is a real-valued
smooth function on M which takes different values at those points.

In ordinary mechanics, only pure states are used. Expectation values of
\emph{all} observables in these states are their precise values
at the relevant points and the theory is deterministic (obtained here as a
special case of a probabilistic theory). More general states are employed in
classical statistical mechanics where, in most applications, they are taken
to be represented by densities on M [$ d \mu = \rho (\xi) d \xi $ where
$ d \xi = dq dp $ is the Liouville volume element on M].

The state evolution equation (65) of supmech gives, in the present
context \begin{eqnarray} \int_M (\frac{\partial \rho
(\xi,t)}{\partial t})(\xi)f(\xi) d \xi = \int_M \rho (\xi,t) \{H, f
\}_{cl}(\xi) d \xi.
\end{eqnarray}
For the remainder of this subsection, we take $ M= R^{2n}$ . To
satisfy the normalization condition, the density $\rho$ must vanish
at infinity. Now, using the Poisson bracket of section 3.8 in
Eq.(194), performing a partial integration and discarding the
(vanishing) surface term, the right hand side of Eq.(194) becomes
\begin{eqnarray*}
- \int_M \omega_{cl}^{ab}(\xi)\frac{\partial H}{\partial \xi^a}
\frac{\partial \rho}{\partial \xi^b}f d \xi = \int_M \{ \rho,H \}_{cl}f d \xi.
\end{eqnarray*}
Since f is arbitrary, Eq.(194) now gives the traditional Liouville
equation \begin{eqnarray} \frac{\partial \rho}{\partial t} = \{
\rho,H \}_{cl}. \end{eqnarray}

A classical Galilean elementary system is a system
characterized/labelled by the three Galilean invariants
m,$\sigma$,u. Its fundamental observables other than the invariants
are the position, momentum and spin vectors \textbf{X,P,S}
satisfying the PB relations of section 5.2 (where the symbols now
represent phase space variables). The observable \textbf{X} has the
interpretation of the position vector of the center of mass of the
system. A particle is an elementary system with the internal energy
u = 0 and negligible size so that \textbf{X} now refers to the
particle position. The free particle Hamiltonian for for a spinless
particle is given by Eq.(110) and the one with interaction in
Eq.(111). For a detailed treatment of classical Galilean systems we
refer to the literature (SM; Alonso 1979).

\vspace{.12in} \noindent \textbf{7.7. Superclasical systems}

\vspace{.12in} Superclassical mechanics is an extension of classical
mechanics which employs, besides the traditional phase space
variables, Grassmann variables $\theta^{\alpha} (\alpha = 1,..n$,
say) satisfying the relations
\begin{eqnarray*}
\theta^{\alpha}\theta^{\beta} + \theta^{\beta}\theta^{\alpha} = 0
\hspace{.2in} \textnormal{for all} \ \alpha, \beta;
\end{eqnarray*}
[in particular $(\theta^{\alpha})^2 = 0$ for all $ \alpha$]. These objects
generate the so -called Grassmann algebra (with n generators) $ \mathcal{G}_n$
whose elements are functions of the form
\begin{eqnarray}
f(\theta) = a_0 + a_{\alpha} \theta^{\alpha} + a_{\alpha \beta}
\theta^{\beta} \theta^{\alpha} + ...
\end{eqnarray}
where the coefficients $ a_{..}$ are complex numbers. If the
coefficients in Eq.(196) are taken to be smooth functions on, say,
$R^m$, the resulting functions $ f(x,\theta) $ are referred to as
smooth functions on the superspace $ R^{m \mid n}; $ the algebra of
these functions is denoted as $ C^{\infty}(R^{m \mid n}). $ With
parity zero assigned to the variables $x^a$ (a = 1,..,m) and one to
the $\theta^{\alpha}, C^{\infty}(R^{m \mid n})$ is a
supercommutative superalgebra. Restricting the variables $x^a$ to an
open subset U of $R^m$, one obtains the superdomain $U^{m \mid n}$
and the superalgebra $C^{\infty}(U^{m \mid n})$ in the
above-mentioned sense. Gluing such superdomains appropriately, one
obtains the objects called supermanifolds (Leites 1980; Berezin
1987; Vorontov 1992; DeWitt 1984). These are the objects serving as
phase spaces in superclassical mechanics. We shall, for simplicity,
restrict ourselves to the simplest supermanifolds $R^{m \mid n}$ and
take, for the development of supermechanics in the present context,
$ \sca = C^{\infty}(R^{m \mid n})$. The `coordinate variables' $x^a,
\theta^{\alpha}$ will be jointly referred to as $\xi^A$. We shall
write $ \epsilon(\xi^A) = \epsilon_A$. A
*-operation is assumed to be defined on \sca \ for which $ (\xi^A)^*
= \xi^A$.

Left and right differentiations with respect to the odd variables
are defined as follows :
\begin{eqnarray}
\frac{\partial_l}{\partial \theta^{\alpha}}
(\theta^{\alpha_1}..\theta^{\alpha_s})
  =     \delta_{\alpha}^{\alpha_1}\theta^{\alpha_2}..\theta^{\alpha_s}
          - \delta_{\alpha}^{\alpha_2}\theta^{\alpha_1}\theta^{\alpha_3}..
      \theta^{\alpha_s} + .. \nonumber \\
      (-1)^{s-1}\delta_{\alpha}^{\alpha_s}
      \theta^{\alpha_1}..\theta^{\alpha_{s-1}}
\end{eqnarray}
\begin{eqnarray}
\frac{\partial_r}{\partial \theta^{\alpha}}(\theta^{\alpha_1}..\theta^{\alpha_s})
   =       \delta_{\alpha}^{\alpha_s}\theta^{\alpha_1}..\theta^{\alpha_{s-1}}
             - \delta_{\alpha}^{\alpha_{s-1}} \theta^{\alpha_1}..
         \theta^{\alpha_{s-2}} \theta^{\alpha_s} + .. \nonumber \\
         (-1)^{s-1} \delta_{\alpha}^{\alpha_1}
         \theta^{\alpha_2}..\theta^{\alpha_s}
\end{eqnarray}
and extended by linearity to general elements of \sca. Taking

\[ \frac{\partial_l f}{\partial x^a} = \frac{\partial_r f}{\partial x^a}
\equiv \frac{\partial f}{\partial x^a}, \]
we now have left and right derivatives with respect to $ \xi^A $ defined on
\sca. Defining
\[  _Ae = \frac{\partial_l}{\partial \xi^A}, \hspace{.2in}
e_A = \frac{\partial_r}{\partial \xi^A} \]
we have, for any two homogeneous elements f,g of \sca,
\begin{eqnarray}
_Ae (fg) = (_Ae f)g + (-1)^{\epsilon_f \epsilon_A} f (_Ae g);
\nonumber \\
 e_A (fg) = f (e_A g) + (-1)^{\epsilon_g \epsilon_A}
(e_A f) g.
\end{eqnarray}
The objects $_Ae$ (but not $e_A$) are superderivations of the superalgebra
\sca. A general element of \sdera \ (called a supervectorfield) is of the form
\begin{eqnarray}
X = X^A(\xi)  \ _Ae = e_A \  ^AX(\xi).
\end{eqnarray}
The differential of a function $f \in \sca$ can be written as
\begin{eqnarray}
df = d \xi^A \frac{\partial_l f}{\partial \xi^A} =
\frac{\partial_r f}{\partial \xi^A} d \xi^A
\end{eqnarray}
where $ d\xi^A$ are symbols serving as basis vectors in the space of 1-forms.
We have
\begin{eqnarray}
\frac{\partial_r f}{\partial \xi^A} =
(-1)^{\epsilon_A (\epsilon_f +\epsilon_A)}
\frac{\partial_l f}{\partial \xi^A}.
\end{eqnarray}
A general 1-form $\omega^{(1)}$ and a 2-form $\omega^{(2)}$ can be written as
\begin{eqnarray}
\omega^{(1)} = \omega^{(1)}_A d\xi^A = d \xi^A \   _A\omega^{(1)}
\end{eqnarray} \begin{eqnarray}
\omega^{(2)}  =   \omega^{(2)}_{AB}d\xi^B d \xi^A
                   = d \xi^B d \xi^A \   _{AB}\omega
          =  d \xi^A \  _A\omega_B  d\xi^B.
\end{eqnarray}
Note that, when A,B are odd, $\omega^{(2)}_{AB} =
\omega^{(2)}_{BA}$. It follows that, the odd dimension n (in $R^{m
\mid n}$) need not be an even number for a symplectic form to exist;
the number m must, of course, be even.

Given a symplectic form $\omega$ on \sca, we have, for any $ f \in \sca, $
\begin{eqnarray}
\omega(X_f,Y) = - (df)(Y)
\end{eqnarray}
where $X_f$ is the Hamiltonian supervector field corresponding to f
and $ Y \in \sdera. $ Since $ \omega $ is even, Equations (204,205)
give (writing $ f,_A $ for $ \frac{\partial_r f}{\partial \xi^A}$)
\begin{eqnarray}
X_f ^{A}\   _A\omega_B = - f,_B.
\end{eqnarray}
On $R^{m \mid n}$, the symplectic form can be chosen so that the coefficients
$_A\omega_B$ are independent of $\xi$. Assuming this and introducing the
inverse $(^A\omega^B)$ of the matrix $(_A\omega_B)$ :
$_A\omega_B \  ^B \omega^C = \delta_A^C,$ we have
\begin{eqnarray}
X_f^A = - f,_B \  ^B \omega^A.
\end{eqnarray}
The Poisson bracket of $f,g \in \sca$ is
\begin{eqnarray}
\{ f, g \}  =  X_f(g) = X_f^{A}\frac{\partial_l g}{\partial \xi^A}
        =  - \frac{\partial_r f}{\partial \xi^B} \  ^B\omega^A
             \frac{\partial_lg}{\partial \xi^A}.
\end{eqnarray}
The dynamics is governed by the supmech Hamilton equation (64) with
H an even Hermitian element of \sca \ and the Poisson bracket of
Eq.(208).

States in superclassical mechanics are normalized positive linear
functionals on $ \sca = C^{\infty}(R^{m \mid n})$; they are
generalizations of the states (193)  given by
\begin{eqnarray}
\phi(f) = \int_{R^{m \mid n}} f(x, \theta) d \mu (x, \theta)
\end{eqnarray}
where the measure $\mu$ satisfies the normalization and positivity
conditions
\begin{eqnarray}
1 = \phi (1) = \int d \mu (x, \theta)
\end{eqnarray}
\begin{eqnarray}
0 \leq \int f f^* d \mu \hspace{.2in}
\textnormal{for all} \ f \in \sca.
\end{eqnarray}
In the rest of this subsection, we shall consider only states
represented by a density function :
\begin{eqnarray}
d \mu (x, \theta) = \rho (x, \theta) d\theta^1...d\theta^n d^mx.
\end{eqnarray}
To ensure real expectation values for observables, $ \rho(.,.)$ must
be even(odd) for n even(odd). The condition (210) implies that
\begin{eqnarray}
\rho(x, \theta) = \rho_0(x) \theta^n...\theta^1 + \textnormal{terms
of lower order in} \ \theta
\end{eqnarray}
where $ \rho_0$ is a probability density on $ R^m$.

The inequality (211) implies inequalities involving the coefficient
functions on the right in Eq.(213). They eventually determine a
convex domain $ \mathcal{D}$ in a real vector space. Pure states
correspond to points on the boundary  of $ \mathcal{D} $ (which is
generally not a manifold).

The CC condition  is, unfortunately, not generally satisfied by the
pair $ (\oa, \sone(\sca))$ in super-classical mechanics. To show
this, it is  adequate to give an example (Berezin 1987). Taking $
\sca = C^{\infty}(R^{0 \mid 3}) \equiv \mathcal{G}_3$, we have
\begin{eqnarray}
\rho (\theta) = \theta^3 \theta^2 \theta^1 + c_{\alpha}\theta^{\alpha}.
\end{eqnarray}
The inequality (211) with $ f = a \theta^1 + b \theta^2 $ (with a
and b arbitrary complex numbers) implies $c_3 = 0;$ similarly, $ c_1
= c_2 = 0, $ giving, finally
\begin{eqnarray}
\rho(\theta) = \theta^3 \theta^2 \theta^1.
\end{eqnarray}
There is only one possible state which must be pure. This state does not
distinguish, for example, observables $ f = a + b \theta^1 \theta^2 $ with
the same `a' but different `b', thus verifying the assertion made above.

The fermionic extension of classical  mechanics, therefore,
 appears to have a fundamental inadequacy; no wonder,
therefore, that it is not realized by systems in nature.

The argument presented above, however, does not apply to  the $ n =
\infty$ case.

\vspace{.12in} \noindent \textbf{7.8. Quantum-Classical
Correspondence}

\vspace{.12in} \noindent It will now be shown that supmech permits a
transparent treatment of quantum-classical correspondence. In
contrast to the general practice in this domain, we shall be careful
about the domains of operators and avoid some usual pitfalls in the
treatment of the $ \hbar \rightarrow 0 $ limit.

Our strategy will be to start with a quantum Hamiltonian system,
transform it to an isomorphic supmech Hamiltonian system involving
phase space functions and $ \star $-products [Weyl-Wigner-Moyal
formalism (Weyl 1949; Wigner 1932; Moyal 1949)] and show that, in
this latter Hamiltonian system, the subclass of phase space
functions in the system algebra which go over to smooth functions in
the $\hbar \rightarrow 0$ limit yield the corresponding classical
Hamiltonian system. For simplicity, we restrict ourselves to the
case of a spinless nonrelativistic particle though the results
obtained admit trivial generalization to systems with phase space $
R^{2n}$.

 In the existing literature, the works on quantum-classical
correspondence  closest to the present treatment are those of Liu
(1975,1976), Gracia-Bond$\acute{i}$a and V$\acute{a}$rilly (1988)
and H$\ddot{o}$rmander (1979); some results from these works,
especially Liu (1975,1976), are used below. The reference
(Bellissard and Vitot 1990) is a comprehensive work reporting on
some detailed features of quantum-classical correspondence employing
some techniques of noncommutative geometry; its theme, however, is
very different from ours.

In the case at hand, we have the quantum triple $(\sch, \mathcal{D},
\sca)$ where $ \sch = L^2(R^3), \mathcal{D} = \scs(R^3)$ and \sca \
is the algebra of the spinless Galilean particle treated in section
7.4 as a standard quantum system. As in Eq.(190), we shall take the
potential function V to be a function of $\mathbf{X}$ only. For $ A
\in \sca $ and $ \phi,\psi $ normalized elements in $\mathcal{D}$,
 we have the well defined  quantity
\begin{eqnarray}
(\phi,A\psi) = \int \int \phi^*(y) K_{A}(y,y^\prime) \psi(y^\prime)
dydy^\prime
\end{eqnarray}
where the kernel $ K_A$ is a (tempered) distribution. Recalling the
definition of Wigner function (Wigner 1932; Wong 1998) corresponding
to the wave function $ \psi $ :
\begin{eqnarray}
W_\psi(x,p) = \int_{R^3} exp[-ip.y/\hbar] \psi(x + \frac{y}{2})
\psi^*(x-\frac{y}{2})dy
\end{eqnarray}
and defining the quantity  $A_W(x,p)$ by
\begin{eqnarray}
A_W(x,p) = \int exp[-ip.y/\hbar]K_A(x + \frac{y}{2}, x - \frac{y}{2})dy
\end{eqnarray}
(note that $W_\psi$ is nothing but the quantity $P_W$ where P is the
projection operator $|\psi><\psi|$ corresponding to  $\psi$) we have
\begin{eqnarray}
(\psi,A\psi) = \int \int A_W(x,p) W_\psi(x,p) dx dp.
\end{eqnarray}
Whereas the kernels $K_A$ are distributions, the objects $A_W$ are well
defined functions. For example,
\begin{eqnarray*}
A = I : \hspace{.15in} K_A(y,y^\prime) = \delta(y-y^\prime) \hspace{.15in}
A_W(x,p) = 1 \\
A = X_j  : \hspace{.15in} K_A(y,y^\prime) = y_j\delta(y-y^\prime)
\hspace{.15in} A_W(x,p) = x_j \\
A = P_j : \hspace{.15in} K_A(y,y^\prime) = -i\hbar\frac{\partial}
{\partial y_j}\delta(y-y^\prime) \hspace{.15in} A_W(x,p) = p_j.
 \end{eqnarray*}

The Wigner functions $ W_{\psi}$ are generally well-behaved
functions. We shall use Eq.(219) to characterize the class of
functions $A_W$ and call them Wigner-Schwartz integrable (WSI)
functions [i.e. functions integrable with respect to the Wigner
functions corresponding to the Schwartz functions in the sense of
Eq.(217)]. For the relation of this class to an appropriate class of
symbols in the theory of pseudodifferential operators, we refer to
Wong (1998) and references therein.

The operator A can be reconstructed (as an element of \sca) from the function
$A_W$ :
\begin{equation}
\begin{array}{l}
(\phi,A\psi) = \\
\displaystyle (2\pi\hbar)^{-3}\int \int \int exp[ip.(x-y)/\hbar]
\phi^*(x)A_W(\frac{x+y}{2},p)\psi(y)dpdxdy.
\end{array}
\end{equation}

 Replacing, on the right hand side of Eq.(217), the quantity $\psi(x
 +\frac{y}{2})\psi^*(x-\frac{y}{2})$ by $K_{\rho}(x+\frac{y}{2},
 x-\frac{y}{2})$ where $K_{\rho}(.,.)$ is the kernel of the density
 operator $\rho$, we obtain the Wigner function $\rho_W(x,p)$
 corresponding to $\rho$. Eq.(219) then goes over to the more
 general equation
 \begin{eqnarray} Tr(A \rho)  = \int \int A_W(x,p) \rho_W(x,p)dxdp.
 \end{eqnarray} The Wigner function $\rho_W$ is real but generally
 not non-negative.

  Introducing, in $ R^{6}, $ the notations $ \xi $ = (x,p), $ d\xi = dxdp $
and $ \sigma ( \xi, \xi^{'} )
 = p.x^{'} - x.p^{'} $ (the symplectic form in $ R^{6} $ ), we have, for
 A,B $ \in \sca $

\vspace{2mm}
\noindent
\begin{eqnarray}
(AB)_{W}(\xi)
& =  &     (2\pi)^{-6} \int \int  exp [ -i \sigma
                               (\xi - \eta, \tau)] A_{W} ( \eta +
                               \frac{\hbar \tau}{4} ). \nonumber \\
   & \ & . B_{W} (\eta - \frac{\hbar \tau }{4})d\eta d\tau \nonumber \\
                        &  \equiv &  (A_{W} \star B_{W}) ( \xi).
\end{eqnarray}

The product $ \star $ of Eq.(222) is the \emph{twisted product} of
Liu (1975,1976) and the \emph{$\star$- product} of Bayen et al
(1978). The associativity condition $ A(BC) = (AB)C $ implies the
corresponding condition $ A_W \star (B_W \star C_W) = (A_W \star
B_W) \star C_W $ in the space $ \sca_W$ of WSI functions which is a
complex associative non-commutative, unital *-algebra (with the
star-product as product and complex conjugation as involution).
There is an isomorphism between the two star-algebras \sca \ and
$\mathcal{A}_W$ as can be verified from equations (220) and (218).

Recalling that, in the quantum Hamiltonian system $(\sca,
\omega_Q,H)$ the form $\omega_Q$ is fixed by the algebraic structure
of \sca \ and noting that, for the Hamiltonian H of Eq.(111)[with V
= V(\textbf{X})],
\begin{eqnarray}
H_W(x,p) = \frac{p^2}{2m} + V(x),
\end{eqnarray}
we have an isomorphism between the supmech Hamiltonian systems
$(\sca,\omega_Q,H) $ and $ (\sca_W,  \omega_W, H_W)$ where $\omega_W
= -i\hbar \omega_c. $
 Under this isomorphism, the quantum mechanical
PB (160) is mapped to the Moyal bracket
\begin{eqnarray}
\{ A_{W}, B_{W} \}_{M} \equiv
(-i\hbar)^{-1} ( A_{W} \star B_{W} - B_{W} \star A_{W} ).
\end{eqnarray}

  For  functions f,g in $\mathcal{A}_{W} $ which are smooth and
such that   $ f(\xi)$ and $ g(\xi)$ have no $ \hbar-$dependence, we
have, from Eq.(222),
\begin{eqnarray}
f \star g = fg - (i\hbar/2) \{ f, g \}_{cl} + O ( \hbar^{2} ).
\end{eqnarray}
The functions $ A_{W} (\xi) $ will have, in general, some $ \hbar $
dependence and the $ \hbar \rightarrow 0 $ limit may be singular for
some of them (Berry 1991). We denote by $(\mathcal{A}_{W})_{reg}$
the subclass of functions in $ \mathcal{A}_{W} $ whose $ \hbar
\rightarrow 0 $ limits exist and are smooth (i.e. $ C^{\infty} $ )
functions; moreover, we demand that the Moyal bracket of every pair
of functions in this subclass also have smooth limits. This class is
easily seen to be a subalgebra of $ \mathcal{A}_{W}$ closed under
Moyal brackets.  Now, given two functions $A_W$ and $B_W$ in this
class, if $ A_{W} \rightarrow A_{cl} $
 and $ B_{W} \rightarrow B_{cl} $  as $ \hbar \rightarrow 0 $ then
 $ A_{W}  \star B_{W}
\rightarrow A_{cl} B_{cl} $; the subalgebra
$(\mathcal{A}_{W})_{reg}$, therefore, goes over, in the $ \hbar
\rightarrow 0 $ limit , to a subalgebra $\mathcal{A}_{cl}$ of    the
commutative algebra $ C^{\infty}(R^{6}) $ (with pointwise product as
multiplication). The Moyal bracket of Eq.(224) goes over to the
classical PB $\{ A_{cl}, B_{cl} \}_{cl}$; the subalgebra
$\mathcal{A}_{cl}$, therefore, is closed under the classical Poisson
brackets. The classical PB $\{, \}_{cl}$ determines the
nondegenerate classical symplectic form $\omega_{cl}$. When $ H_{W}
\in (\mathcal{A}_{W})_{reg} $[which is the case for the $H_W$ of
Eq.(223)],  the subsystem $ (\mathcal{A}_{W}, \omega_{W}, H_{W}
)_{reg}$ goes over to the supmech Hamiltonian system $
(\mathcal{A}_{cl}, \omega_{cl}, H_{cl})$.

When the $\hbar \rightarrow 0$ limits of $A_W$ and $\rho_W$ on the
right hand side of Eq.(221) exist (call them $A_{cl}$ and
$\rho_{cl}$), we have
\begin{eqnarray} Tr(A \rho) \rightarrow \int \int A_{cl}(x,p)
\rho_{cl}(x,p) dxdp. \end{eqnarray} The quantity $\rho_{cl}$ must be
non-negative (and, therefore, a genuine density function). To see
this, note that, for any operator $A \in \sca$ such that $A_W \in
(\sca_W)_{reg}$, the object $A^*A$ goes over to $\bar{A}_W* A_W$ in
the Weyl-Wigner-Moyal formalism which, in turn, goes to
$\bar{A}_{cl}A_{cl}$ in the $\hbar \rightarrow 0$ limit; this limit,
therefore, maps non-negative operators to non-negative functions.
Now if, in Eq.(226), A is a non-negative operator, the left hand
side is non-negative for an arbitrarily small value of $\hbar$ and,
therefore, the limiting value on the right hand side must also be
non-negative. This will prove the non-negativity of $\rho_{cl}$ if
the objects $A_{cl}$ in Eq.(226) realizable as classical limits
constitute a dense set of non-negative functions in $C^{\infty}(M)$.
This class is easily seen to include non-negative polynomials; good
enough.

In situations where the $\hbar \rightarrow 0$ limit of the time
derivative equals the time derivative of the classical limit [i.e.
we have $A(t) \rightarrow A_{cl}(t)$ and $\frac{dA(t)}{dt}
\rightarrow \frac{d A_{cl}(t)}{dt}$], the Heisenberg equation of
motion for A(t) goes over to the classical Hamilton's equation for
$A_{cl}(t)$. With a similar proviso, one obtains the classical
Liouville equation for $\rho_{cl}$ as the classical limit of the von
Neumann equation.

\vspace{.1in} Before closing this section, we briefly discuss an
interesting point :

  For commutative algebras, the inner derivations vanish and one can
have only outer derivations. Classical mechanics employs a subclass
of such algebras (those of smooth functions on  manifolds). It is an
interesting contrast to note that, while the standard quantum
systems have system algebras with only inner derivations, classical
system algebras have only outer derivations. The deeper significance
of this is related to the fact that the noncommutativity of standard
quantum algebras is tied to the nonvanishing of the Planck constant
$\hbar$. [This is seen most transparently in the star  product of
Eq.(222) above.] In the limit $\hbar \rightarrow 0$, the algebra
becomes commutative (the star product of functions reduces to
ordinary product)and the inner derivations become outer derivations
(commutators go over to classical Poisson brackets implying that an
inner derivation $D_A$ goes over to the Hamiltonian vector field
$X_{A_{cl}}$).

\vspace{.15in} \noindent \textbf{8. MEASUREMENTS IN QUANTUM
MECHANICS}

In this section we shall employ the formalism of section 6.2 to the
treatment of measurements in QM taking both the measured system and
the apparatus to be quantum systems. We shall, however, not adopt
the von Neumann procedure (von Neumann 1955; Wheeler and Zurek 1983)
of introducing vector states for the pointer positions; instead, we
shall assign density operators to the pointer states and exploit the
fact that the apparatus admits a classical description to a very
good approximation. We shall do this by using the phase space
description of the QM of the apparatus (the Weyl-Wigner-Moyal
formalism) and then go to the classical approximation (exploiting
the fact that supmech accommodates both classical and quantum
mechanics as special subdisciplines). The undesirable macroscopic
superpositions (of system + apparatus pure states) are shown to be
suppressed when observations on the apparatus are restricted to
macroscopically distinguishable pointer readings.

We shall start by putting the measurement problem in proper
perspective.

\vspace{.12in} \noindent \textbf{8.1. The measurement problem in
quantum mechanics}

\vspace{.12in} A measurement is an activity in which a system [about
which some information is desired --- to be called the `measured
system' (denoted here as S); it may be microscopic or macroscopic],
prepared in a specified state, is made to interact with a
(generally, but not necessarily, macroscopic) system called the
`apparatus' (denoted here as A) so as to eventually produce a
phenomenon accessible through sensory perception (typically a
pointer reading) or a permanent record (which may be noted at
convenience). The pointer reading or record (the `measurement
outcome') is a numerical value which is interpreted (by employing an
underlying theory and some common sense logic formalizable in terms
of classical physics) as the value of some physical quantity (an
observable). Thus one can talk about measurement of an observable of
a system prepared in a given state.

In quantum theoretic treatments, a \emph{value} of an observable (a
self-adjoint operator) is understood to be a real number in its
spectrum (Omnes 1994, p.115). Supmech events of the type $\nu(E)$ of
section 4.1 can be associated with domains in the spectrum of an
observable A by employing the resolution of identity (172)
corresponding to A :
\begin{eqnarray} \nu(\Delta) = \int_{\Delta} d \mu(\lambda)
|\lambda><\lambda| \end{eqnarray} where $\Delta$ is a measurable
subset of $\Omega = \sigma(A).$ These $\nu(\Delta)$s should, more
appropriately, be called \emph{quantum events}. We have already used
objects of the form (227) [see Eq.(184)]. Given a state $\phi$ in
which the system S is prepared, we have, recalling Eq.(56),
\begin{eqnarray} p_{\phi}(\Delta) = \phi(\nu(\Delta)) \end{eqnarray}
as the probability that, on measurement of the observable A, its
value will be found in the domain $\Delta$. These probabilities are
the  predictions of the underlying theory. Verification of the
theory consists in comparing these probabilities with the
appropriate relative frequencies in repeated measurements with the
system prepared in the same state.

We consider, for simplicity, the measurement of an observable (of a
quantum system S) represented by a self-adjoint operator F (acting
in an appropriate domain in the Hilbert space $\sch_S$ of S) having
a nondegenerate spectrum with the eigenvalue equations $F |\psi_j> =
\lambda_j |\psi_j>$ (j = 1,2,...). The apparatus A is chosen such
that, to each of the eigenvalues $\lambda_j$ corresponds a pointer
position $M_j$. If the system is initially in an eigenstate
$|\psi_j>$, the apparatus is supposedly designed to give, after the
measurement interaction, the pointer reading $M_j$; the outcome of
the measurement is then understood as $\lambda_j$. A question
immediately arises : `What is the measurement outcome when the
initial state of the system S is a superposition state $|\psi> =
\sum_j c_j |\psi_j> $ ?' To find the answer, we must consider the
dynamics of the coupled system (S + A) with an appropriate
measurement interaction.

 The standard treatment of measurements in QM (von Neumann 1955;
 Wheeler and Zurek 1983; Jauch 1968; Omnes 1994; Dass 2005) is due to
von Neumann who emphasized that quantum mechanics being, supposedly,
 a universally applicable  theory, every system is basically
quantum mechanical; to have a consistent theory of measurement, we
must, therefore, treat the apparatus A also quantum mechanically.
Accordingly, one introduces a Hilbert space $\sch_A$ for the
apparatus A; the pointer positions $M_j$ are assumed to be
represented by the state vectors $|\mu_j>$ in this space. The
Hilbert space for the coupled system (S + A) is
 taken to be $\sch = \sch_S \otimes \sch_A$.

The measurement interaction is elegantly described (Omnes 1994; Dass
2005) by a unitary operator U on \sch \ which, acting on the initial
state  of (S+A) (with the system S in the initial state in which it
is prepared for the experiment and the apparatus in the `ready'
state which we denote as $|\mu_0>$) gives an appropriate final
state. We shall assume the measurement to be \emph{ideal} which is
supposedly such that (Omnes 1999) `when the measured system is
initially in an eigenstate of the measured observable, the
measurement  leaves it in the same state.' In this case, the
measurement outcome must be the corresponding eigenvalue which must
be indicated by the final pointer position. This implies
\begin{eqnarray}
U(|\psi_j> \otimes |\mu_0>) = |\psi_j> \otimes |\mu_j>.
\end{eqnarray}
For S in the initial state $ |\psi> = \sum c_j|\psi_j>$, the final (S + A)-
state must be, by linearity of U,
\begin{eqnarray}
|\Psi_f> \  \equiv U[(\sum_j c_j |\psi_j>) \otimes |\mu_0>] = \sum_j
c_j [|\psi_j> \otimes |\mu_j>].
\end{eqnarray}
Note that the right hand side of Eq.(230) is a superposition of the
quantum states of the (generally \emph{macroscopic}) system (S + A).

Experimentally, however, one does not observe such superpositions.
Instead, one obtains, in each measurement, a definite outcome
$\lambda_j$ corresponding to the final (S + A)-state $|\psi_j>
\otimes |\mu_j>$. Repetitions of the experiment, with system in the
same initial state, yield various outcomes randomly such that, when
the number of repetitions becomes large, the relative frequencies of
various outcomes tend to have fixed values. To account for this, von
Neumann postulated that, after the operation of the measurement
interaction as above, a discontinuous, noncausal and instantaneous
process takes place which changes the state $ |\Psi_f>$ to the
state represented by the density operator
\begin{eqnarray}
\rho_f & = &\sum_i \tilde{P}_i |\Psi_f><\Psi_f| \tilde{P}_i \\
       & = & \sum_j |c_j|^2 [|\psi_j><\psi_j| \otimes
       |\mu_j><\mu_j|];
\end{eqnarray}
here $\tilde{P}_i = |\psi_i><\psi_i| \otimes I_A$ where $I_A$ is the
identity operator on $\sch_A$. This is referred to as von Neumann's
\emph{projection postulate} and the phenomenon with the above
process as the underlying process the \emph{state vector reduction}
or \emph{wave function collapse}. Eq.(232) represents, in the von
Neumann scheme, the (S +A)-state on the completion of the
measurement. It represents an ensemble of (S + A)-systems in which a
fraction $p_j = |c_j|^2$ appears in the j th product state in the
summand. With the projection postulate incorporated, the von Neumann
formalism, therefore, predicts that, in a measurement with the
system S initially in  the superposition state as above,

\noindent
(i) the measured values of the observable F are the random numbers
$\lambda_j$ with respective probabilities $|c_j|^2$;

\noindent (ii) when the measurement outcome is $\lambda_j$, the
final state of the system is $|\psi_j>$.

\noindent Both the predictions are in excellent agreement with
experiment.

The main problem with the treatment of a quantum measurement given
above is the ad-hoc nature of the projection postulate. Moreover,
having to invoke a discontinuous, acausal and instantaneous process
is a very unpleasant feature of the formalism. The so-called
measurement problem in QM is essentially the problem of explaining
the final state (232) without introducing anything ad-hoc and/or
physically unappealing in the theoretical treatment. This means that
one should either give a convincing dynamical explanation of the
reduction process or else circumvent it.

A critical account of various attempts to solve the measurement
problem and related detailed references may be found in the author's
article (Dass 2005); none of them can be claimed to have provided a
satisfactory solution. In the following subsection, a
straightforward treatment of a quantum measurement in the supmech
framework is presented, making up for some deficiencies in the von
Neumann treatment.

\vspace{.12in} \noindent 8.2. \emph{Supmech treatment of a quantum
measurement}

\vspace{.12in} We shall now treat the (S +A) system in the supmech
framework of section 6.2 treating both, the system S and the
apparatus A, as quantum Hamiltonian systems. Given the two quantum
triples $(\sch_S, \mathcal{D}_S, \sca_S)$ and $(\sch_A,
\mathcal{D}_A, \sca_A)$ corresponding to S and A, the quantum triple
corresponding to (S+A) is $(\sch_S \otimes \sch_A, \mathcal{D}_S
\otimes \mathcal{D}_A, \sca_S \otimes \sca_A )$.

 Von Neumann's treatment does not do adequate justice to the physics of the
apparatus and needs some improvements. We propose to take into
consideration the following points :

\vspace{.1in} \noindent (i) The apparatus A is a quantum mechanical
system admitting, to a very good approximation, a classical
description. Even when the number of the effective apparatus degrees
of freedom is not large (for example, the Stern-Gerlach experiment,
treated in the next section, where the center of mass position
vector of a silver atom acts as the effective apparatus variable), a
classical description of the relevant variables is adequate. This
feature is of more than academic interest and must be incorporated
in the theoretical treatment.

\vspace{.1in} \noindent (ii) Introduction of vector states for the
pointer positions is neither necessary nor desirable. A better
procedure is to introduce density operators for them and take into
consideration the fact that the Wigner functions corresponding to
them are approximated well by classical density functions.

\vspace{.1in} \noindent (iii) The pointer states have a stability
property : After the measurement interaction is over, the apparatus,
left to itself, settles quickly into one of the pointer positions.
It is this process which should replace von Neumann's
`instantaneous, non-causal and discontinuous' process.

\vspace{.1in} \noindent \emph{Note.} A detailed mathematical
treatment of this process, as we shall see below, is not necessary;
it is adequate to take its effect correctly into account. To get a
feel for this, note that, in, for example, the Stern-Gerlach
experiment, treated in the next subsection, the measurement
interaction is over (ignoring fringe effects) after the atom is out
of the region between the magnetic pole pieces. In this case, by
`the apparatus settling to a pointer position' one means the
movement of the atom from just outside the pole pieces to a
detector. In this case, the choice of the detector is decided by the
location of the atom just after the measurement interaction. Details
of motion of the atom from the magnets to the detector is of no
practical interest in the present context. In the case of a
macroscopic apparatus, the `settling ...' refers to the process of
the apparatus reattaining thermal equilibrium (disturbed slightly
during the measurement interaction) after the measurement
interaction; again, details of this process are not important in the
present context (the eventual pointer position is decided by the
region of the apparatus phase space in which the apparatus happens
to be immediately after the measurement interaction).

\vspace{.1in} \noindent (iv) Observations relating to the apparatus
are restricted to the pointer positions $M_j$. A properly formulated
Hamiltonian dynamics (classical or quantum) which takes this into
consideration (treating the apparatus `respectfully' as a
\emph{system}) would involve, at appropriate stage, averaging over
the inoperative part of the phase space of the apparatus [the
`internal environment' (Dass 2005) of the apparatus]. It is this
averaging, as we shall see below, which produces the needful
decoherence effects (Zurek 2003) to wipe out undesirable quantum
interferences.

\vspace{.1in} \noindent (v) Different pointer positions are
macroscopically distinguishable. We shall take this into
consideration by employing an appropriate energy-time uncertainty
inequality. This condition will be seen to play a crucial role in
the demonstration of the demolition of the unwanted quantum
interference terms.

 A general pointer observable for A is of the form
\begin{eqnarray}
J = \sum_j b_j P_j
\end{eqnarray}
where $P_j$ is the projection operator onto the space of states in
$\sch_A$ corresponding to the pointer position $M_j$ [considered as
an apparatus property; for a detailed treatment of the relationship
between classical properties and quantum mechanical projectors, see
Omnes(1994,1999) and references therein] and $b_j$s are real numbers
such that $b_j \neq b_k$ for $j \neq k$. In purely quantum
mechanical terms, the projector $P_j$ represents the question (von
Neumann 1955; Jauch 1968): `Is the pointer at position $M_j$?' The
observable J has different `values' at different pointer positions.
Since one needs only to distinguish between different pointer
positions, any observable J of the above mentioned specifications
can serve as a pointer observable.

The phase space function $P^W_j$ corresponding to the projector
$P_j$ is  supposedly approximated well by a function $P_j^{cl}$ on
the phase space $\Gamma$ of the apparatus A (the $\hbar \rightarrow
0$ limit of $P^W_j$). Now, in $\Gamma$, there must be
non-overlapping domains $D_j$ corresponding to the pointer positions
$M_j$. In view of the point (iv) above, different points in a single
domain $D_j$ are not distinguished by the experiment. We can,
therefore, take $P_j^{cl}$ to be proportional to the characteristic
function $\chi_{D_j}$ of the domain $D_j$; it follows that the phase
space function $J^W$ corresponding to the operator J above is
approximated well by the classical pointer observable
\begin{eqnarray} J^{cl} = \sum_j b_j^{\prime} \chi_{D_j}
\end{eqnarray} where $b_j^{\prime}$s have properties similar to the
$b_j$s above.

The pointer states $\phi_j^{(A)}$ corresponding to the  pointer
positions $M_j$ are density operators of the form (constant)$P_j$;
the phase space functions corresponding to these states are
approximated well by the classical phase space density functions
$\rho_j^{cl} = V(D_j)^{-1} \chi_{D_j}$ where V(D) is the phase space
volume of the domain D. [Note. If $V(D_j)$ is infinite, one can
treat $\rho_j^{cl}$ as a function on $\Gamma$ which vanishes outside
$D_j$ and varies very slowly in $D_j$.]

We shall take $ H_{int} = F \otimes K$ (absorbing the coupling
constant in K) where F is the measured quantum observable and K is a
suitably chosen apparatus variable . We shall make the usual
assumption that, during the measurement interaction, $H_{int}$ is
the dominant part of the total Hamiltonian $(H \simeq H_{int})$. The
unitary operator U of subsection 8.1 describing the measurement
interaction in the von Neumann scheme must now be replaced by the
measurement operator in supmech [which implements the appropriate
canonical transformation on the states of the (S +A) system] given
by $ M \equiv exp[\tau \tilde{\partial}_H]$ where $\tau = t_f-t_i$
is the time interval of measurement interaction and
$\tilde{\partial}_H$ is the evolution generator in the supmech
Liouville equation (65) [it is the transpose of the operator $
\partial_H \equiv Y_H$, the evolution generator in Eq.(64)].

Assuming, again, that the measurement is ideal and denoting the
`ready state' of the apparatus by $\phi_0^{(A)}$, we have the
following  analogue of Eq.(229):
\begin{eqnarray}
M (|\psi_j><\psi_j| \times \phi_0^{(A)}) =
|\psi_j><\psi_j| \times \phi_j^{(A)}.
\end{eqnarray}
Here and in the following paras, we have identified the quantum
states of the system S with the corresponding density operators.
When the system is initially in the superposition state $|\psi>$ as
above, the initial and final (S+A)- states are
\begin{eqnarray}
\Phi_{in} = |\psi><\psi| \times \phi_0^{(A)}; \ \ \Phi_f = M(\Phi_{in}).
\end{eqnarray}

Note that the `ready' state may or may not correspond to one of the
pointer readings. (In a voltage type measurement,it does; in the
Stern-Gerlach experiment with spin half particles, it does not.) For
the assignment of the $\Gamma$-domain to the `ready' state, the
proper interpretation (which covers both the situations above) of
the ready state is `not being in any of the (other) pointer states'.
Accordingly, we assign, to this state, the domain
\begin{eqnarray}
\tilde{D}_0 \equiv \Gamma - \cup_{j \neq 0}D_j
\end{eqnarray}
where the condition $j \neq 0$ on the right is to be ignored when
the `ready' state is not a pointer state.

We must now take care of the point (iii) above. When the measurement
interaction is over, the apparatus, left to itself, will quickly
occupy, in any single experiment, a pointer position $M_j$
(depending on the region of the phase space $\Gamma$  it happens to
be in after the measurement interaction). For the ensemble of (S +A)
systems described by the initial state $\Phi_{in}$, the final state
(after `settling down') must be of the form
\begin{eqnarray} \hat{\Phi}_f = \sum_j p_j \rho^{(S)}_j \times
\phi_j^{(A)} \end{eqnarray} where $\rho_j^{(S)}$ are some states of
S. Since, during the transition from the state $\Phi_f$ to
$\hat{\Phi}_f$, the system and the apparatus are left to themselves
and only some internal changes take place in the apparatus (which,
in view of the stability property (iii) above, are not expected to
change the expectation value of a pointer observable J), we must
have, for an arbitrary system observable A,
\begin{eqnarray} \hat{\Phi}_f (A \otimes J) = \Phi_f( A \otimes J).
\end{eqnarray}

Now, $\Phi_f = \Phi_f^{\prime} + \Phi_f^{\prime \prime}$ where
\begin{eqnarray}
\Phi_f^{\prime} & = & M \left( \sum_j |c_j|^2 [|\psi_j><\psi_j|
\times
              \phi_0^{(A)}] \right) \nonumber \\
& = & \sum_j |c_j|^2 [|\psi_j><\psi_j| \times
                       \phi_j^{(A)}] \end{eqnarray}
(where we have used the fact that, in supmech,  a canonical
transformation on states preserves convex combinations) and
\begin{eqnarray} \Phi_f^{\prime \prime} = M \left( [\sum_{j \neq k}
c_k^* c_j |\psi_j><\psi_k|] \times \phi_0^{(A)} \right) \equiv M
(R).
\end{eqnarray}
[Note that R, the operand of M, is not an (S +A)-state; here M has
been implicitly extended by linearity to the dual space of the
algebra $\sca_S \otimes \sca_A$.]

We shall now prove that \begin{eqnarray} W \equiv \Phi_f^{\prime
\prime} (A \otimes J ) \simeq 0. \end{eqnarray} Transposing the M
operation to the observables, we have
\begin{eqnarray}
W & = & R[exp(\tau \partial_H) (A \otimes J)] \nonumber \\
  & = & \int_{\Gamma} d \Gamma \rho^{(A)W}_0 \sum_{j \neq k}
        c_k^* c_j <\psi_k|exp(\tau \partial_{H^{\prime}})
        (A \otimes J^W)|\psi_j>
\end{eqnarray}
where we have adopted the phase space description of the QM of the
apparatus, $d \Gamma$ is the phase space volume element,
$\rho^{(A)W}_0$ is the Wigner function corresponding to the state
$\phi^{(A)}_0$ and $H^{\prime} = F\otimes K^W$. Using equations
(154) and (224), we have
\begin{eqnarray}
\partial_{H^{\prime}}(A \otimes J^W)
& = & \{ F \otimes K^W, A \otimes J^W \} \nonumber \\
& = & (-i\hbar)^{-1} \left( [F,A] \otimes \frac{K^W *
                      J^W + J^W * K^W}{2} \right.\nonumber \\
& \ & + \left. \frac{FA + A F}{2}
                      \otimes (K^W * J^W - J^W * K^W) \right).
\end{eqnarray}
Given the fact that the apparatus is well described classically, we
have $K^W \simeq K^{cl}$ and $J^W \simeq J^{cl}$ to a very good
approximation. This gives
\[ \partial_{H^{\prime}} (A \otimes J^W) \simeq
(-i\hbar)^{-1} K^{cl} J^{cl} [F,A] \] which, in turn, implies
\begin{eqnarray*} <\psi_k| exp(\tau \partial_{H^{\prime}})
(A \otimes J^W) |\psi_j> \simeq exp[\frac{i}{\hbar}(\lambda_k -
\lambda_j)K^{cl} \tau] J^{cl} <\psi_k|A|\psi_j>. \end{eqnarray*} We
now have, replacing $\rho^{(A)W}_0$ by its classical approximation
$\rho_0^{(A)cl}$,
\begin{eqnarray}
W \simeq  \int_{\tilde{D}_0} d \Gamma \rho^{(A)cl}_0
    \sum_{j \neq k} c_k^* c_j exp[\frac{i}{\hbar}
    (\lambda_k - \lambda_j)K^{cl}\tau] J^{cl} <\psi_k|A|\psi_j>.
\end{eqnarray}
Let
\begin{eqnarray} <K^{cl}>_0 \  \equiv  \int_{\tilde{D}_0} K^{cl}
\rho_0^{(A)cl} d \Gamma \end{eqnarray} (the mean value of $K^{cl}$
in the domain $\tilde{D}_0$). Putting $K^{cl} = <K^{cl}>_0 s$,
taking s to be one of the integration variables and writing $d
\Gamma = ds d\Gamma^{\prime}$, we have
\begin{eqnarray} W \simeq \int_{\tilde{D}_0} ds d\Gamma^{\prime}
\rho_0^{(A)cl} \sum _{j \neq k} c_k^*c_j
exp[\frac{i}{\hbar}\eta_{jk}s]J^{cl} <\psi_k|A|\psi_j>
\end{eqnarray} where
\begin{eqnarray} \eta_{jk} = (\lambda_k - \lambda_j)<K^{cl}>_0 \tau.
\end{eqnarray}
Note that s is a real dimensionless variable with domain of
integration of order unit length.

We shall now argue that, for $j \neq k,$
\begin{eqnarray}  | \eta_{jk}| > > \hbar. \end{eqnarray}
(This is not obvious; when F is a component of spin, for example,
the $\lambda$s are scalar multiples of $\hbar$.) To this end, we
invoke the apparatus feature (v) above. A fairly straightforward way
of formulating a criterion for macroscopic distinguishability of
different pointer positions would be to identify a quantity of the
dimension of action which could be taken as characterizing the
physical separation between two different pointer positions and show
that its magnitude is much larger than $\hbar$. The objects
$\eta_{jk}$ (for $j \neq k$)are quantities of this type.  The
inequality (249) then follows from the assumed macroscopic
distinguishability of different pointer positions. Another,
essentially equivalent, way of seeing this is to treat Eq.(249) as
the time-energy uncertainty inequality $ |\Delta E \Delta t| > >
\hbar $ where $ \Delta t = \tau$ and $\Delta E$ is the difference
between the energy values corresponding to the apparatus locations
in two different domains $D_j$ and $D_k$ in $\Gamma$. Recalling that
$H \simeq H_{int}$ during the relevant time interval, we have
$\Delta E \simeq (\lambda_k - \lambda_j)<K^{cl}>_0$.

The large fluctuations implied by Eq.(249) wipe out the integral
above giving $ W \simeq 0 $ as desired.

Equations (239), (238) and (242) now give \begin{eqnarray*} 0 & = &
(\hat{\Phi}_f - \Phi_f^{\prime}) ( A \otimes J ) \\
& = & \sum_j \phi_j^{(A)} (J) Tr ([p_j \rho^{(S)}_j - |c_j|^2
|\psi_j><\psi_j|]A) \end{eqnarray*} which must be true for all J
[with arbitrary $b_j$ in Eq.(233) satisfying the stated condition].
This gives \[ Tr([ p_j \rho^{(S)}_j - |c_j|^2 |\psi_j><\psi_j|]A) =
0
\] for all system observables A and, therefore, \[ p_j \rho^{(S)}_j =
|c_j|^2 |\psi_j><\psi_j|. \] Finally, therefore, we have
$\hat{\Phi}_f = \Phi_f^{\prime} $ which is precisely the state
obtained from $\Phi_f$ by applying the von Neumann projection.

This completes the derivation of the von Neumann projection rule.
This has been obtained through straightforward physics; there is no
need to give any ad hoc prescriptions. The derivation makes it clear
as to the sense in which this reduction rule should be understood :
it is a prescription to correctly take into consideration the effect
of the `settling down' of the apparatus after the measurement
interaction for obtaining the final state of the observationally
constrained (S + A) system.

Eq.(247), followed by the reasoning above, represents,  in a
\emph{live} form, the operation of  environment-induced decoherence.
To see this, note that, the domain $\tilde{D}_0$  may be taken to
represent the internal environment of the apparatus. With this
understanding, the mechanism wiping out the unwanted quantum
interference terms is, indeed, the environment-induced decoherence.
In the treatment presented here this mechanism becomes automatically
operative. (Even the external environment can be trivially included
by merely saying that the system A above represents `the apparatus
and its environment'.)

\vspace{.12in} \noindent \textbf{8.3. Example : the Stern-Gerlach
experiment}

\vspace{.12in} As an illustration, we consider the Stern-Gerlach
experiment (Busch, Grabowski and Lahti 1995; Omnes 1994;
Cohen-Tannoudji, Diu and Lalo$\ddot{e}$ 2005) with, say, silver
atoms (which means spin s = $\frac{1}{2}$). A collimated beam of
(unpolarized) silver atoms is made to pass through inhomogeneous
magnetic field after which the beam splits into two beams
corresponding to atoms with $S_z = \pm \frac{\hbar}{2}$. The spin
and magnetic moment operators of an atom are $\mathbf{S} =
\frac{\hbar}{2} \mathbf{\sigma}$ and $\mathbf{\mu} = g \mathbf{S}$
(where g is the magnetogyric ratio). Let the magnetic field be
$\mathbf{B}(\mathbf{r}) = B(z) \mathbf{e_3}$ (in obvious notation).
[Refinements (Potel et al. 2004) introduced to ensure the condition
$ \mathbf{\bigtriangledown.B} =0$ do not affect the essential
results obtained below.] We have
\begin{eqnarray}
H_{int} = -\mathbf{\mu . B} = -g B(z) S_3.
\end{eqnarray}
The force on an atom, according to Ehrenfest's theorem, is
\begin{eqnarray}
\mathbf{F} = - \mathbf{\bigtriangledown} < - \mathbf{\mu . B}> = g
\frac{d B(z)}{dz}<S_3> \mathbf{e_3}
\end{eqnarray}
where the average is taken in the quantum state of the atom. During
the experiment, the internal state of the atom remains unchanged (to
a very good approximation); only its center of mass \textbf{r} and
spin \textbf{S} have significant dynamics. In this experiment, $S_3$
is the measured quantum observable  and \textbf{r} acts as the
operative apparatus variable.

\begin{sloppypar}Let us assume that the beam initially moves in the
positive x-direction, the pole pieces are located in the region $x_1
\leq x \leq x_2$ and the detectors located in the plane $ x = x_3 >
x_2$ (one each in the regions $z >0$ and $z < 0;$ these regions
contain the emergent beams of silver atoms corresponding,
respectively, to $S_3 = +\frac{\hbar}{2}$ and $S_3 = -
\frac{\hbar}{2}$). We have, in the notation used above, $F = S_3$
and K = - g B(z). Assuming the experiment to start when the beam
reaches at $ x = x_1$, the phase space of the apparatus is
\begin{eqnarray}
\Gamma = \{ (x, y, z, p_x, p_y, p_z) \in R^{6}; x \geq x_1, * \}
\end{eqnarray}
where * indicates the restriction that, for $x_1 \leq x \leq x_2, $
the space available for the movement of atoms is the one between the
two pole pieces. For the order of magnitude calculation below, we
shall ignore the shape of the pole pieces and take * to imply $z_1
\leq z \leq z_2$. \end{sloppypar}

The domains $D_1$ and $D_2$ corresponding to the two pointer positions are
\begin{eqnarray*}
D_1 & = & \{ (x, y, z, p_x, p_y, p_z) \in \Gamma; x > x_2, p_z >0 \} \\
D_2 & = & \{ (x, y, z, p_x, p_y, p_z) \in \Gamma; x > x_2, p_z < 0
\};
\end{eqnarray*}
the domain $\tilde{D}_0 = \Gamma -(D_1 \cup D_2)$. For simplicity,
let us take $ B(z) = b_0 + b_1 z$ where $b_0$ and $b_1$ are
constants. For $j \neq k$, we have $\lambda_j - \lambda_k = \pm
\hbar$. The relevant integral is [see Eq.(247) above]
\begin{eqnarray}
I = \int_{z_1}^{z_2} dz (...)exp[\pm \frac{i}{\hbar}\mu b_1 z \tau]
\end{eqnarray}
where $\mu = g \hbar$. Putting $z = (z_2 - z_1) u$, the new
integration variable u is a dimensionless variable taking values in
a domain of length of order one. The quantity of interest is
\begin{eqnarray}
|\eta| = \mu |b_1| (z_2 - z_1) \tau.
\end{eqnarray}
According to the data in (Cohen-Tannoudji, Diu and Lalo$\ddot{e}$
2005) and (Goswami 1992, problem 4.6), we have ($v_x$ is the x-
component of velocity of  the silver atom)
\begin{eqnarray*}
|b_1| \sim |\frac{dB}{dz}| \sim 10^5 \ gauss/cm \\
z_2 - z_1 \simeq 1 mm, \ \ v_x \sim 500 \ m/sec \\
x_2- x_1 = 3 cm, \ \ x_3 - x_2 = 20 \ cm
\end{eqnarray*}
This gives
\begin{eqnarray*}
\tau \sim \frac{x_3 - x_1}{v_x} \sim 5 \times 10^{-4} sec.
\end{eqnarray*}
Denoting the Bohr magneton by $\mu_b$ and putting $ \mu \sim \mu_b
\simeq 0.9 \times 10^{-20}$ erg/gauss, we have $|\eta| \sim 10^{-19}$erg-sec.
With $ \hbar \simeq 1.1 \times 10^{-27}$ erg-sec, we have, finally
$(|\eta|/\hbar) \sim 10^8$, confirming the strong suppression of the
undesirable quantum interferences.

\vspace{.15in} \noindent \textbf{9. AXIOMS}

\vspace{.12in} We shall now write down a set of axioms covering the
work presented in the preceding sections. Before  the statement of
axioms, a few points are in order :

\vspace{.1in} \noindent (i) These axioms are meant to be
provisional; the `final' axioms will, hopefully, be formulated (not
necessarily by the present author) after a reasonably satisfactory
treatment of quantum theory of fields and space-time geometry in an
appropriately augmented supmech type framework  has been given.

\noindent (ii) The terms `system', `observation', `experiment' and a
few other `commonly used' terms will be assumed to be understood.
The term `relativity scheme' employed below will be understood to
have its meaning as explained in section 5.1.

\noindent (iii) The `universe' will be understood as the largest
possible observable system containing every other observable system
as a subsystem.

\noindent (iv) By an \emph{experimentally accessible system} we
shall mean one whose `identical' (for all practical purposes) copies
are reasonably freely available for repeated trials of an
experiment. Note that the universe and its `large' subsystems are
not included in this class.

\noindent (v) The term `system' will, henceforth will normally mean
an experimentally accessible one. Whenever it is intended to cover
the universe and/or its large subsystems (this will be the case in
the first two axioms only), the term system$^*$ will be used.

\vspace{.1in} The axioms will be labeled as \textbf{A1,..., A8}.

\vspace{.1in} \noindent \textbf{A1.}(Probabilistic framework; System
algebra and
states) \\
(a) \emph{System algebra; Observables}. A system$^*$ S has
associated with it a supmech-admissible (as defined in section 4.1)
superalgebra $\sca = \sca^{(S)}.$ (Its elements will be denoted as
A,B,...). Observables of S are elements of the subset  \oa \ of even
Hermitian elements of \sca.

\noindent (b) \emph{States}. States of the system$^*$, also referred
to as the states of the system algebra \sca (denoted by the letters
$\phi, \psi,..$), are defined as (continuous) positive linear
functionals on \sca \ which are normalized [i.e. $\phi(I) =1$ where
I is the unit element of \sca]. The set of states of \sca \ will be
denoted as \scs(\sca) and the subset of pure states by $\sone
(\sca)$. For any $A \in \mathcal{O}(\sca)$ and $ \phi \in \scs
(\sca)$, the quantity $\phi(A)$ is to be interpreted as the
expectation value of A when the system is in the state $\phi$.

\noindent (c) Expectation values of odd elements of \sca \ vanish in
every pure state (hence in every state).

\noindent (d) \emph{Compatible completeness of observables and pure
states}. The pair $(\oa,\  \sone(\sca))$ satisfies the CC condition
described in section 4.1.

\noindent (e) \emph{Experimental situations and probabilities}. An
experimental situation (relating to observations on the system$^*$
S) has associated with it a positive observable-valued measure
(PObVM) as defined in section 4.1; it associates, with measurable
subset of a measurable space, objects called supmech events which
have measure-like properties. Given the system prepared in a state
$\phi$, the probability of realization of a supmech event $\nu (E)$
is $\phi(\nu (E))$.

 \vspace{.12in} \noindent \textbf{A2.} \emph{Differential calculus;
 Symplectic structure}. The system algebra \sca \ of a system$^*$ S is such as
to permit the development of derivation-based differential calculus
based on it; moreover, it is equipped with a real symplectic form
$\omega$ thus constituting a symplectic superalgebra $ (\sca,
\omega)$ [more generally, a generalized symplectic superalgebra
$(\sca, \scx, \omega)$ when the derivations are restricted to a
distinguished Lie sub-superalgera \scx \ of \sdera].

\vspace{.12in} \noindent \textbf{A3.} \emph{Dynamics}. The dynamics
of a system S is described by a one-parameter family of  canonical
transformations generated by an even Hermitian element H (the
Hamiltonian) of \sca \ which is bounded below in the sense of
section 4.2.

The mechanics described by the above-stated axioms will be referred
to  as supmech. The triple $(\sca, \omega, H)$ [or the quadruple
$(\sca, \sone(\sca),\omega, H)]$ will be said to constitute a
supmech Hamiltonian system.

\vspace{.12in} \noindent \textbf{A4.} \emph{Relativity scheme}. (a)
For systems admitting space-time description, the `principle of
relativity' as described in section 5.1,will be operative. \\
(b)The admissible relativity schemes will be restricted to (i)
Galilean relativity, (ii) special relativity.

\vspace{.12in} \noindent \textbf{A5.} \emph{Elementary systems;
Material particles}. (a) In each of the admitted relativity schemes,
material particles will  be understood to be localizable elementary
systems (as defined in sections 4.7 and 5.1) with positive mass. \\
(b) The system algebra for a material particle will be (the
topological completion of) the one generated by its fundamental
observables (as defined in section 5.1) and the identity element.

\vspace{.12in} \noindent \textbf{A6.} \emph{Coupled systems}. Given
two systems $S_1$ and $S_2$ described as supmech Hamiltonian systems
$(\sca^{(i)}, \scs_1^{(i)}, \omega^{(i)}, H^{(i)})$ (i=1,2), the
coupled system $(S_1 + S_2)$ will be described as a supmech
Hamiltonian system $(\sca, \scs_1, \omega, H)$ with
\[ \sca = \sca^{(1)} \otimes \sca^{(2)}, \ \ \scs_1 = \scs_1(\sca),
 \ \ \omega = \tilde{\omega}^{(1)} + \tilde{\omega}^{(2)}
\] [see Eq.(134)] and H as in Eq.(156).

\vspace{.12in} \noindent \textbf{A7.} \emph{Quantum systems}.  All
(experimentally accessible) systems in nature have noncommutative
system algebras (and hence are quantum systems); they have a quantum
symplectic structure (as defined in section 3.9) with the universal
parameter $b = -i \hbar$.

\vspace{.1in} \noindent \emph{Note.} (i) The subclass of quantum
systems having finitely generated system algebras were shown (in
section 7.3) to have Hilbert space based realizations (without
introducing  additional postulates --- explicitly or implicitly).

\vspace{.1in} \noindent (ii) Axioms A7 and A5(a) imply that all
material particles are localizable elementary quantum systems. Since
they have finitely generated system algebras, the corresponding
supmech Hamiltonian systems are guaranteed to have Hilbert space
based realizations. They can be treated as in section 7.4 without
introducing any extra postulates; in particular, introduction of the
Schr$\ddot{o}$dinger wave functions with the traditional Born
interpretation and the Schr$\ddot{o}$dinger dynamics are automatic.

\vspace{.1in} \noindent (iii) General quantum systems were shown in
section 7.5 to admit superselection rules.

\vspace{.12in} \noindent \textbf{A8.} \emph{Measurements}. In a
measurement involving a `measured system' S and apparatus A \\
(a) both S and A are standard quantum systems (as defined in section 7.1); \\
(b) the Hamiltonian system $(\sca^{(A)}, \sone^{(A)}, \omega^{(A)},
H^{(A)})$ corresponding to the apparatus admits an equivalent (in
the sense of section 4.3) phase space realization (in the
Weyl-Wigner-Moyal scheme) $(\sca_W^{(A)}, \mathcal{S}_{1W}^{(A)},
\omega_W^{(A)}, H_W^{(A)})$; \\
(c) elements of $\sca_W^{(A)}$ and $\mathcal{S}^{(A)}_{1W}$
appearing in the description of dynamics of the coupled system (S+A)
admit $\hbar
\rightarrow 0$ limits and are approximated well by these limits; \\
(d) the various pointer positions of the apparatus (i) have the
stability property as stated in section 8.2 and (ii) are
macroscopically distinguishable [the macroscopic distinguishability
can be interpreted, for example, in terms of an energy-time
uncertainty product inequality ($\Delta E \Delta t > > \hbar$)
relevant to the experimental situation]; \\
(e) observations on the apparatus are restricted to  readings of the
output devices (pointers).

\vspace{.1in} \noindent \emph{Note}. Part (b) in this axiom is
expected to be redundant and is included `to be on the safe side'.
(A redundancy is excusable if it serves to bring some extra clarity
without introducing any inconsistency.)

 \vspace{.15in} \noindent \textbf{10. CONCLUDING REMARKS}

\vspace{.12in} \noindent 1. Sometimes the question is raised : `Why
algebras ?' The answer emerging from the present work is : `Because
they provide the right framework for noncommutative symplectic
geometry as well as for noncommutative probability and, therefore,
are natural objects to employ in the construction of a formalism
integrating the two [in the spirit of unification of physics and
probability theory envisaged in the formulation of (augmented)
Hilbert's sixth problem]'.

\vspace{.1in}  Indeed, as we have seen, for an autonomous
development of quantum mechanics, the fundamental objects are
algebras and not Hilbert spaces.

\vspace{.12in} \noindent 2. A contribution of the present work
expected to be of some significance to the algebraic schemes in
theoretical physics and probability theory is  the introduction of
the condition of compatible completeness for observables and pure
states [axiom A1(d)] which plays an important role in ensuring that
the so-called `standard quantum systems' defined algebraically in
section 7.1, have faithful Hilbert space-based realizations. It is
desirable to formulate necessary and/or sufficient conditions on the
superalgebra \sca \ alone (i.e. without reference to states) so that
the CC condition is automatically satisfied.

An interesting result, obtained in section 7.7, is that the
superclassical systems with a finite number of fermionic generators
generally do not satisfy the CC condition. This probably explains
their non-occurrence in nature. It is worth investigating whether
the CC condition is related to some stability property of dynamics.

\vspace{.12in} \noindent
3. Some features of the development of QM
in the present work (apart from the fact that it is autonomous)
should please theoreticians : there is a fairly broad-based
algebraic formalism connected smoothly to the Hilbert space QM;
there is a natural place for commutative superselection rules and
for the Dirac's bra-ket formalism; the Planck constant is introduced
`by hand' at only one place (at just the right place : the quantum
symplectic form) and it appears at all conventional places
automatically. Moreover, once the concepts of localization,
elementary system and standard quantum system  are introduced at
appropriate places, it is adequate to define a material particle as
a localizable elementary quantum system ; `everything else' ---
including the interpretation of the Schr$\ddot{o}$dinger wave
function and the Schr$\ddot{o}$dinger equation --- is automatic.

\vspace{.12in} \noindent 4. The treatment of quantum-classical
correspondence in section 7.8, illustrated with the example of a
nonrelativistic spinless particle, makes clear as to how the subject
should be treated in the general case :  go from the traditional
Hilbert space -based description of the quantum system to an
equivalent (in the sense of a supmech hamiltonian system) phase
space description in the Weyl-Wigner-Moyal formalism, pick up the
appropriate subsets in the observables and states having smooth
$\hbar \rightarrow 0$ limits and verify that the limit gives a
commutative  supmech Hamiltonian system (which is generally a
traditional classical hamiltonian system).

\vspace{.12in} \noindent 5. It is of some interest to note that, in
the formalization of a measurement situation in supmech in section
4.1, the concept of PObVM naturally appears; it is the abstract
counterpart (in fact, a generalization) of the concept of POVM which
has been dealt with by quantum theorists in various contexts during
the past three decades.

\vspace{.12in} \noindent 6. The treatment of measurements in QM in
section 8 provides a justification for the collapse postulate
through straightforward physics : it is the prescription for
obtaining the effective final state of an observationally
constrained coupled system. The sight of Eq.(247) where one can see
the operation of the decohering  effect of `internal environment' of
the apparatus in live action, should please theoreticians. The
incorporation of the external environment (for the restricted
purpose of realizing von Neumann reduction) has been reduced to a
matter of less than two lines : just saying that the symbol `A' now
stands for `the apparatus and its environment'.

\vspace{.12in} \noindent 7. The title of the present paper was
originally intended to be, as announced in Ref.[47,49], `Supmech: a
unified symplectic view of physics'; while preparing the final
version of it, however, the author felt that the present title would
represent the contents better.

\vspace{.2in} \noindent \textbf{ACKNOWLEDGEMENTS}

 The work relating to the present article was spread
over about one and a half decade and six institutions: Indian
Institute of Technology, Kanpur, Inter-University Centre for
Astronomy and Astrophysics, Pune, Chennai Mathematical Institute,
Chennai, Institute of Mathematical Sciences, Chennai, Indian
Statistical Institute, Delhi Centre, New Delhi and Centre for
Theoretical Physics, Jamia Millia Islamia, New Delhi. The author
thanks these institutions for support and research facilities. Among
individuals, the author would like to thank M. Dubois-Violette  and
M.J.W. Hall for their critical comments on Ref.[56] and [47]
respectively, to K.R. Parthasarathy, R. Sridharan and V. Balaji for
helpful discussions and to J.V. Narlikar, N.K. Dadhich, C.S.
Seshadri , R. Balasubramanian, K.R. Parthasarathy , R. Bhatia, A.G.
Bhatt, P. Sharan and M. Sami for facilitating his post-retirement
research work.

\vspace{.2in} \noindent \textbf{REFERENCES} {\footnotesize
\begin{description}
\item  Abraham R,  Marsden, J.E. (with the assistance of T.Ratiu
and R. Cushman): Foundations of Mechanics, second edition. Benjamin,
Massachusetts (1978).
\item Accardi, L.,\emph{Phys. Rep.} \textbf{77}, 169 (1981).
\item Alonso, L.M.,\emph{J. Math. Phys.} \textbf{18}, 1577 (1977).
\item Alonso, L.M. \emph{J. Math. Phys.} \textbf{20}, 219 (1979).
\item $\acute{A}$lvarez, M.C. :`Variations on a theme of Gel'fand
and Naimark', arXiv : math.FA/0402150 (2004).
\item Antoine, A.P., \emph{J. Math. Phys.} \textbf{10}, 53,2276 (1969).
\item Araki, H.: ``Mathematical theory of quantized fields'',
Oxford Univ. Press (1999).
\item Arnold, V.I. :``Mathematical Methods of Classical Mechanics'',
Springer-Verlag, New York (1978).
\item Bacry, H. :``Localizability and Space in Quantum Physics'',
Lecture Notes in Physics, vol 308, Springer-Verlag, Berlin (1988).
\item Bargmann, V.,\emph{Ann. Math.} \textbf{59},1 (1954).
\item Bargmann, V, Wigner, E.P.,\emph{Proc. Nat. Acad. Sc.}
\textbf{34}, 211 (1948).
\item Bayen, F. et al, \emph{Annals of Phys.} \textbf{110}, 61,111 (1978).
\item Bell, J. : ``Speakable and Unspeakable in Quantum Mechanics'',
Cambridge University Press (1988).
\item Bellissard, J., Vitot, M., \emph{Ann. Inst. Henri
Poincar$\acute{e}$} \textbf{52}, 175 (1990).
\item Berezin, F.A. :``Superanalysis'', edited by A.A. Kirillov,
 D. Reidel Pub. Co., Dordrecht (1987).
\item Berezin, F.A., Marinov, M.S.,\emph{Ann. Phys.}
\textbf{104}, 336 (1977).
\item Berndt, R. : ``An Introduction to Symplectic Geometry'',
American Mathematical Society, Providence (2001).
\item Berry, M. :`Some Quantum-Classical Asymptotics' in
``Chaos and Quantum Physics'',  Les Houches, session LII, 1989, J.
Elsevier Science Publishers (1991).
\item Bogolubov, N.N., Logunov,A.A., Oksak, A.I., Todorov,
I.T. :``General Principles of Quantum Field Theory'', Kluwer
Academic Publishers (1990).
\item B$\ddot{o}$hm, A. :``The Rigged Hilbert Space and Quantum
Mechanics'', Lecture Notes in Physics, vol 78, Springer, Berlin
(1978).
\item Bohm,D., Hiley, B.J. :``The Undivided Universe: An
Ontological Interpretation of Quantum Theory'', Routledge, Chapman
and Hall, London (1993).
\item Born,M., \emph{Z. Physik} \textbf{37}, 863 (1926).
\item Born, M., Heisenberg, W., Jordan, P.,\emph{Zs. f. Phys.}
\textbf{35}, 557 (1926).
\item Born M., Jordan, P.,\emph{Zs. f. Phys.} \textbf{34}, 858 (1925).
\item Bratteli, O., Robinson, D.W. : ``Operator algebras and
quantum statistical mechanics I. $C^*-$ and $W^*$-algebras, symmetry
groups, decomposition of states'', Springer, New York (1979; II.
``Equilibrium States; Models in Quantum Statistical Mechanics'',
ibid (1981).
\item Busch, P., Grabowski, M., Lahti, P.J. :``Operational Quantum
Physics'', Springer-Verlag, Berlin (1995).
\item Cari$\tilde{n}$ena, J.F., Santander, M.,\emph{J. Math.
Phys.} \textbf{16}, 1416 (1975).
\item Caro,J., Salcedo, L.L. :``Impediments to mixing
classical and quantum dynamics'' arXiv : quant-ph/98120402 v2
(April, 1999).
\item Cartan, H., Eilenberg, S. : ``Homological Algebra'',
Princeton University Press (1956).
\item Cohen-Tannoudji, C., Diu B., Lalo$\ddot{e}$ F :
``Quantum mechanics'', vol I, John Wiley and Sons (2005).
\item Connes, A. :``Noncommutative  geometry'', Academic Press, New
York (1994).
\item Dass T. : ``Noncommutative geometry and unified formalism for
classical and quantum mechanics'', Indian Institute of Technology
Kanpur preprint, unpublished (1993).
\item Dass T. :``Symmetries, gauge fields, strings and
fundamental interactions'', vol. I: ``Mathematical techniques in
gauge and string theories'', Wiley Eastern Limited, New Delhi
(1993).
\item Dass T. :``Towards an autonomous formalism for quantum
mechanics'', arXiv: quant-ph/0207104 (2002).
\item Dass T. : ``Histories approach to quantum mechanics''
arXiv : quant-ph/0501161 (2005).
\item Dass T. : `Measurements and Decoherence', arXiv:
quant-ph/0505070 (2005).
\item Dass T. :  ``Consistent quantum-classical interaction
and solution of the measurement problem'' arXiv : quant-ph/0612224
(2006).
\item Dass T. : ``Universality of quantum symplectic structure'',
 \newline arXiv : 0709.4312 [math.SG] (2007).
\item Dass T., Joglekar, Y.N., \emph{Annals of Physics}
\textbf{287} (2001), 191.
\item Dass T., Sharma S.K. : ``Mathematical Methods in
Classical and Quantum Physics'', Universities Press, Hyderabad
(1998).
\item de la Madrid R. : ``The role of the rigged Hilbert space in
Quantum Mechanics'',  ArXiv : quant-ph/0502053 (1905).
\item DeWitt B. : ``Supermanifolds'', Cambridge University Press (1984).
\item Dirac P.A.M., \emph{Proc. Roy. Soc.} \textbf{A 109}, 642 (1926).
\item Dirac P.A.M. : ``Principles of Quantum Mechanics'',
 Oxford University Press, London (1958).
\item Djemai A.E.F. : ``Introduction to Dubois-Violette's Noncommutative
Differential Geometry'' , \emph{Int. J. Theor. Phys.} \textbf{34},
801 (1995).
\item Dubin D.A., Hennings M.A. : ``Quantum Mechanics, Algebras
and Distributions'' Longman Scientific and Technical, Harlow
(1990).
\item Dubois-Violette M. : ``Noncommutative differential geometry,
quantum mechanics and gauge theory'', in Lecture Notes in Physics,
vol. 375 Springer, Berlin (1991).
\item Dubois-Violette M. : ``Some aspects of noncommutative
differential geometry'', arXiv: q-alg/9511027 (1995). \item
Dubois-Violette M. : ``Lectures on graded differential algebras and
noncommutative geometry'', arXiv: math.QA/9912017 (1999).
\item Dubois-Violette M., Kerner R., Madore J., Class. Quant. Grav.
\textbf{8}, 1077 (1991).
\item Dubois-Violette M., Kerner R., Madore, J.,\emph{J. Math.
Phys.} \textbf{31}, 316 (1994).
\item Dubois-Violette M., Kriegl A., Maeda Y., Michor P.W. :
``Smooth algebras'', arXiv : math.QA/0106150 (2001).
\item Emch G.E. : ``Algebraic Methods in Statistical
Mechanics and Quantum Field Theory'', Wiley, New York (1972).
\item Emch G.E. : ``Mathematical and Conceptual Foundations of
20th Century Physics'', North-Holland, Amsterdam (1984).
\item Gelfand I.M., Vilenkin, N.J. : ``Generalized Functions'',
vol. IV, Academic Press, New York (1964).
\item  Glimm J., Jaffe A. : ``Quantum Physics: a Functional
Integral Point of View'', Springer Verlag, New York (1981).
\item Goswami A. : ``Quantum mechanics'', Wm. C. Brown Publishers
(1992).
\item Gracia-Bond$\acute{i}$a J.M., V$\acute{a}$rilly J.C.,
\emph{J.Phys.A: Math.Gen.}\textbf{21}, L879 (1988).
\item Gracia Bondia J.M., Varilly J.C., Figuerra H. : ``Elements
of Noncommutative Geometry'', Birkha$\ddot{u}$ser, Boston (2001).
\item Greub W. : ``Multilinear Algebra'', Springer-Verlag, New York (1978).
\item Griffiths R.B., \emph{J. Stat. Phys.} \textbf{36},219 (1984).
\item Griffiths R.B. : ``Consistent Quantum Theory'',
Cambridge University Press (2002).
\item Grosse H., Reiter G. : ``Graded differential geometry of
graded matrix algebras'',  arXiv: math-ph/9905018 (1999).
\item Guillemin V., Sternberg S. : ``Symplectic Techniques in
Physics'', Cambridge University Press (1984).
\item Haag R. : ``Local Quantum Physics'', Springer, Berlin, 1992.
\item Haag R., Kastler D., \emph{J. Math. Phys.} \textbf{5}, 848
(1964).
\item Heisenberg W., \emph{Zs. f. Phys.} \textbf{33}, 879 (1925).
\item Ya. Helemskii A. : ``The Homology of Banach and Topological
Algebras'', Kluwer Academic Publishers (1989).
\item Hilbert D. : ``Mathematical Problems'', lectures delivered before
the International Congress of Mathematicians in Paris in 1900,
translated by M.W. Newson, Bull. Amer. Math. Soc. \textbf{8}, 437
(1902).
\item Hochscild G., Serre J-P., \emph{Ann. of Math.} \textbf{57} (1953), 591.
\item Holevo A.S. : ``Probabilistic and Statistical Aspects of
Quantum Theory'', North Holland Publishing Corporation, Amsterdam
(1982).
\item H$\ddot{o}$rmander L., \emph{Comm. Pure Appl. Math.}
\textbf{32}, 359 (1979).
\item Horuzhy S.S. : ``Introduction to Algebraic Quantum Field
Theory'', Kluwer Academic Publishers, Dordrecht (1990).
\item Houtappel R.M.F., Van Dam H., Wigner E.P., \emph{Rev. Mod.
Phys.} \textbf{37}, 595 (1965).
\item Iguri S., Castagnino M., \emph{Int. J. Theor. Phys.}
\textbf{38}, 143 (1999).
\item Inoue A. : ``Tomita-Takesaki Theory in Algebras of Unbounded
Operators'', Springer, Berlin (1998).
\item Jammer M: \emph{The Philosophy of Quantum Mechanics},
New York: John Wiley and Sons (1974).
\item Jauch J.M. : ``Foundations of Quantum Mechanics'', Addison
Wesley, Reading, Mass., 1968.
\item Jauch J.M., Misra B., \emph{Helv. Phys. Acta}
\textbf{34}, 699 (1961).
\item Jordan P., von Neumann J., Wigner E., \emph{Ann. Math.}
\textbf{35} (1934), 29.
\item Kerner R., \emph{J. Geom. Phys.} \textbf{11} (1993), 325.
\item Kristensen P., Mejlbo L., Thue Poulsen E., \emph{Comm.
Math. Phys.} \textbf{1}, 175 (1965).
\item Landi G. : ``An introduction to noncommutative spaces and
their geometries'', Springer, Berlin (1997).
\item Lassner G., \emph{Physica} \textbf{124A} (1984), 471.
\item Leites D.A. : ``Introduction to the theory of supermanifolds'',
\emph{Russ. Math. Surveys} \textbf{35},1 (1980).
\item Liu K.C., \emph{J. Math. Phys.}  \textbf{16}, 2054 (1975).
\item Liu K.C., \emph{J. Math. Phys.}\textbf{17}, 859 (1976).
\item Ludwig G. : ``An Axiomatic basis for Quantum Mechanics'',
Springer-Verlag, Berlin (1985).
\item Mackey G.W., Proc. Nat. Acad. Sci. U.S. \textbf{35}, 537 (1949).
\item Mackey G.W. : ``Mathematical Foundations of Quantum
Mechanics'', Benjamin-Cummings, Reading, Mass. (1963).
\item Madore J. : ``An Introduction to Noncommutative Geometry and
its Applications'', Cambridge University Press (1995).
\item Manin Y.I. : ``Gauge Fields and Complex geometry'',
 Springer, Berlin (1988).
\item P -A Meyer P. -A. : ``Quantum Probability for Probabilists'', second
edition, Springer-Verlag, Berlin (1995).
\item Moyal J.E., \emph{Proc. Camb. Phil. Soc.} \textbf{45}, 99 (1949).
\item Newton T.D., Wigner E.P., \emph{Rev. Mod. Phys.}
\textbf{21}, 400 (1949).
\item Omn$\grave{e}$s R., \emph{Rev. Mod. Phys.} \textbf{64}, 339
(1992).
\item Omnes R. : ``The Interpretation of Quantum Mechanics'',
Princeton University Press (1994).
\item Omnes R. : ``Understanding Quantum Mechanics'', Princeton
University Press (1999).
\item Parthasarathy K.R.: ``An Introduction to Quantum Stochastic
Calculus'', Birkha$\ddot{u}$ser, Basel (1992).
\item Peres A. : ``Quantum Theory : Concepts and Methods'',
Kluwer  Academic, Dordrecht (1993).
\item Potel G. et al, arXiv : quant-ph/0409206 (2004).
\item Powers R.T. : ``Self-adjoint algebras of unbounded
operators'', \emph{Comm. Math. Phys.} \textbf{21}, 85 (1971).
\item Roberts J.E., \emph{J. Math. Phys.} \textbf{7}, 1097 (1966).
\item Rudin W. : ``Functional Analysis'', Tata
McGraw-Hill, New Delhi (1974).
\item Scheunert M. : ``The theory of Lie superalgebras'', Lecture
Notes in Mathematics, 716, Springer-Verlag, Berlin (1979).
\item Scheunert M., \emph{J. Math. Phys.} \textbf{20}, 712 (1979);
 \textbf{24}, 2658 (1983);  \textbf{39}, 5024 (1998).
\item Schr$\ddot{o}$dinger E., \emph{Annalen der Physik} \textbf{79}
 (1926), 361.
\item Segal I.E., \emph{Ann. of Math.} \textbf{48}, 930 (1947).
\item Segal I.E. : ``Mathematical Problems of Relativistic Physics''
(with an appendix by G.W. Mackey), Amer. Math. Soc., Providence
(1963).
\item Souriau J.-M. : ``Structure of Dynamical Systems,
a Symplectic View of Physics'', Birkh$\ddot{a}$user, Boston (1997).
\item E.C.G. Sudarshan E.C.G., Mukunda N. : ``Classical Dynamics : A
Modern Perspective'' Wiley, New York (1974).
\item Varadarajan V.S. : ``Geometry of Quantum Theory'', 2nd ed.
 Springer-Verlag, New York (1985).
\item von Neumann J. : ``Mathematical Foundations of Quantum
Mechanics'', Princeton University Press (1955).
\item Vorontov T. : ``Geometric Integration Theory on
Supermanifolds'', \emph{Sov. Sci. Rev. C Mat. Phys.} \textbf{9},
1-138 (1992).
\item Weibel C.A. : ``An Introduction
to Homological Algebra'', Cambridge University Press (1994).
\item Weyl H. : ``Theory of Groups and Quantum Mechanics'',
Dover, New York (1949).
\item Wheeler J.A., Zurek W.H. : ``Quantum Theory and
Measurement'', Princeton University Press (1983).
\item Wightman A.S., Rev. Mod. Phys. \textbf{34}, 84 (1962).
\item Wightman A.S. : ``Hilbert's sixth problem: mathematical
treatment of the axioms of physics'', \emph{Proceedings of Symposia
in Pure Mathematics} \textbf{28}, 147 (1976).
\item Wigner E.P., \emph{Phys. Rev.} \textbf{40}, 749 (1932).
\item Wigner E.P., \emph{Ann. Math.(N.Y.)} \textbf{40}, 149 (1939).
\item Wigner E.P. : ``Group Theory and its Application to
the Quantum Mechanic of Atomic Spectra'', Academic Press, New York
(1959).
\item Wong M.W. : ``Weyl Transforms'', Springer, New York (1998).
\item Woodhouse N. : ``Geometric Quantization'' Clarendon Press,
Oxford (1980).
\item Zurek W. H. : `Decohernce, Einselection and Quantum Origin of
the Classical', \emph{Rev. Mod. Phys.} \textbf{75}, 715 (2003).

\end{description}}

\end{document}